\documentclass[useAMS,usenatbib,usegraphicx]{mn2e}

\usepackage{graphicx}
\usepackage[latin1]{inputenc}
\usepackage{color}
\usepackage{times}
\usepackage{natbib}
\usepackage{setspace}
\newif\ifAMStwofonts
\AMStwofontstrue
\definecolor{red}{rgb}{1,0.,0.}

\newcommand{\be}{\begin{equation}}
\newcommand{\ee}{\end{equation}}
\newcommand{\bea}{\begin{eqnarray}}
\newcommand{\eea}{\end{eqnarray}}
\newcommand{\msunyr}{M$_\odot$ yr$^{-1}$}
\newcommand{\msun}{M$_\odot$}
\newcommand{\kms}{km s$^{-1}$}

\title[Faint Lyman-Break galaxies] {Faint Lyman-Break galaxies as a crucial test
  for galaxy formation models}
\author[Lo Faro et al.]{Barbara Lo Faro$^{1,3}$, 
  Pierluigi Monaco$^{2,3}$, Eros Vanzella$^2$, Fabio Fontanot$^4$, Laura~Silva$^2$,
  \newauthor{Stefano~Cristiani$^2$}\\
  $^1$Max-Planck-Institut for Extraterrestrial Physics, Giessenbachstrasse, 85748 Garching, Germany\\
  $^2$INAF-Osservatorio Astronomico di Trieste, via Tiepolo 11, 34131 Trieste, Italy \\
  $^3$Dipartimento di Astronomia, Universit\`a di Trieste, via Tiepolo 11, 34131 Trieste, Italy \\
  $^4$Max-Planck-Institute for Astronomy, K\"onigstuhl 17, D-69117, Heidelberg, Germany \\
  email: lofaro@mpe.mpg.de; monaco, vanzella, silva, cristiani@oats.inaf.it; 
  fontanot@mpia-hd.mpg.de}

\begin{document}

\date{Accepted ... Received ...}

\pagerange{\pageref{firstpage}--\pageref{lastpage}} \pubyear{}

\maketitle
\label{firstpage}

\begin{abstract}

It has recently been shown that galaxy formation models within
  the $\Lambda$CDM cosmology predict that, compared to the observed
  population, small galaxies (with stellar masses $<10^{11}$ {\msun})
  form too early, are too passive since $z\sim3$ and host too old
  stellar populations at $z=0$.  We then expect an overproduction of
  small galaxies at $z\ga4$ that should be visible as an excess of
  faint Lyman-break galaxies. To check whether this excess is
  present, we use the {\sc morgana} galaxy formation model and {\sc
  grasil} spectro-photometric $+$ radiative transfer code to generate
mock catalogues of deep fields observed with HST-ACS.  We add
observational noise and the effect of Lyman-alpha emission, and
perform color-color selections to identify Lyman-break galaxies.  The
resulting mock candidates have plausible properties that closely
resemble those of observed galaxies.  We are able to reproduce the
evolution of the bright tail of the luminosity function of Lyman-break
galaxies (with a possible underestimate of the number of the brightest
$i$-dropouts), but uncertainties and degeneracies in dust absorption
parameters do not allow to give strong constraints to the model.
Besides, our model shows a clear excess with respect to observations
of faint Lyman-break galaxies, especially of $z_{850}\sim27$
$V$-dropouts at $z\sim5$.  We quantify the properties of these
``excess'' galaxies and discuss the implications: these galaxies are
hosted in dark matter halos with circular velocities in excess of 100
km s$^{-1}$, and their suppression may require a deep re-thinking
  of stellar feedback processes taking place in galaxy formation.

\end{abstract}


\section{Introduction}
\label{s:introduction}

Despite the large observational and theoretical efforts of the last
decade, the formation of galaxies in the context of the $\Lambda$CDM
cosmology is still an open problem.  In fact, the wealth of
multi-wavelength data on deep fields, coupled with wide local surveys,
has produced new constraints on number densities, stellar masses, star
formation rates and metallicities of galaxies in the redshift range
from 0 to $\sim7$.  From the phenomenological point of view, a
remarkable behaviour of observed galaxies is their ``downsizing''
trend in star formation rates, stellar masses, chemical enrichment,
reconstructed age of stellar populations, reconstructed duration of
the main star formation episode \citep[see references
in][]{Fontanot09b}.  For ``downsizing'' here we mean that more massive
galaxies tend to be older than less massive ones in terms of either
stellar ages or mass assembly, a behaviour that is at variance with
that of dark matter halos.

These constraints are a
strong challenge to theoretical models of galaxy formation. In 
a recent paper, \cite{Fontanot09b} compared the predictions
of three galaxy formation models \citep{Wang08, Monaco07,
  Somerville08} with observational data which showed
downsizing trends.  This analysis allowed the authors to highlight a significant
discrepancy between models and data that pertains to relatively small
galaxies, with stellar masses in the range $10^9-10^{10.5}$ {\msun}.
In all three models these galaxies are formed at high redshift and are
rather old and passive at late times, while the population of real
galaxies in the same stellar mass range builds up at low $z$, growing
in number by a factor of $\sim3$ from $z\sim1$ to the present time, and
shows higher specific star formation rates up to $z\sim3-4$.  Similar
conclusions are obtained considering the age of stellar populations in
local galaxies: real galaxies host much younger stars on average than
model galaxies.  The consistency of results coming from stellar
masses, star formation rates and reconstructed stellar ages makes this
trend robust.  Previous comparisons, focused on stellar mass
downsizing, showed that this discrepancy is not limited to the three
models tested in that paper but extends to many semi-analytic and
N-body codes \citep[see the comparisons to models in][]{Fontana06,
  Cirasuolo08, Marchesini08}.

This is a severe discrepancy because, as we will discuss later in this
paper, it cannot be easily solved by a simple fine-tuning of
parameters.  Also, it is not directly related to the well-known
tendency of semi-analytic models to produce steep luminosity
functions, because all the model stellar mass functions have roughly the
correct slope at $z=0$, or to the well-known problem of overquenching
of satellite galaxies, which are too red and passive in models
\citep[see, e.g., the recent analysis of][]{Kimm09}.  Clearly the
solution of this problem amounts to understanding why do small
galaxies form so early and why do they stop forming stars later.  The
second problem is most likey a byproduct of the first one: once too
many small galaxies are produced at high redshift, strong stellar
feedback must be invoked to suppress their star formation at lower
redshift, so as to recover the correct number density at $z=0$.
Then, the crucial objects that are at the origin of this discrepancy
are the small star-forming galaxies at $z>3$.  Unfortunately, with
presently available facilities only the rest-frame UV radiation of
such galaxies is accessible; these are visible as
distant faint Lyman-break galaxies. Their name takes origin from the
technique used to select them, also known as ``dropout
technique''. This technique selects galaxies over a specific redshift
interval by detecting the drop of the UV flux observed in
correspondence with the Lyman limit at $\lambda$ $<$ 912 {\AA}, the
wavelength below which the ground state of neutral hydrogen may be
ionized.

In this paper we use the {\sc morgana} + {\sc grasil} model of galaxy
formation to generate predictions of $B$-, $V$- and $i$-dropouts to be
compared with the results of (i) the GOODS-S photometric and
spectroscopic data by \cite{Giavalisco04} and \cite{Vanzella08}, (ii)
the GOODS-MUSIC sample \citep{Grazian06} with photometric redshifts,
(iii) the results of \cite{Bouwens07}, where many HST-ACS deep fields
were analysed and scaled to the depth of the Hubble Ultra Deep Field
\citep[HUDF,][]{Beckwith07}.

Several previous theoretical studies addressed the properties of
Lyman-break galaxies using both semi-analytical models
\citep{Somerville01, Poli01, Poli03, Blaizot04, Baugh05, Mao07, Gao08,
  Overzier08,Lee08,Guo08} and cosmological simulations \citep{Nagamine02,
  Weinberg02, Harford03, Nagamine04, Finlator06, Night06, Nagamine08}.  These
papers checked how models within the $\Lambda$CDM cosmology are able
to reproduce the properties of Lyman-break galaxies and their
clustering, often reporting satisfactory agreement.
\noindent
Despite some authors \citep[namely][]{Poli01,Poli03,Night06,Finlator06} noticed that models tend to overproduce faint high-redshift Lyman Break Galaxies with respect to observations, no quantitative claim had been possible at that time, given the uncertainties on the observational constraints. As an example, \cite{Finlator06} found that their predicted steep faint-ends of the luminosity function were still in agreement with the available data, which spanned a range of values from $-2$.2 to $-1.6$. This wide range has recently been revised by \citep{Bouwens07}, so that the most recent determinations give a slope $\sim1.7 \pm 0.1$ from $z\sim2$ to $z\sim6$, which implies a disagreement with the theoretical predictions.

The aim of this paper is a careful revision of this comparison, in the light of the improved observational constraints, based on the generation of reliable and improved mock galaxy catalogues through state-of-the-art semi-analytical and spectro-photometric models. This approach will allow us to test the impact of the selection criteria, as well as different theoretical ingredients (with particular emphasis on the modeling of dust attenuation). Moreover, we aim to include this comparison in a cosmological context, and to discuss its implications for current models of galaxy formation and the so-called "downsizing" scenario (see e.g. \cite{Fontanot09a}).

The plan of the paper is the following. Section~\ref{s:model} presents
the model, which contains some improvements with respect to previous
versions. Section~\ref{s:deepfields} describes the adopted procedure
to select Lyman-break galaxies from the simulated galaxy catalogues
and to tune the free parameters connected to dust extinction.
Section~\ref{s:results} presents the comparison of model and data in
terms of number counts, redshift and color distributions 
and luminosity functions of $B$,$V$ and $i$-dropouts, while
Section~\ref{s:inxs} quantifies the properties of the model ``excess''
galaxies and discusses future observational tests that can help
shedding light on this population. Section~\ref{s:conclusions} gives
the conclusions.

\section{The {\sc morgana} + {\sc grasil} model}
\label{s:model}

\subsection{\sc morgana}
\label{s:morgana}

The MOdel for the Rise of GAlaxies aNd Agns ({\sc morgana}) is
described in full detail in \cite{Monaco07}.
The model follows the
general scheme of semi-analytic models \citep[see][for a
  review]{Baugh06}; here we only mention how the main processes are
implemented.
(i) Each dark matter halo hosts a central galaxy and many satellite
galaxies, contained within dark matter substructure that has not yet
been destroyed by orbital decay due to dynamical friction.  Baryons
are subdivided into stellar halo, bulge and disc components.  Each component
has cold, hot and stellar phases.
(ii) Merger trees of dark matter halos are obtained using the {\sc
  pinocchio} tool (\citeauthor{Monaco02}\ 2002;\ \citeauthor*{MTT02}\
2002;\ \citeauthor{Taffoni02}\ 2002).
(iii) After a merging of dark matter halos, galaxy merging times are
computed using the results of \cite{Taffoni03}, which are tuned on
N-body simulations that include dynamical friction, tidal stripping
and tidal shocks.
(iv) At each merger a fraction of the satellite stellar mass is moved
to the diffuse stellar component of the stellar halo, as explained in
\cite{Monaco06}.
(iv) The evolution of the baryonic components and phases is performed
by numerically integrating a system of equations for all the mass,
energy and metal flows.
(v) The intergalactic medium infalling on a dark matter halo is shock-heated to a
specific thermal energy equal to 1.2 times that of dark matter.
(vi) At each time-step of the numerical integration, the hot gas
component is assumed to be in hydrostatic equilibrium with the dark
matter halo.
(vii) Cooling flows are computed with a new cooling model, tested
against simulations of isolated halos in \cite{Viola08}, which takes
into account mass and energy input from feedback sources.
(viii) The cooling gas is let infall on the central galaxy on a
dynamical time-scale.  It is divided between disc and bulge according
to the fraction of the disc that lies within the half-mass radius of
the bulge.
(ix) The gas infalling on the disc keeps its angular momentum; disc sizes
are computed with an extension of the Mo, Mao \& White (1998) model that
includes the contribution of the bulge to the disc rotation curve.
(x) Disc instabilities and major mergers of galaxies lead to the
formation of bulges.  In minor mergers the satellite mass is given to
the bulge component of the larger galaxy.
(xi) Star formation and feedback in discs are inserted following the
model of \cite{Monaco04}, while in bulge components the
\cite{Kennicutt98} law is used.  In both cases, hot gas is ejected to
the halo (in a hot galactic wind) at a rate equal to the
star-formation rate.
(xii) Feedback from star formation heats the hot halo phase.  When its
temperature gets significantly higher than the virial one, it leaves
the dark matter halo in a galactic hot super-wind.  The matter is
re-acquired later, when the virial temperature of the descendant halo
gets as high as that of the hot superwind gas.
(xiii) In star-forming bulges cold gas is ejected in a cold galactic
wind by kinetic feedback due to the predicted high level of turbulence
driven by SNe.  Analogously to the hot superwind, cold halo gas can be
ejected out of the dark matter halo if its kinetic energy is high
enough.
(xiv) Accretion of gas onto massive black holes (starting from small
seeds present in all galaxies) is connected to the ability of cold gas
to loose angular momentum by some (unspecified) mechanisms driven by
star formation.  This is explained in full detail in
\cite{Fontanot06}.
(xv) The kind of feedback from the AGN depends on the accretion rate
in Eddington units: whenever this is higher than 1 per cent the AGN can
trigger a massive galaxy wind, \cite[see][]{Monaco05} which leads to
the complete removal of Inter-Stellar Medium (ISM) from the galaxy, while at lower accretion
rates the energy is ejected through jets that feed back on the hot
halo gas; this can lead to quenching of the cooling flow.

\subsection{The improvements with respect to the earlier version}
\label{s:improvements}

The version of the model used here presents a few improvements with
respect to that of \cite{Monaco07}.  The most
important difference lies in the treatment of star formation in the
bulge component.  In the earlier version, the size of the starburst,
used to compute the gas surface density and then the star formation
rate through the \cite{Kennicutt98} relation, was assumed to be equal
to the bulge effective radius $R_b$. As a consequence, a starburst
declines very slowly because the residual gas has a decreasingly low
surface density, and then a long star formation time-scale.  This is a
clearly naive estimate, because the dissipational gas settles in the
centre of the galaxy.  In this version we assume that the size of the
starburst is determined by the level of turbulence that the gas
acquires.  The velocity dispersion $\sigma_{\rm cold}$ due to
turbulence is:

\be \sigma_{\rm cold} = \sigma_0 \left( \frac{t_\star}{1\ {\rm Gyr}}
\right)^{-1/3} \label{eq:turb}\ee

\noindent
where the parameter $\sigma_0$ is set to 60
\kms. \cite{Fontanot06} show that this value allows a good
reproduction of the evolution of the AGN population. The time-scale
of star formation $t_\star$ is:

\be t_\star = 4\;  \left(\frac{\Sigma_{\rm cold}}{1\ {\rm M}_\odot\ 
{\rm pc}^{-2}}\right)^{-0.4} \ {\rm Gyr} \label{eq:schmidt} \ee

\noindent
where $\Sigma_{\rm cold}$ is the (cold) gas surface density.  The size
of the startburst is determined as the radius for which the
turbulence-driven velocity dispersion equals that of the host bulge
computed at the same radius.  For the Young mass profile
used in the model, the bulge velocity curve is roughly constant (assuming a value
$\sim0.707 V_{\rm bulge}$, where $V_{\rm bulge}$ is the circular
velocity of the bulge) down to 0.04 times the effective radius $R_b$, where it
quickly drops to zero.  Consequently, the size of starbursts in bulges
is estimated as:

\be R_{\rm gas} = \left( \frac{\sigma_{\rm cold}(R_b)}
{0.707 V_b} \right)^{3.75} R_b \label{eq:compgas}\ee

\noindent
where $\sigma_{\rm cold}(R_b)$ is computed using $R_b$ to determine
the surface density (equation~\ref{eq:schmidt}).  The radius $R_{\rm
  gas}$ is not allowed to be smaller than $0.04R_b$.

A second difference is that we optimized this version of the model for
a \cite{Chabrier03} mass function instead of the Salpeter one used
before. To achieve the same level of agreement with data as with Salpeter
we perform the following parameter changes (please refer
to \cite{Monaco07} for all details): (i) the stellar mass per
supernova is $M_{\rm \star,SN}=84$ \msun, the restoration rate is
$f_{\rm rest}=0.44$, while the metal yield per generation is set to
$Y=0.016$; (ii) the preferred values of some parameters change to:
$n_{\rm quench}=0.3$, $n_{\rm dyn}=0.3$, $f_{\rm wind}=2.0$, $f_{\rm
  th,B}=0.1$, $f_{\rm th,D}=0.32$, $f_{\rm jet,0}=3.0$.  
Quasar-triggered galaxy winds are present as in \cite{Fontanot06};
dark matter halo concentrations are not rescaled to improve the fit of the
baryonic Tully-Fisher relation.  Moreover, the threshold for disc
instability is relaxed from $\epsilon_{\rm limit}=0.9$ to 0.7; this
allows to remove the excess of small bulges present in the previous
version of the model \citep[see Figure~7 of][]{Monaco07}.

Finally, to slow down the merger-driven evolution of the most massive
galaxies, at each merger a fraction $f_{\rm scatter}=0.8$ of the
stellar mass of each satellite is given to the halo diffuse stellar
component; however, this is done only at $z<1$, as suggested by
\cite{Conroy07}.  This is not a physically acceptable solution of the problem
raised by \cite{Monaco06}, but has the advantage of giving realistic
results.

\subsection{\sc grasil}
\label{s:grasil}

For each galaxy modeled by {\sc morgana} the corrisponding UV-to-radio
SED is computed with the spectro-photometric $+$ radiative transfer
code {\sc grasil} \citep[][please refer to this paper for a detailed
  description of this code]{Silva98}.  According to this model,
stars and dust are distribuited in a bulge (King profile) + disc
(radial and vertical exponential profiles) axisymmetric
geometry. Three different dusty environments are taken into account:
(1) dust in interstellar HI clouds heated by the general interstellar
radiation field of the galaxy (the ``cirrus'' component), (2) dust
associated with star-forming molecular clouds and HII regions, and (3)
circumstellar dust shells produced by the windy final stages of
stellar evolution. These environments have different importance in
different galactic systems at various evolutionary stages. Therefore
the residual gas fraction in the galaxy is divided into two phases:
the dense giant molecular clouds (MCs) and the diffuse medium
(``cirrus''). The fraction of gas in the molecular phase, $f_{\rm
  mol}$, is a model parameter.  The molecular gas is subdivided into
clouds of given mass and radius: the stars are assumed to be born
within the optically thick MCs and gradually to escape from them, as
they get older, on a time-scale $t_{\rm esc}$.  To compute the emitted
spectrum of the star-forming molecular clouds a radiative transfer
code is used. The diffuse dust emission is derived by describing the
galaxy as an axially symmetric system, in which the local dust
emissivity is consistently computed as a function of the local field
intensity due to the stellar component.

For each galaxy modelled by {\sc morgana}, star formation
  histories, together with gas and metal contents at the
  ``observation'' time for both bulge and disc component, are then
  given as an input to {\sc grasil}. 
  Unobscured stellar SEDs are generated by using again a
  \cite{Chabrier03} IMF with stellar masses in the limit [0.15-120] \msun.
  In the radiative transfer computation the effect of dust is limited to absorption;
  dust emission, which is very expensive from the point of view of
  computational time, is irrelevant in the rest-frame UV spectral
  regions sampled here.  Absorption from the Inter-Galactic
  Medium (IGM) is taken into account by applying to the redshifted SEDs the
  \cite{Madau95} attenuation, as explained in \cite{Fontanot07}.

Here below the {\sc grasil} parameters that are not provided by
MORGANA and which play an important role in the dust parametrization
are briefly discussed. We also recall the values used in
  \cite{Fontanot07} to compare with $K$-band and sub-mm selected
  galaxies.
 
 (i) $t_{esc}$ represents the time-scale for young O-B stars to escape
from their parent MCs. In \cite{Fontanot07} we fixed the value of
$t_{esc}$ to $10^{7} yr.$. This is an intermediate value between those
found by \cite{Silva98} to well describe the SED of spirals ($\sim$ a
few Myr) and starburst ($\sim$ a few 10 Myr), and it is of the order
of the estimated destruction time-scale of MCs by massive stars.
(ii) The cold gas mass provided by MORGANA is subdivided
  between the dense and diffuse phases. In \cite{Fontanot07} we fixed
  the fraction of gas in MCs $f_{\rm mol}$ to $0.5.$.
  (iii) The total (i.e., cirrus + MCs) dust content of a galaxy ISM is
  obtained by the residual gas mass and the dust-to-gas mass ratio
  $\delta$ which is assumed to scale linearly with the metallicity
  ($\delta=0.45Z $). The optical depth of MCs depends on their mass
  and radius only through the relation $\tau\propto\delta
  M_{MC}/r^{2}_{MC},$.  In \cite{Fontanot07} we assumed $M_{MC} =
    10^{6} M_{\odot}$, $r_{MC} = 16 pc$.
(iv) The bulge and disc scale radii for stars and gas are given
  by MORGANA, while the disc scale-heights, $h^{*}_{d}$ and
  $h^{d}_{d}$ for stars and dust, respectively, are set as in
  \cite{Fontanot07} to $0.1$ the corresponding scale radii. The dust
  grain size distribution and abundances are set so as to match the
  mean Galactic extinction curve and emissivity as in \cite{Silva98}
  and \cite{Vega05} and are not varied here.

  Because the rest-frame UV emission is especially sensitive to dust
  attenuation, there is no guarantee that the parameters used in
  \cite{Fontanot07} are suitable for Lyman-break galaxies
  \citep[see][for a discussion on dust attenuation at low and high
  redshift]{Fontanot09a}.  In this paper we will explore the effect
  of changing the values of the two parameters $t_{esc}$ and $f_{mol}$
  keeping fixed the other ones. These parameters, in fact, by
  definition, are those which can play a fundamental role in the
  attenuation of the UV emission from young massive stars. This will
  be discussed in more detail in the next section.

\subsection{Lyman-$\alpha$ emission line}
\label{s:lya}

Many Lyman-break galaxies show a prominent Lyman-$\alpha$
  emission line, which may influence galaxy colors.  This is
  especially true at the highest redshift, where the $i$-$z$ color can
  get significantly bluer, so that the selection of a galaxy as an
  $i$-dropout is shifted to higher redshift. Modeling the effect of
  such emission is very difficult due to the resonant scattering of
  Lyman-$\alpha$ photons by neutral hydrogen which implies that even a
  small amount of dust is sufficient to absorb many emitted photons
  and attenuate the Lyman-$\alpha$ line.  The observed lines are
  generally ascribed to the presence of galactic winds which allow
  Lyman-$\alpha$ photons to escape the galaxy after a few scatterings.
  Radiative transfer calculations of Lyman-$\alpha$ through winds have
  shown that this process can explain the observed asymmetric line
  profiles \citep[e.g.][]{Shapley03,Vanzella06,Verhamme08}.  Due to
  this complexity, we prefer to make an empirical estimate of the
  possible influence of the Lyman-$\alpha$ line on galaxy colors and
  number counts.

To this aim, we add a Lyman-$\alpha$ emission line to the produced
spectra, with specified $EW$.  Given the relatively coarse wavelength
sampling used (sufficient to compute magnitudes over standard
filters), this line is added to one single wavelength bin.  With the
VLT/FORS2 spectroscopic campaign of the GOODS-S field the number of
Lyman-break candidates with observed Lyman-$\alpha$ emission lines is
25 over 85 for the $B$-dropouts, 19 over 52 for the $V$-dropouts, 24
over 65 for the $i$-dropouts.  The median $EW$ for the three cases are
$15^{+19}_{-12}$ {\AA}, $22^{+56}_{-12}$ {\AA} and $64^{+100}_{-48}$ {\AA}
\citep{Vanzella09}.  These numbers should be taken with a grain of
salt, because most dropout candidates (mainly the fainter ones) are
safely confirmed only when a Lyman-$\alpha$ emission line is found, so
it is not easy either to compute an unbiased estimate of the fraction
of emitters with respect to {\em confirmed} dropouts or to assess the
completeness limit on the $EW$ distribution.  As a rough guess we
assume that $1/3$ of the galaxies have emission lines, and assume for
the $EW$ either the median values reported above for the three dropout
categories, applying them at redshifts $[3,4.5]$, $[4.5,5.5]$ and
$[5.5,\infty]$, or the $V$-dropout value of 22 {\AA} for all emitters.
These values are roughly consistent with those assumed by
\cite{Bouwens07} in their analysis.

We will show result of this procedure only for one combination of
  {\sc grasil} parameters; analogous results are obtained in other
  cases.

\section{Simulations of deep fields}
\label{s:deepfields}

To run {\sc morgana} we use merger trees from a $1000^3$ particles {\sc
  pinocchio} run in a 200 Mpc box with the WMAP3 \citep{Spergel05}
cosmology $\Omega_0=0.24$, $\Omega_\Lambda=0.76$, $\Omega_b =0.0456$,
$H_0=73$ km/s/Mpc, $\sigma_8=0.8$, $n_s=0.96$.  The power spectrum was
computed using the fit by \cite{Eisenstein98}.  The particle mass in
this run is $2.84 \cdot 10^8$ {\msun}, so the smallest used dark matter halo is
$1.42\cdot10^{10}$ {\msun} (50 particles) and the smallest resolved
progenitor is $2.84\cdot10^9$ {\msun} (10 particles).

To simulate a deep field, we need to pass from a time sampling of
galaxies in the box to a pencil beam where time is translated to
redshift, then compute galaxy SEDs in the observer frame.

We transform the output from {\sc morgana} into a pencil beam as
follows.  The model gives as an output, for each galaxy and in a fixed
time grid, its physical properties like mass and metal content of each
phase in each component, bulge and disc sizes and circular velocities,
bulge and disc star formation rates \citep[see][for
  details]{Fontanot07}. The time grid length has been set to $\Delta t =
10$ Myr, and the run has been stopped at $z=3$.  

As specified in Paper I, the mass function of dark matter halos is
sparse-sampled, computing the evolution of at most 300 halos per bin
of 0.5 dex in log mass. This results in an oversampling of small
satellite galaxies compared to the central galaxies of similar stellar
mass.  To correct this oversampling, satellites are further
sparse-sampled so as to be as abundant, at the final redshift ($z=3$),
as the corresponding central galaxies of the same stellar mass. An
approximately constant number of galaxies in logarithmic intervals of
mass is thus obtained from this procedure. This way, the number of
produced galaxies is anyway very high, because each galaxy can be
``observed'' at each output time, i.e. each 10 Myr. A further sparse
sampling (1 over 100) is then applied to all galaxies, so as to select
a sufficiently large sample of roughly 35,000 galaxies.  Each of these
sparse samplings define a weight for the galaxies, equal to the
inverse of the selection probability.  All the statistical quantities
are then computed by weighted sums over the galaxies. Our procedure,
explained in full detail in \cite{Fontanot07}, is different from,
e.g., \cite{Kitzbichler07}, where information on the spatial position
of the halos in the simulated box is retained.  Our procedure has the
advantadge of being simpler and, by averaging over the box (redshift
does not depend on the position of the galaxy in the box), giving results
insensitive to large scale structure, while the procedure of
\cite{Kitzbichler07} is necessary to address, for instance, the clustering of
high-redshift galaxies.

The galaxy sample is then transformed into a pencil beam by assigning
to each galaxy ``observed'' at time $t$ a redshift $z$ and proper,
luminosity and diameter distances $r(z)$, $d_L$ and $d_D$.  The
comoving volume associated to each galaxy is computed as the comoving
area of the simulation box times the $\Delta r(z)$ distance associated
to the cosmic time interval $\Delta t$ of 10 Myr.  The solid angle to
be associated to the galaxy is that subtended by the box at the
``observed'' redshift $z$.

The apparent (AB) magnitudes of the galaxies are computed by
convolving the resulting redshifted SEDs from GRASIL with the ACS
filters $B_{435}$, $V_{606}$, $i_{775}$ and $z_{850}$ (in the
following for simplicity we drop the pedices from the magnitudes, with
the exception of $z_{850}$ in order to avoid confusion with the
redshift).

In order to better reproduce the observed number counts of Lyman-break
galaxies and their color distributions, photometric scatter has been
associated to each galaxy magnitude.  Modeling this scatter is
important because of the tendency for fainter, lower signal-to-noise
sources to scatter into the selection through a Malmquist-like effect.
To quantify this scatter it is necessary, first of all, to obtain an
estimation of the flux error associated to the apparent magnitude
of the observed galaxies.

Observational errors are assigned to magnitudes on the basis of the
ACS GOODS catalogue, version 2.0.  Details of the ACS observations, as
well as major features of the GOODS project, can be found in
\cite{Giavalisco04}; additional information about the latest v2.0
release of the GOODS ACS images and source catalogues can be found at
{\tt http://www.stsci.edu/science/goods/v2.0} and will be described in detail
in an upcoming paper.  
Here we only mention that we use the best guess for the total
  magnitude for comparison with the model, while galaxy colours of
  real galaxies are always computed using isophotal magnitudes.
Source detection for this catalog has been
performed on the $z_{850}$ band, the isophotes defined by the
$z_{850}$ image have been then used as apertures for all other bands.
Using this catalogue we compute for each band ($B$, $V$, $i$,
$z_{850}$) the average signal-to-noise ratio as a function of the
AB apparent magnitude of the observed galaxies.  For $B$,
$V$ and $i$ bands this has been done down to 1-$\sigma$ detection
limit, because it is important to set proper upper limits in order to
select faint drop-outs.  For the reference band, $z_{850}$, where by
definition the sources are detected above 5-$\sigma$, we use the $i$
band error to have a good guess of the 1-$\sigma$ error.
Then, given the AB magnitude of a model galaxy, its flux $f$ is
perturbed with a Gaussian-distributed random error with width
$\sigma_f$ equal to the corresponding average 1-$\sigma$ error flux
computed above from the GOODS catalogue.  Fluxes are finally
transformed back to magnitudes.  Whenever the theoretical magnitude is
larger (fainter) than that corresponding to $S/N=1$, which means that
such objects are not detectable in that band ($S/N<1$), we put their
magnitude equal to the corresponding 1-$\sigma$ limit. The new
magnitude represents therefore the lower limit, (upper limit if we are
looking at the flux of the object), at 1-$\sigma$ to the real value.
The corresponding 1-$\sigma$ limit magnitudes are 30.45 ($B_{435}$), 30.56
($V_{606}$) and 30.19 ($i_{775}$ and $z_{850}$).

This procedure is tuned to reproduce ACS deep fields, so we will
compare our predictions to the spectroscopic sample of GOODS-S
Lyman-break galaxies \citep{Vanzella09}, to the GOODS-MUSIC sample
with photometric redshifts \citep{Grazian06}, which is based on the version 1.0 of the ACS GOODS catalogue
including deep $J$ and $K$ data of VLT-ISAAC and 
 mid-IR data of Spitzer-IRAC, with
photometric redshifts, and with the results of \cite{Bouwens07}.
The latter authors measured number counts and luminosity functions of
Lyman-break galaxies in various ACS deep fields, homogenizing the
samples and scaling all results to the depth of the HUDF
\citep{Beckwith07}.

Selection of Lyman break galaxies is then applied to the apparent
magnitudes computed as explained above.  We use the following criteria
respectively for the $B$-, $V$- and $i$-dropout \citep{Giavalisco04}:

\bea
\lefteqn{(B-V > 1.1)\  {\rm AND}\  [B-V > (V-z_{850} )+ 1.1]}\nonumber
\\&& {\rm AND}\ (V-z_{850} ) < 1.6 
\eea
\bea
\lefteqn{\{ [V-i > 1.5 + 0.9(i-z_{850} )]\ {\rm OR}\ (V-i > 2 )
  \}}\nonumber \\&&
{\rm AND}\ (V-i > 1.2)\
{\rm AND}\ (i-z_{850} ) < 1.3
\eea
\be
 (i-z_{850}) > 1.3\  {\rm AND}\  S/N(B)<1\  {\rm AND}\  S/N(V)<1
\ee

\subsection{Tuning {\sc grasil}}
\label{s:params}

\begin{figure*}
  \centerline{
    \includegraphics[width=6cm]{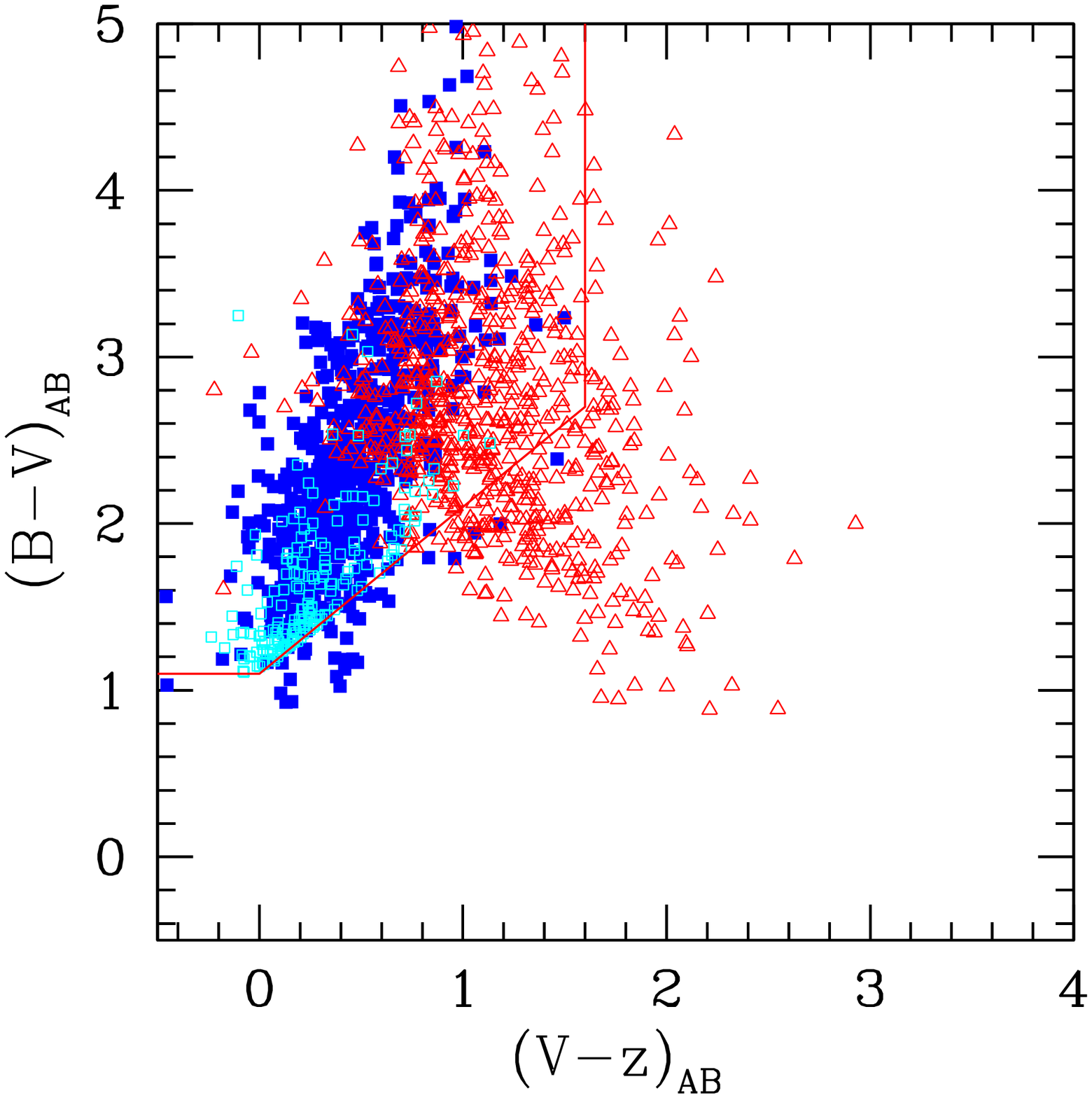} 
    \includegraphics[width=6cm]{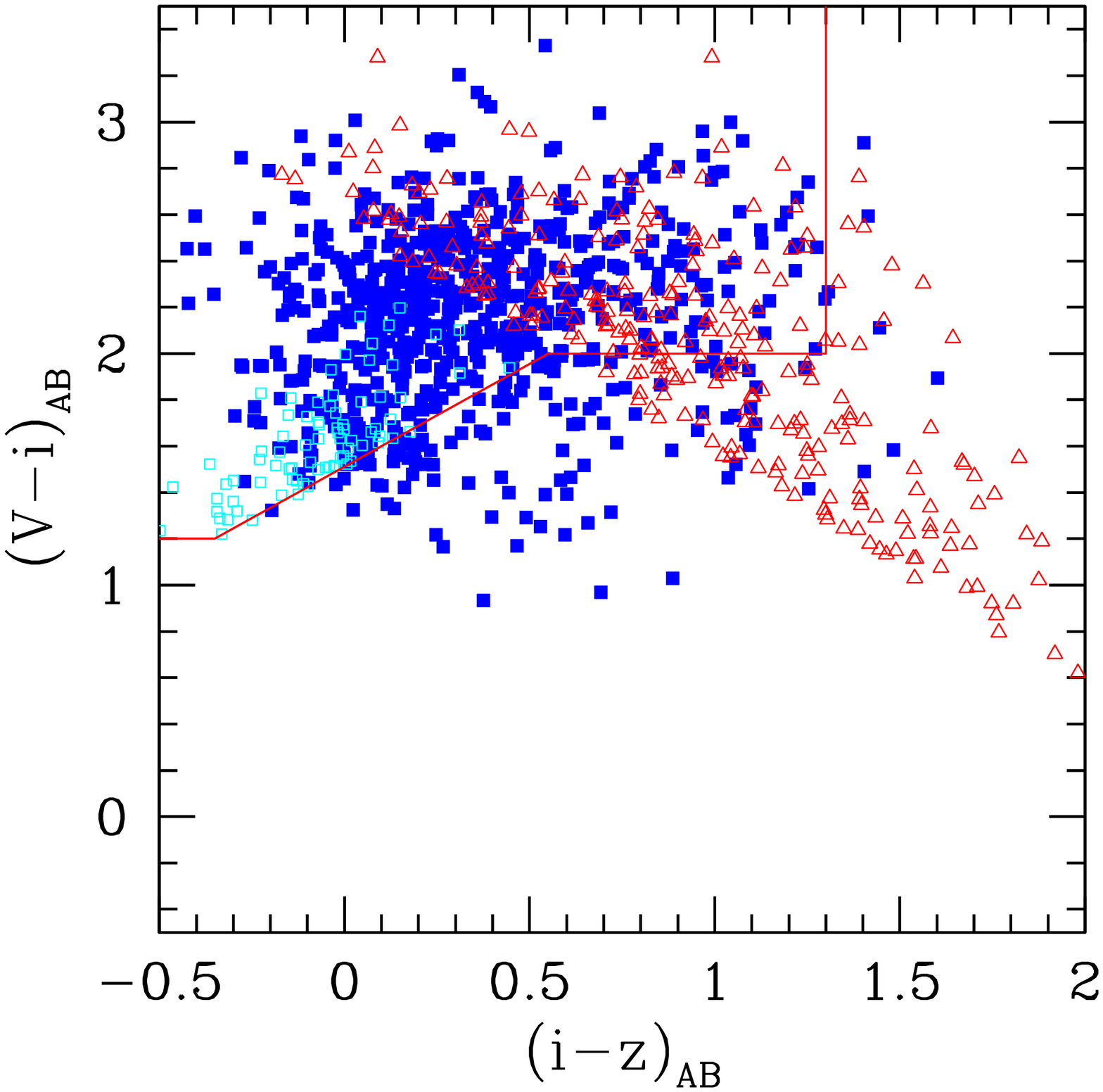} 
    \includegraphics[width=6cm]{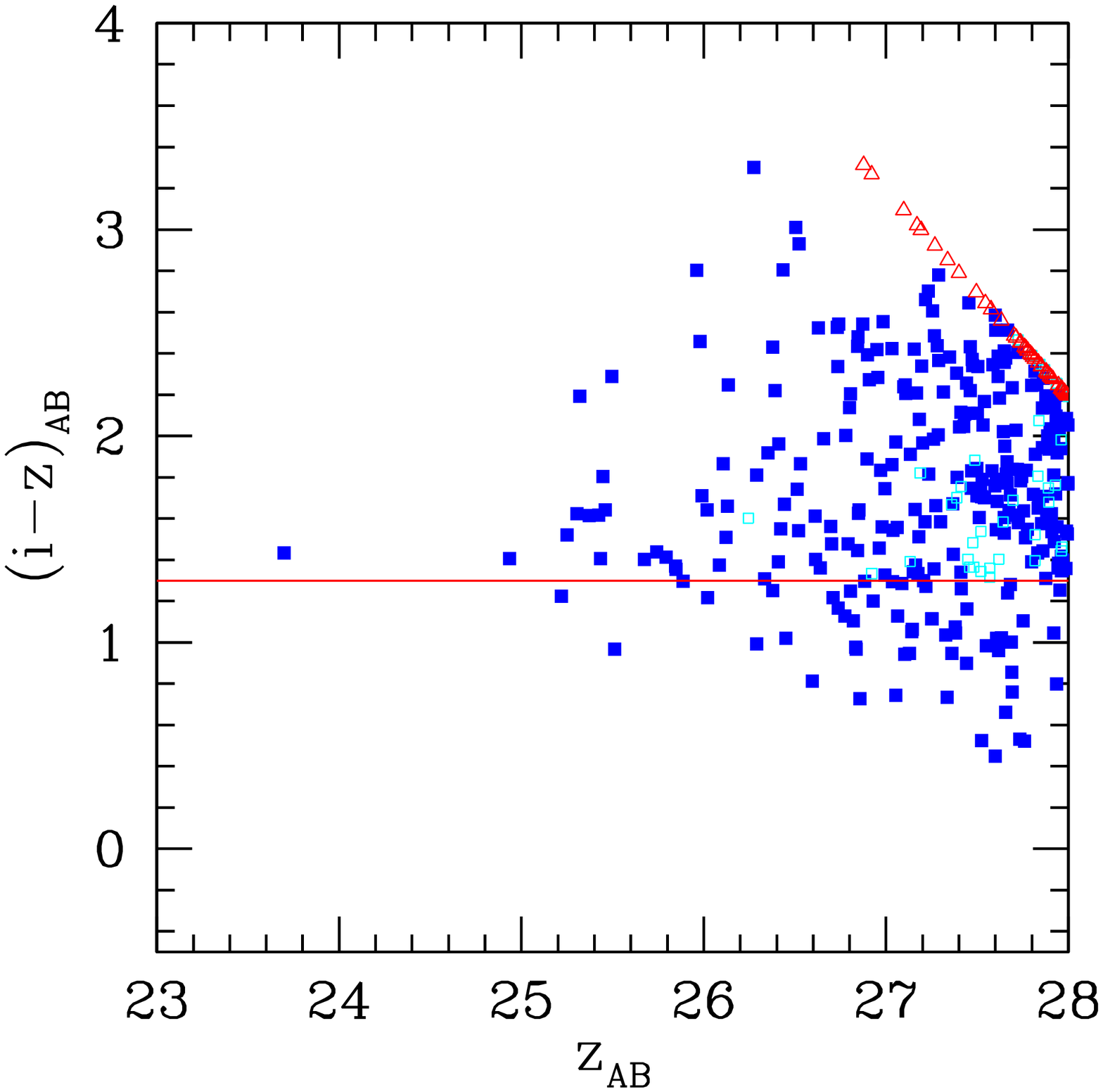} }
    
  \caption{Color-color diagrams of $B$-, $V$- and $i$-dropouts for the
    {\tt std.f095.e03} model. (Blue) filled squares are galaxies
    with redshift in the specific interval of the selection, that is,
    [3.5-4.5], [4.5-5.5] and [5.5-6.5] for $B$-, $V$- and $i$-dropouts
    respectively. The open (cyan) squares represent the interlopers
    (see text for details), while the (red) open triangles are
    upper-limits in $B$,$V$ and $i$. For sake of clarity only the
    central galaxies (see Section \ref{s:morgana}) are here considered,
    (satellite galaxies have very similar colors).}
  \label{fig:colo}
\end{figure*}

The prediction of galaxy apparent magnitudes critically depends on
dust attenuation, for which several uncertain parameters must be
specified in {\sc grasil}.  Among all the parameters, we identify two
of them as the crucial ones, namely $f_{\rm mol}$, the fraction of gas
in molecular clouds, and $t_{\rm esc}$, the time after which OB stars
get out of the highly obscured molecular cloud.  In \cite{Fontanot07}
values of $f_{\rm mol}=0.5$ (typical of the Milky Way) and $t_{\rm
  esc}=10$ Myr were used to make predictions in the $K$ and sub-mm
bands.
However, when applied to our (much more dust-sensitive)
Lyman-break candidates they give dust attenuations of more than 2 mag,
definitely larger, e.g., than the values of 1.4, 1.15 and 0.8 mag
assumed by \cite{Bouwens07} for $B$-, $V$- and $i$-dropouts.  Such
parameter values may not be suitable for compact, gas-rich
starbursting galaxies as our Lyman-break candidates are.  In fact,
higher molecular fractions are observed in galaxies where the
ISM is more pressurized \citep[see,
e.g.,][]{Blitz04}; this would imply much lower attenuation by
diffuse dust in the ISM (cirrus component).

We then run grasil using all combinations of $f_{\rm mol}=0.5$, 0.90,
0.95, 1.0 and $t_{\rm esc}=10$, 3, 1 Myr.  Table \ref{e} gives a list
of all the runs performed.  We use the model described in
Section~\ref{s:morgana} with quasar-triggered galaxy winds ({\tt
  std}), but perform also a few runs without winds ({\tt nowind})
to check the effect of this (very uncertain) ingredient of the model.
For sake of completeness we also show results for the same model ({\tt
  old}) presented in \cite{Monaco07} and \cite{Fontanot07}, with {\sc
  grasil} run with a Salpeter IMF, but with $f_{\rm mol}=0.50$ and
$t_{\rm esc}=3$ Myr, a combination which gives a best-fit of the
Lyman-break luminosity functions.

For one of the models ({\tt std.f095.e03}) we perform two runs with
the addition of the Lyman-$\alpha$ emission line, either assuming a
redshift-dependent $EW$ ({\tt std.f095.e03.lya1}) or assuming a
constant $EW$ ({\tt std.f095.e03.lya2}), as motivated in Section
\ref{s:lya}

\begin{figure*}
  \centerline{
    \includegraphics[width=6cm]{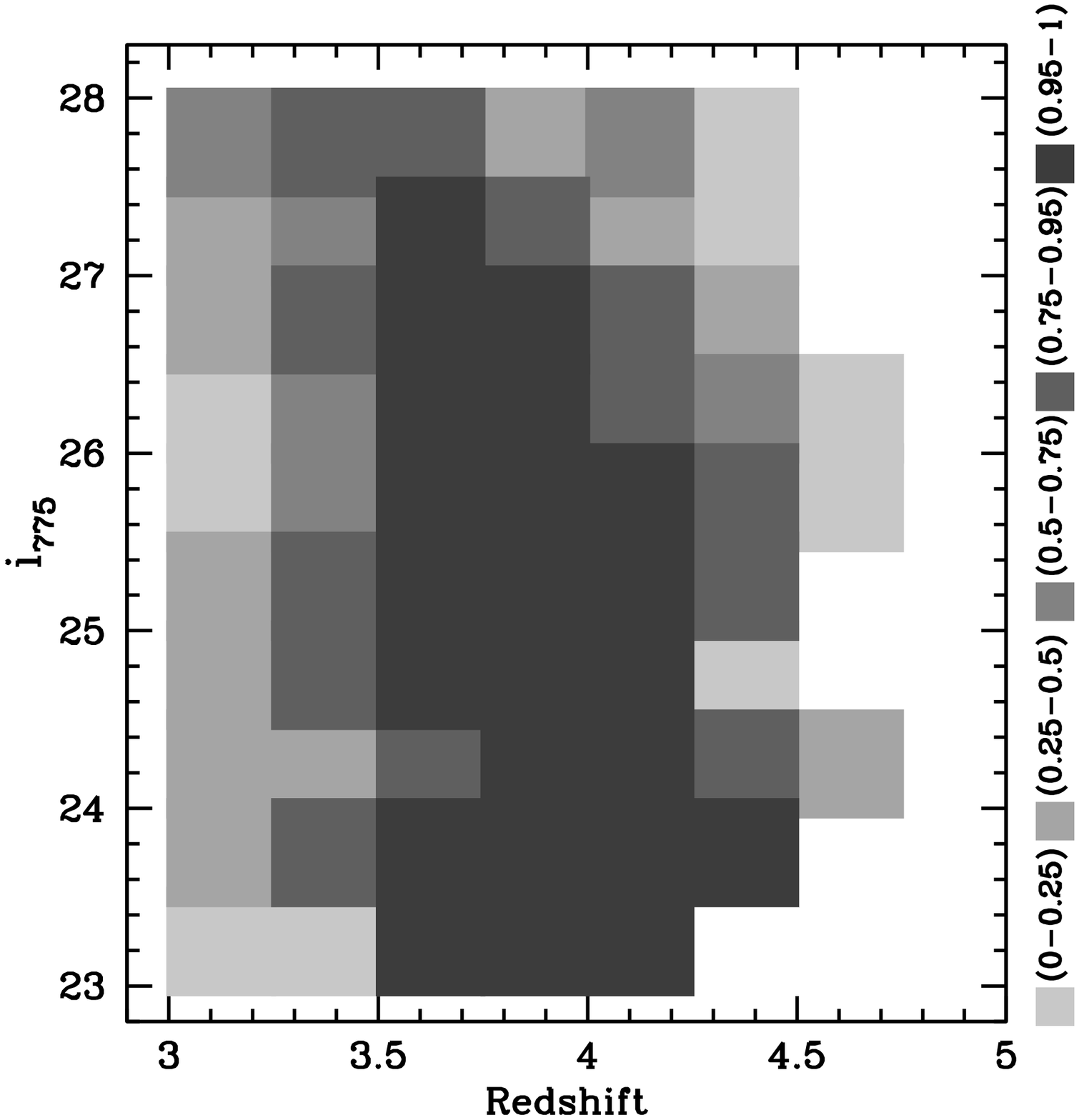} 
    \includegraphics[width=6cm]{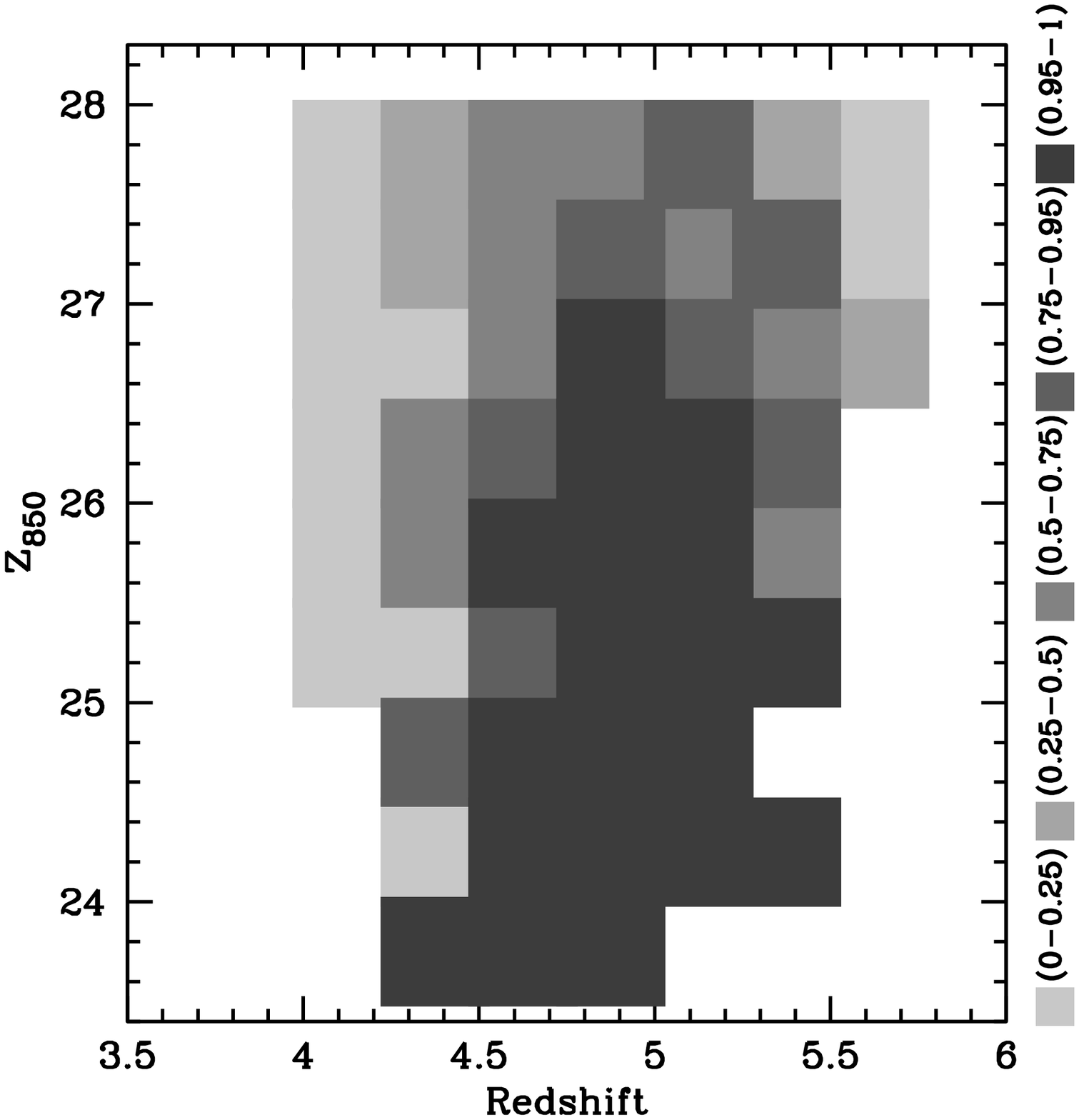} 
    \includegraphics[width=6cm]{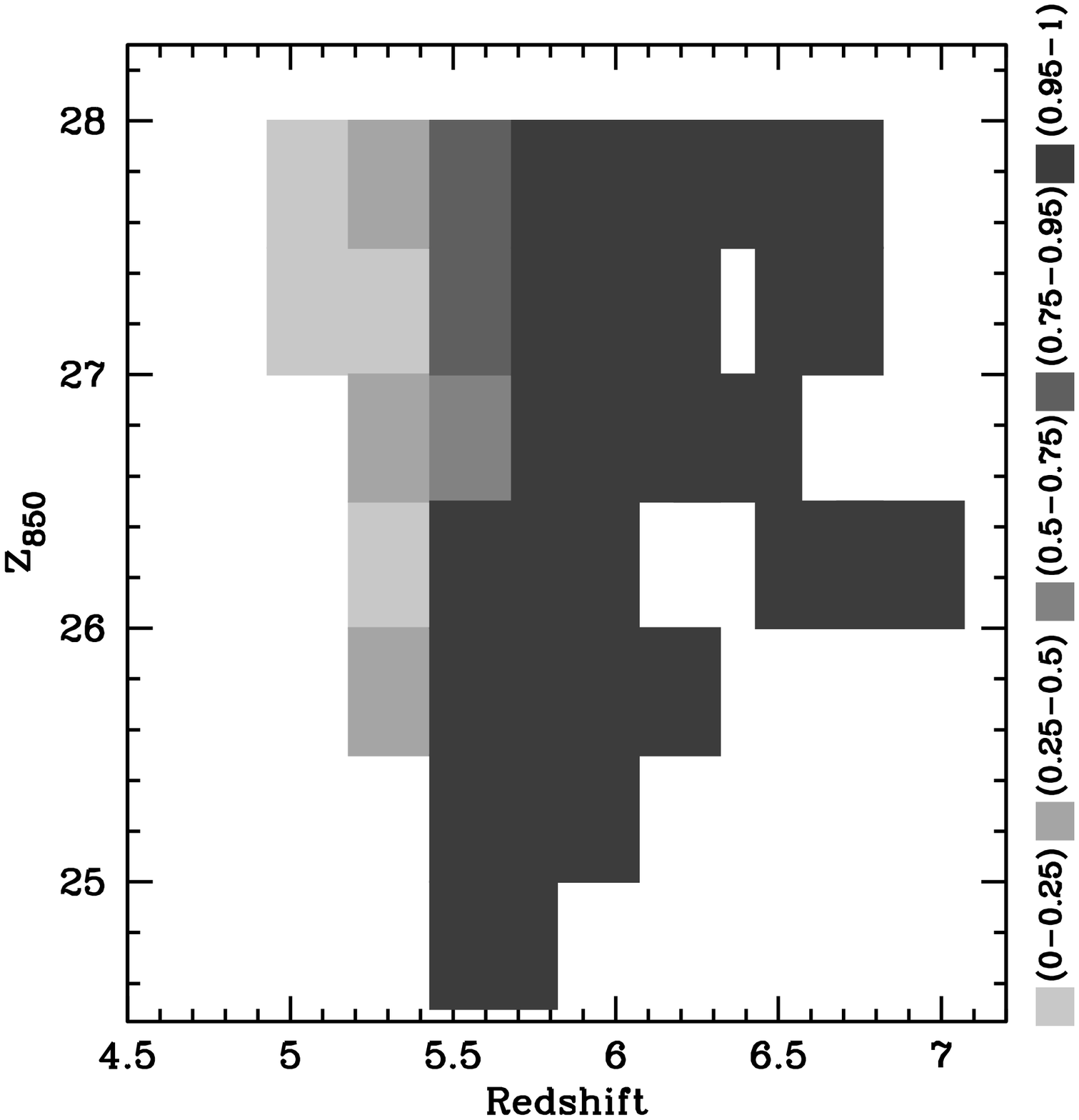}}
    
  \caption{Selection function for $B$-, $V$- and $i$-dropouts for the
    {\tt std.f095.e03} model.  The fraction of selected galaxies is
  computed in bins of redshift and apparent magnitude, and is reported
  as a tone of gray, as indicated in the right of each panel.}
  \label{fig:selection}
\end{figure*}

\begin{table}
\begin{center}
\begin{tabular}{cccc}
\hline
name & $f_{\rm mol}$ & $t_{\rm esc}$ & {\sc morgana} run\\
     &              & (Myr)       &                   \\
\hline
{\tt std.f050.e10} & 0.50 & 10 & with quasar winds \\
{\tt std.f090.e10} & 0.90 & 10 &        ''         \\
{\tt std.f095.e10} & 0.95 & 10 &        ''         \\
{\tt std.f100.e10} & 1.00 & 10 &        ''         \\
{\tt std.f050.e03} & 0.50 &  3 &        ''         \\
{\tt std.f090.e03} & 0.90 &  3 &        ''         \\
{\tt std.f095.e03} & 0.95 &  3 &        ''         \\
{\tt std.f100.e03} & 1.00 &  3 &        ''         \\
{\tt std.f050.e01} & 0.50 &  1 &        ''         \\
{\tt std.f090.e01} & 0.90 &  1 &        ''         \\
{\tt std.f095.e01} & 0.95 &  1 &        ''         \\
{\tt std.f100.e01} & 1.00 &  1 &        ''         \\
\hline
{\tt std.f095.e03.lya1} & 0.95 & 3 & with Ly$\alpha$ in emission  \\
{\tt std.f095.e03.lya2} & 0.95 & 3 &    ''         \\
\hline
{\tt nowind.f090.e03} & 0.90 &  3 & without quasar winds\\
{\tt nowind.f090.e01} & 0.90 &  1 &        ''         \\
{\tt nowind.f095.e03} & 0.95 &  3 &        ''         \\
\hline
{\tt old.f050.e03}  &  0.50  &  3  &  2007 model \\
\hline
\end{tabular}
\caption{Runs performed for this paper, with values of {\sc grasil}
  parameters (see Section \ref{s:grasil} for a brief description of
  the parameters) and indication of the {\sc morgana} model used.}
\label{e}
\end{center}
\end{table}

\section{Results}
\label{s:results}

\begin{figure*}
  \centerline{
    \includegraphics[width=6cm]{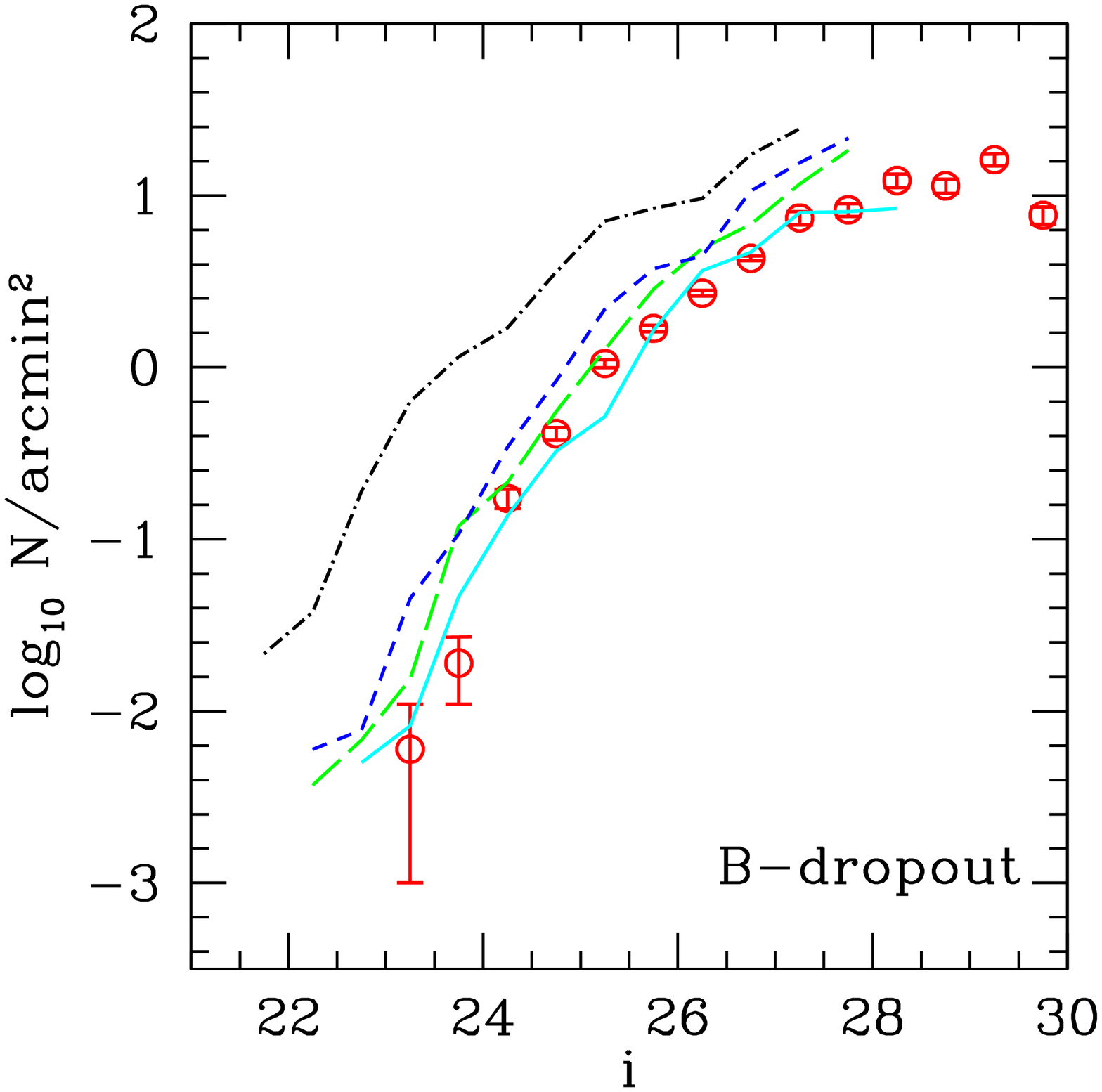} 
    \includegraphics[width=6cm]{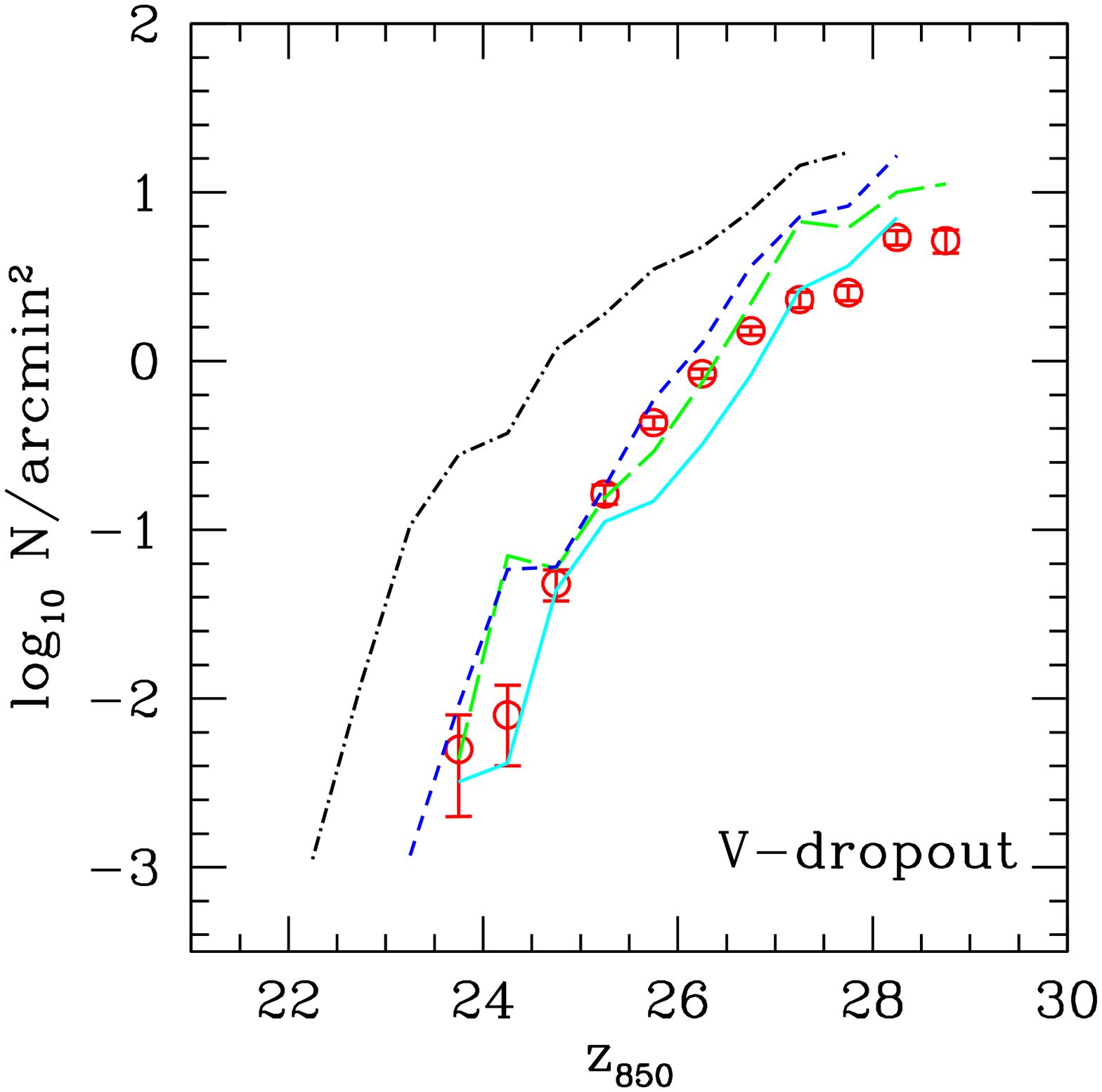} 
    \includegraphics[width=6cm]{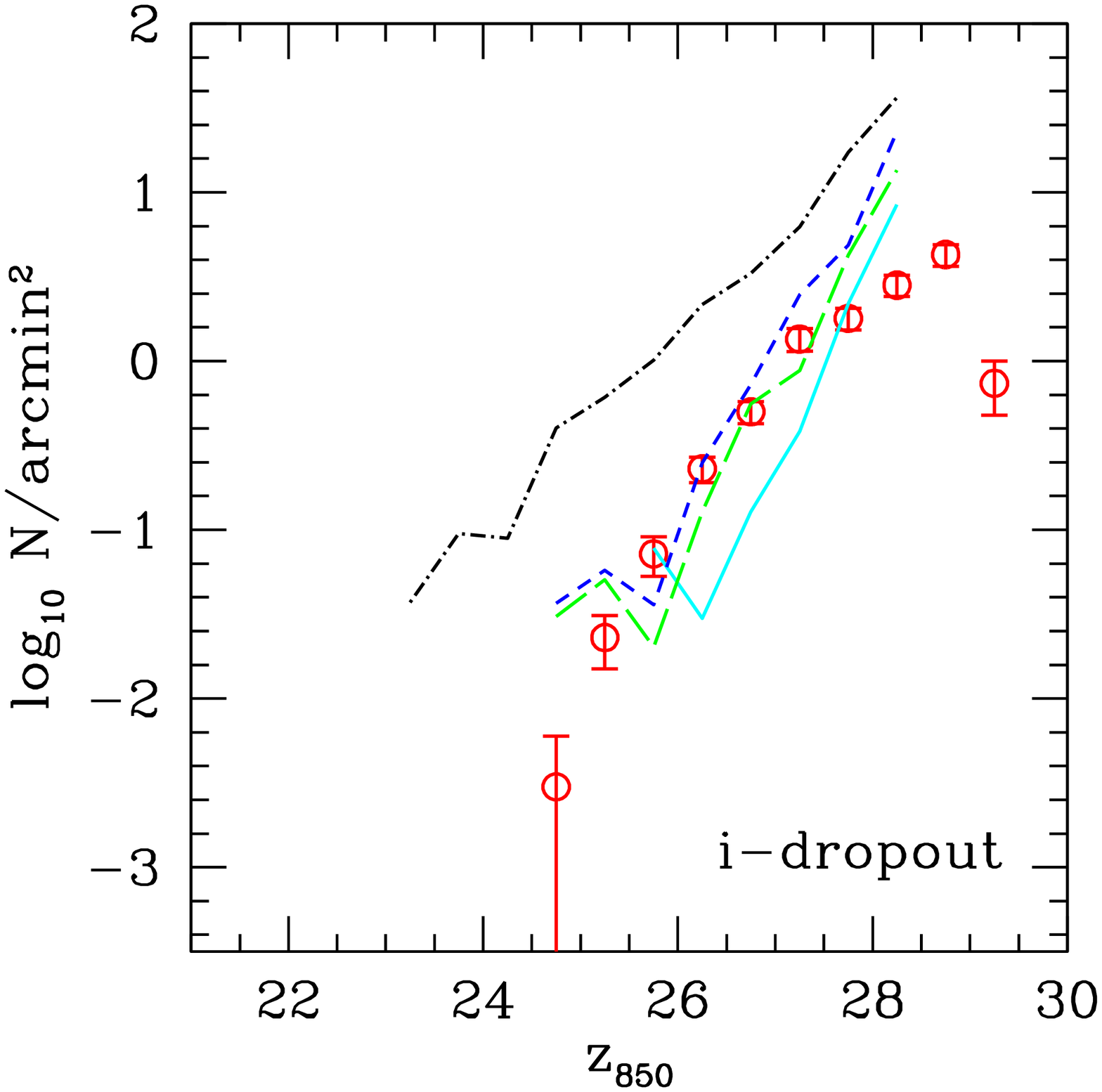} }
    
  \centerline{
    \includegraphics[width=6cm]{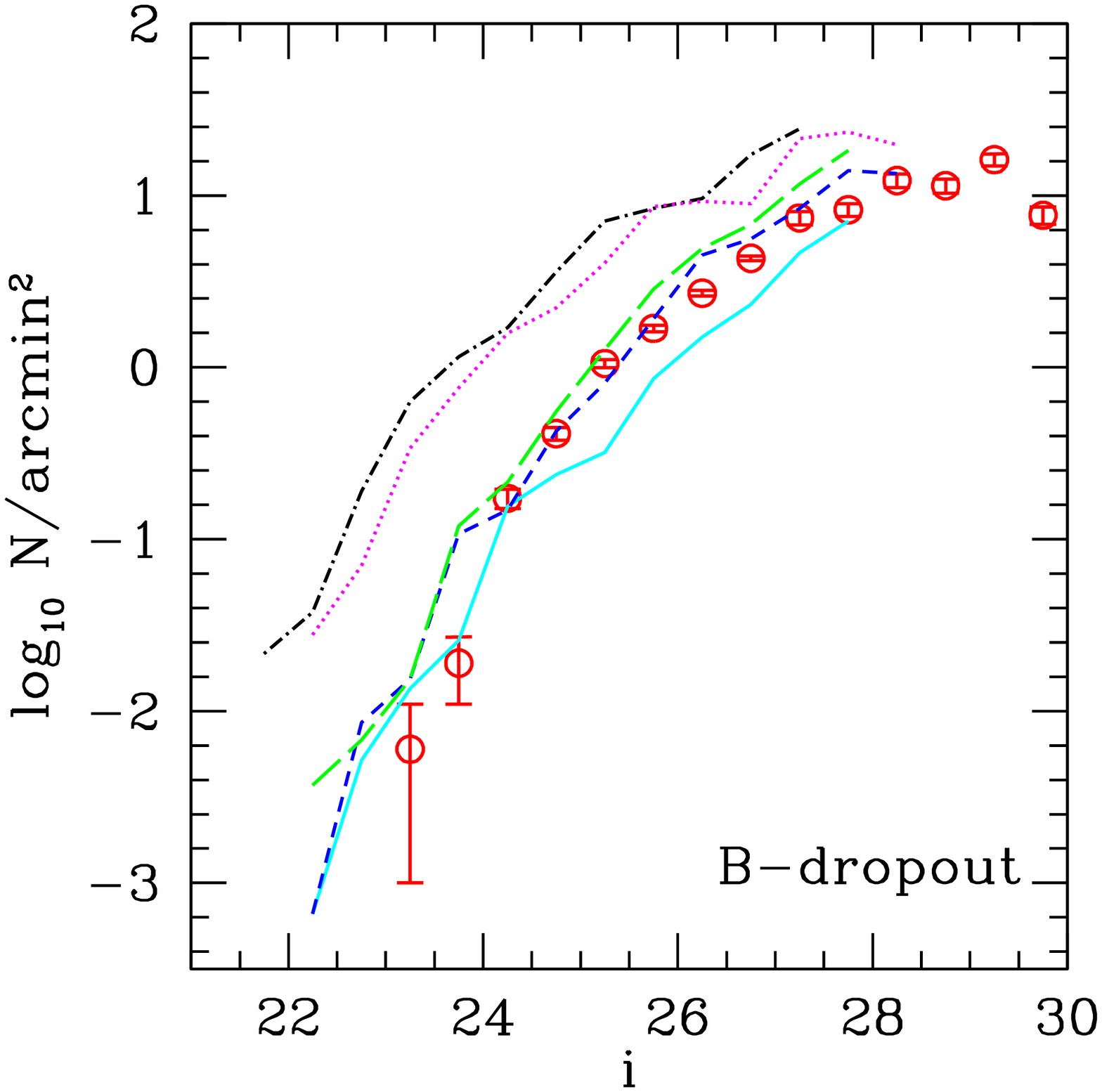} 
    \includegraphics[width=6cm]{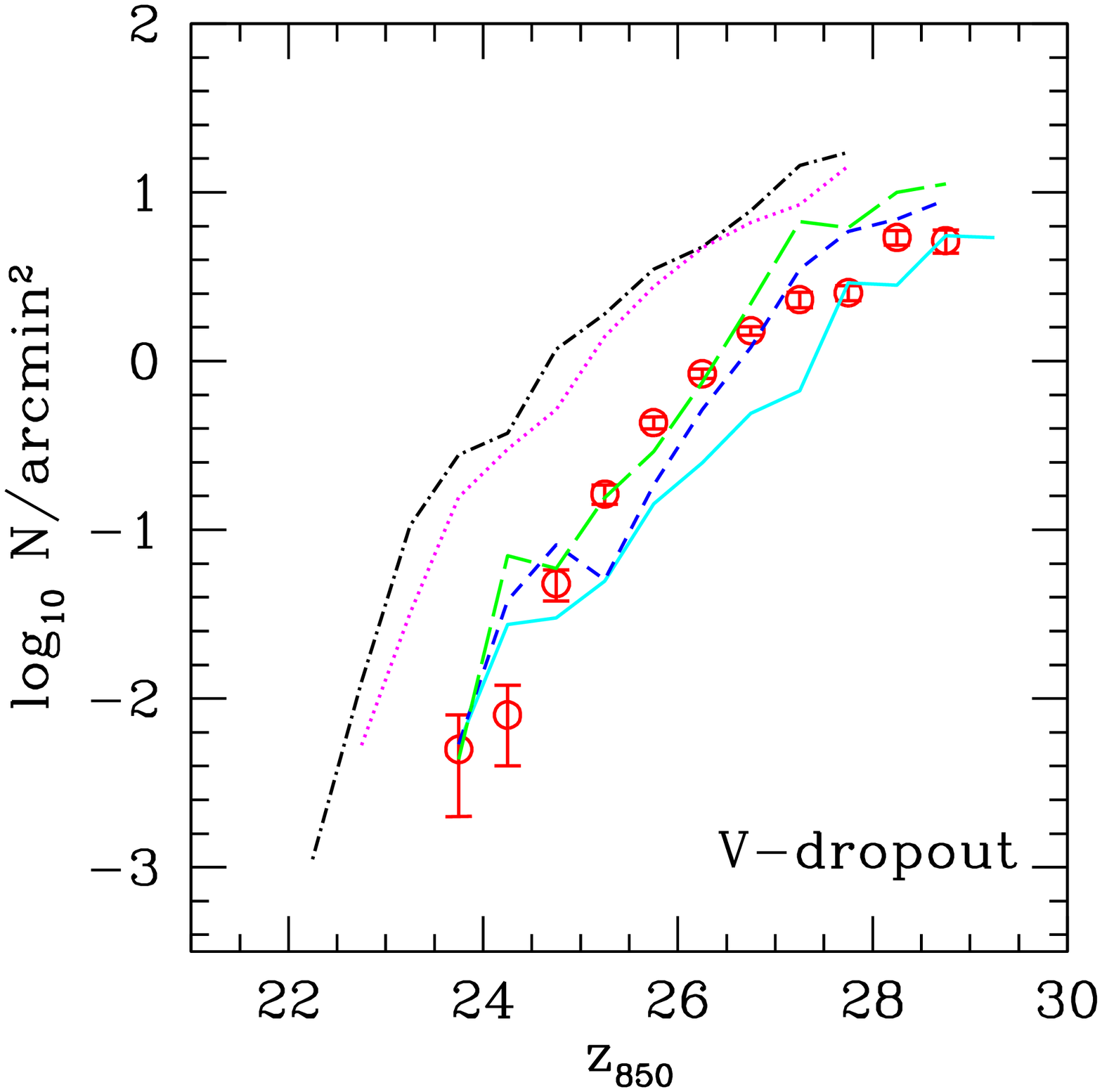} 
    \includegraphics[width=6cm]{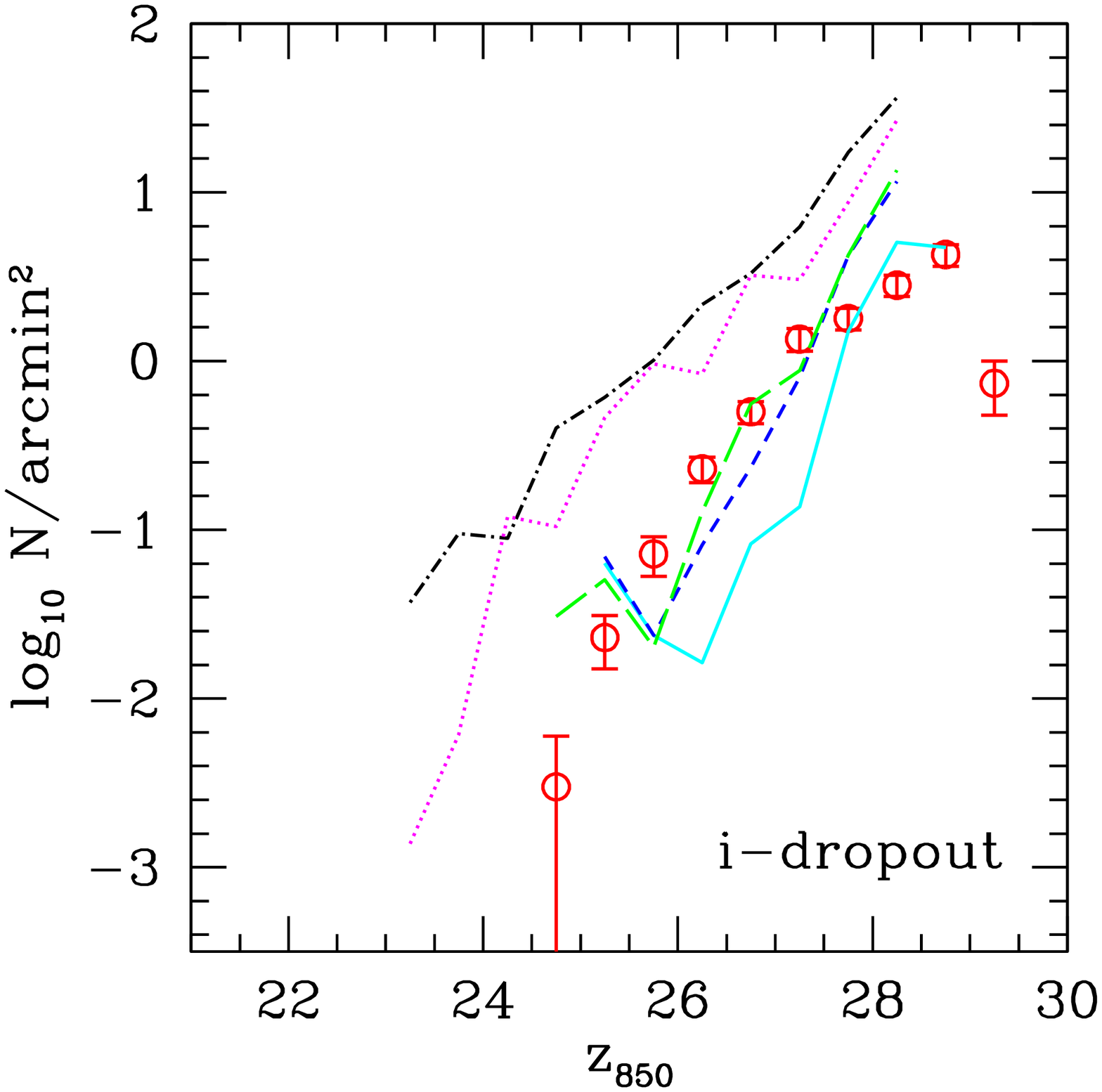} }
    
  \caption{Number counts of $B$-, $V$- and $i$-dropouts.  (Red) points
    are from Bouwens et al. (2007). Upper panels: models with $f_{\rm
      mol}=0.95$, {\tt std.f095.e10} (cyan solid lines), {\tt
      std.f095.e03} (green long-dashed lines) and {\tt std.f095.e01}
    (blue dashed lines) compared to the model without dust attenuation
    (black dot-dashed lines). Lower panels: models with $t_{\rm
      esc}=0.3$, {\tt std.f050.e03} (cyan solid lines), {\tt
      std.f090.e03} (blue dashed lines) and {\tt std.f095.e03} (green
    long-dashed lines) and {\tt std.f100.e03} (magenta dotted lines)
    also compared to the model without dust attenuation (black
    dot-dashed lines).}
  \label{fig:counts2}
\end{figure*}

Figure \ref{fig:colo} shows the color-color plots for the selection of
the model $B$-, $V$- and $i$-dropouts, according to the criteria
defined above. Here and in the next figure we use the {\tt
  std.f095.e03} model; here for sake of clarity only central
galaxies with $z_{850}<28$, the resolution limit of the model, are
used (satellite galaxies have very similar colors). All galaxies in the
redshift intervals of the selection [3.5-4.5], [4.5-5.5] and [5.5-6.5]
for $B$-,$V$- and $i$-dropouts respectively, are shown as full blue
squares, or red triangles if they are upper limits. These are the same
redshift intervals identified by \cite{Bouwens07} as typical of
Lyman-break galaxies at mean redshift of $z\sim4$, $z\sim5$ and
$z\sim6$. 
The few interlopers falling in the selection areas are marked as cyan
open squares. For interlopers we mean galaxies with redshift in the
interval [3.0-3.4] for the $B$-dropouts, [4.0-4.4] for the
$V$-dropouts and [5.0-5.4] for the $i$-dropouts. Clearly, most
galaxies in the relevant redshift interval are selected.  Those
that scatter out of the selection region are mostly faint galaxies
with upper limits in the band from which they drop out; their
colors are made less red by the upper limits.
Figure~\ref{fig:selection} shows the selection function as a function
of redshift and $i$ ($B$-dropouts) or $z_{850}$ ($V$- and $i$-dropouts) magnitude. Selected
galaxies clearly tend to reside in the expected redshift range, though
some $B$- and $V$-dropouts are lost at magnitudes fainter than
$\sim26$.

We first compare model predictions of $B$-, $V$- and $i$-dropouts with
observed number counts, which are the directly observable quantity
when complete spectroscopic samples are not available.

Figure ~\ref{fig:counts2} shows the effect of changing the {\sc
  grasil} parameters $t_{\rm esc}$ and $f_{\rm mol}$ compared to the
standard model without dust attenuation (black dot-dashed
lines). We consider both the effect of changing $t_{\rm esc}$ keeping
fixed the value of $f_{\rm mol}$ (upper panel), and the effect of
varying $f_{\rm mol}$ at fixed $t_{\rm esc}$ (bottom panel).
Model predictions for unabsorbed spectra drop at a
magnitude $\sim27.5$ (fainter for the $i$-dropouts).
This is the
typical magnitude of the central galaxy contained in the smallest resolved
dark matter halo ($1.42\cdot10^{10}$ {\msun}), 
fainter galaxies would be hosted in smaller, unresolved halos and are thus 
under-represented.  It is worth noting that 
HUDF data are deeper than this limit\footnote{
It would be easy too reach deeper magnitudes by running the model on a
smaller box.  However, here we are interested in understanding whether
the {\it same} model that produces small galaxies that are too old at $z=0$
\citep{Fontanot09b} is also over-producing faint star-forming galaxies
at high redshift. To achieve this, the requirement is that observations probe at
least as deep as the model's limiting magnitude.
}
Clearly this limit magnitude increases when
dust attenuation is considered.  When compared to Bouwens et al.'s
observations (open red points in the Figure), number counts in absence of
dust attenuation clearly overshoot the data (by about 2 mag).
This shows how a correct modeling of attenuation is fundamental to
best-fit the data. This feature clearly limits the predictive
power of the model, because small variations of dust parameters within
the allowed range give very large differences in the results.

The figure shows that the two parameters are degenerate; we identify
the two combinations $(t_{\rm esc},f_{\rm mol})=(3\ {\rm Myr}, 0.95)$
(model {\tt std.f095.e03}) and $(1\ {\rm Myr}, 0.90)$ (model {\tt
  std.f090.e01}) as the best-fitting ones. This degeneration concerns
the extinction, we in fact, expect to find differences in the IR due
to the different distributions of temperature and optical depth of the
two phases.
The attenuations given by these best-fit models are roughly
  similar to the ones used by Bouwens et al. (1.4, 1.15 and 0.8 mag
  for $B$-, $V$- and $i$-dropouts) but are systematically higher for brighter
  galaxies. This is in line with the findings of \cite{Shapley01}:
  these authors reported for a large sample of Lyman-break galaxies at
  $z\sim3$ an average $E(B-V)$ attenuation of 0.15 that correlates
  with magnitude, brighter objects being more absorbed.  Expressed in
  terms of the same $E(B-V)$ quantity, our attenuation is $~0.1$ at
  the bright end and declines to $\sim0.05$ for the faint objects.

\begin{figure*}
  \centerline{
    \includegraphics[width=6cm]{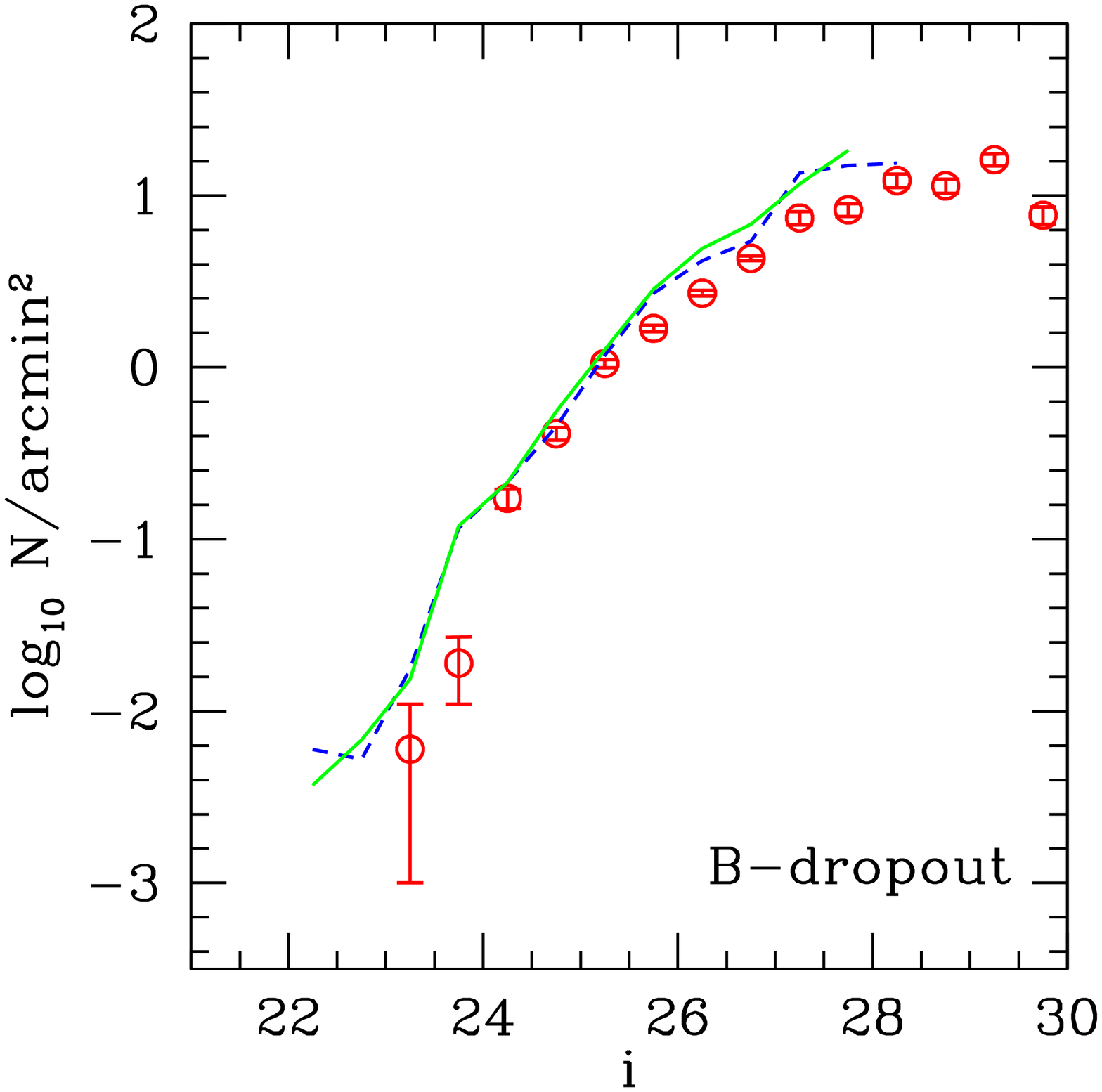} 
    \includegraphics[width=6cm]{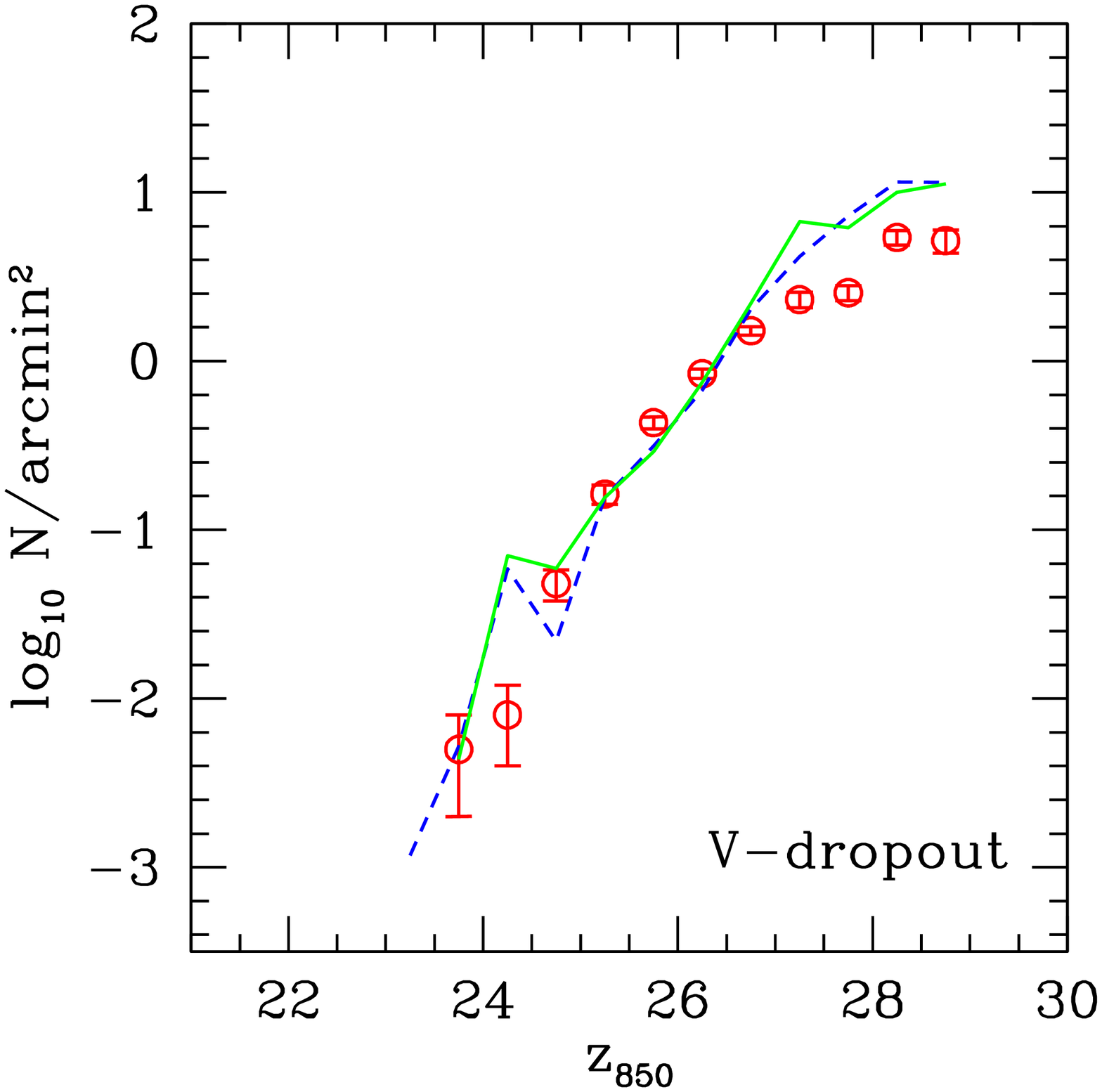} 
    \includegraphics[width=6cm]{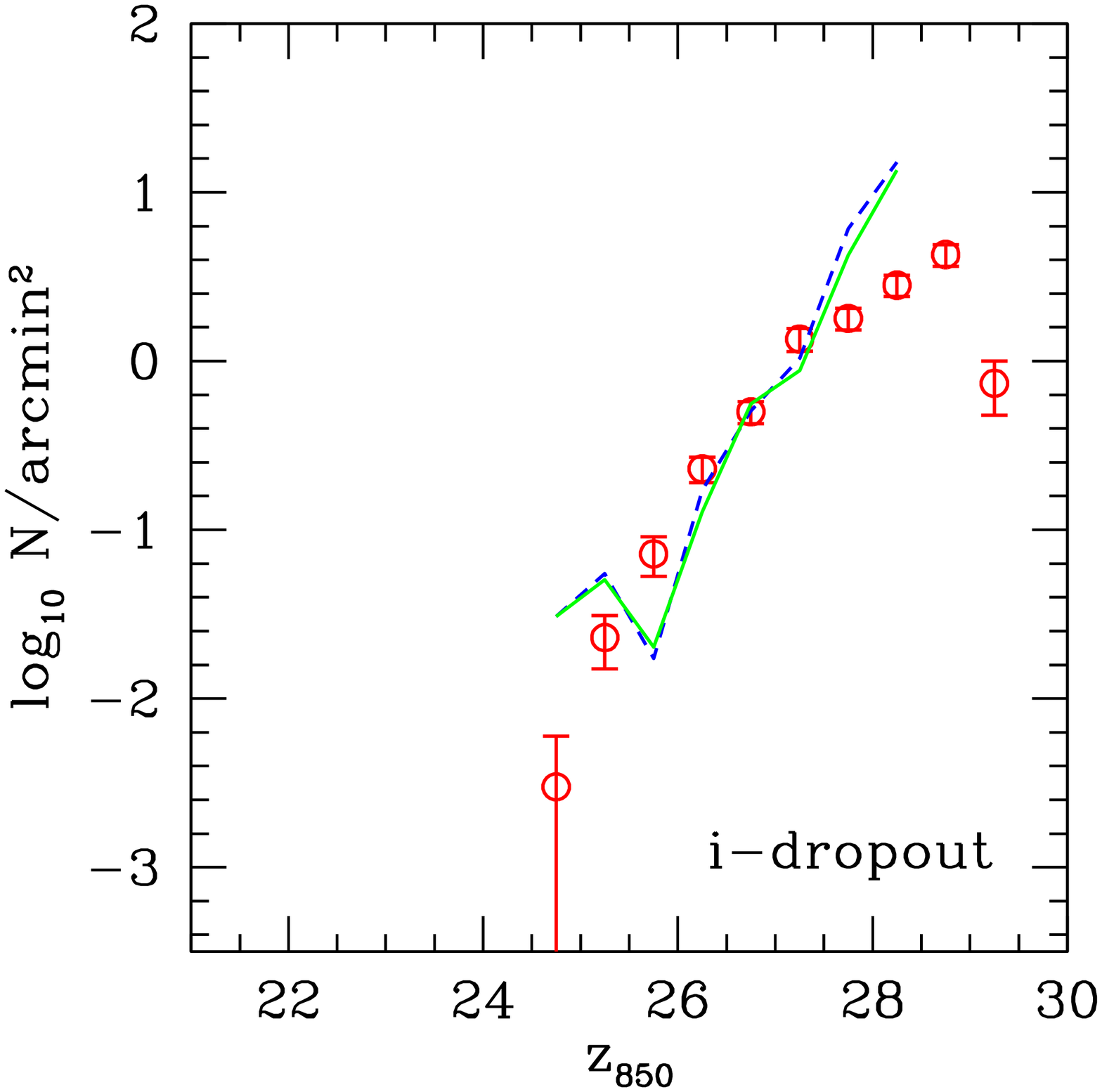} }
    
  \caption{Number counts of $B$-, $V$- and $i$-dropouts. (Red) points
    are from Bouwens et al. (2007), the (blue) dashed and (green) solid lines give the
    predictions of the two best-fit models, respectively 
    {\tt std.f090.e01} and {\tt std.f095.e03}.}
  \label{fig:best-fit}
\end{figure*}

\begin{figure}
  \centerline{
       \includegraphics[width=8cm]{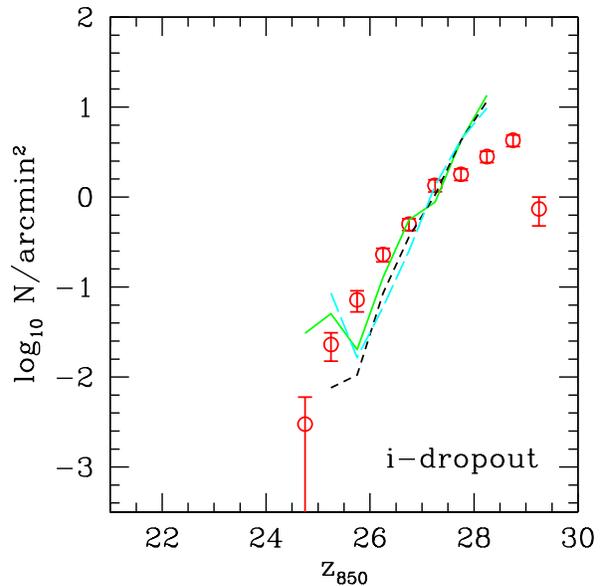} }
     \caption{Number counts of $i$-dropouts. (Red)
       points are from Bouwens et al. (2007), the (green) solid line
       gives the predictions of the best-fit model {\tt
         std.f095.e03}, the (black) dashed and the (cyan)
       long-dashed lines correspond to the same combination of parameters
       but with the contribution of
       Ly$\alpha$ emission line in the two different prescription
       discussed in Section \ref{s:lya} (black: {\tt
         std.f095.e03.lya1}, cyan:
       {\tt std.f095.e03.lya2}).}
  \label{fig:lya}
\end{figure}

Figure~\ref{fig:best-fit} shows number counts for the two best-fit
models.  As expected, while the bright end is well reproduced with its
redshift evolution (with a modest excess for the $B$-dropouts and possibly a dearth of bright $i$-dropouts), 
a clear excess of faint objects is present at
$z_{850}\ga27$, especially for the $V$-dropouts.

\begin{figure*}
  \centerline{
    \includegraphics[width=6cm]{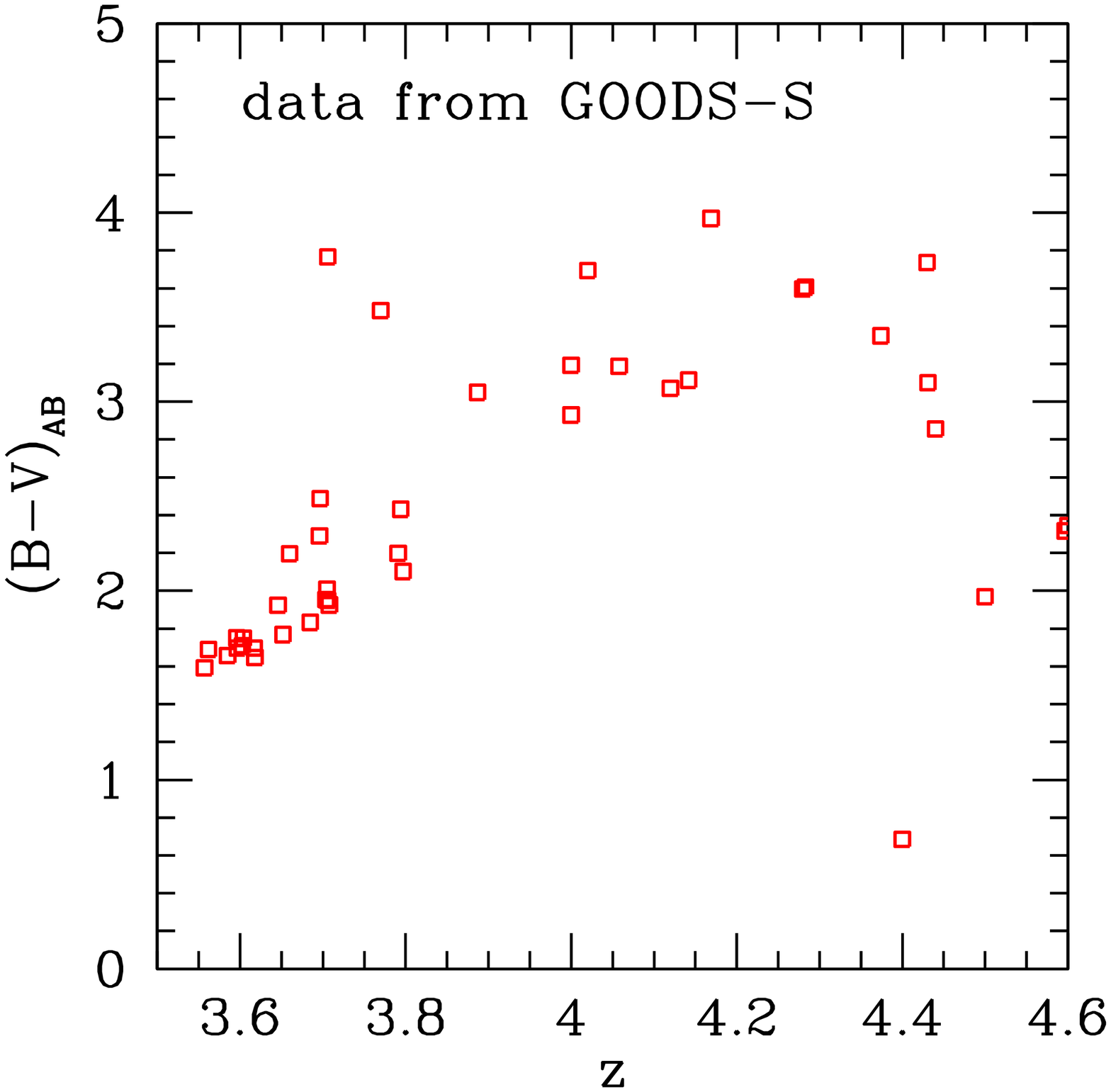} 
    \includegraphics[width=6cm]{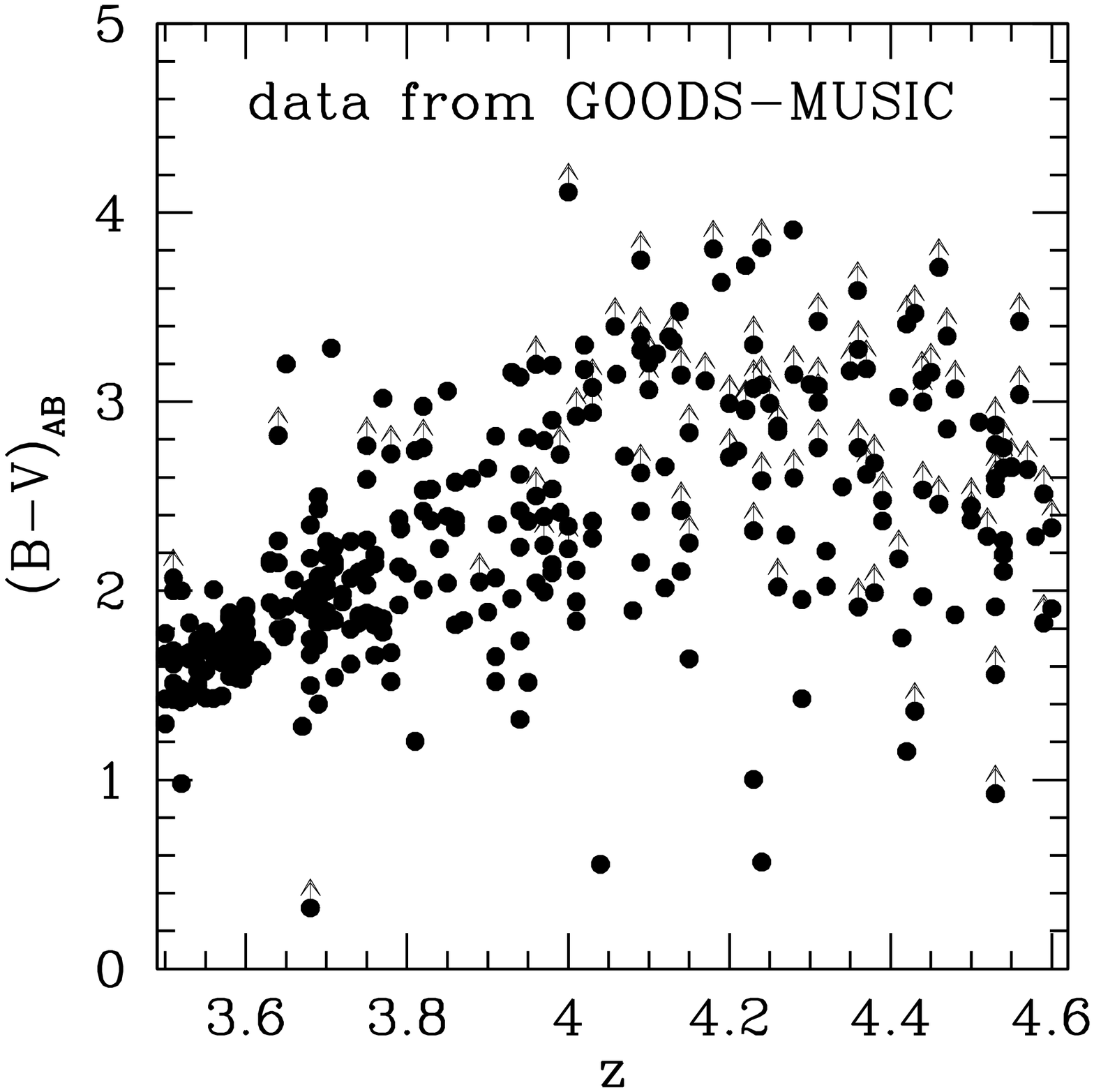} 
    \includegraphics[width=6cm]{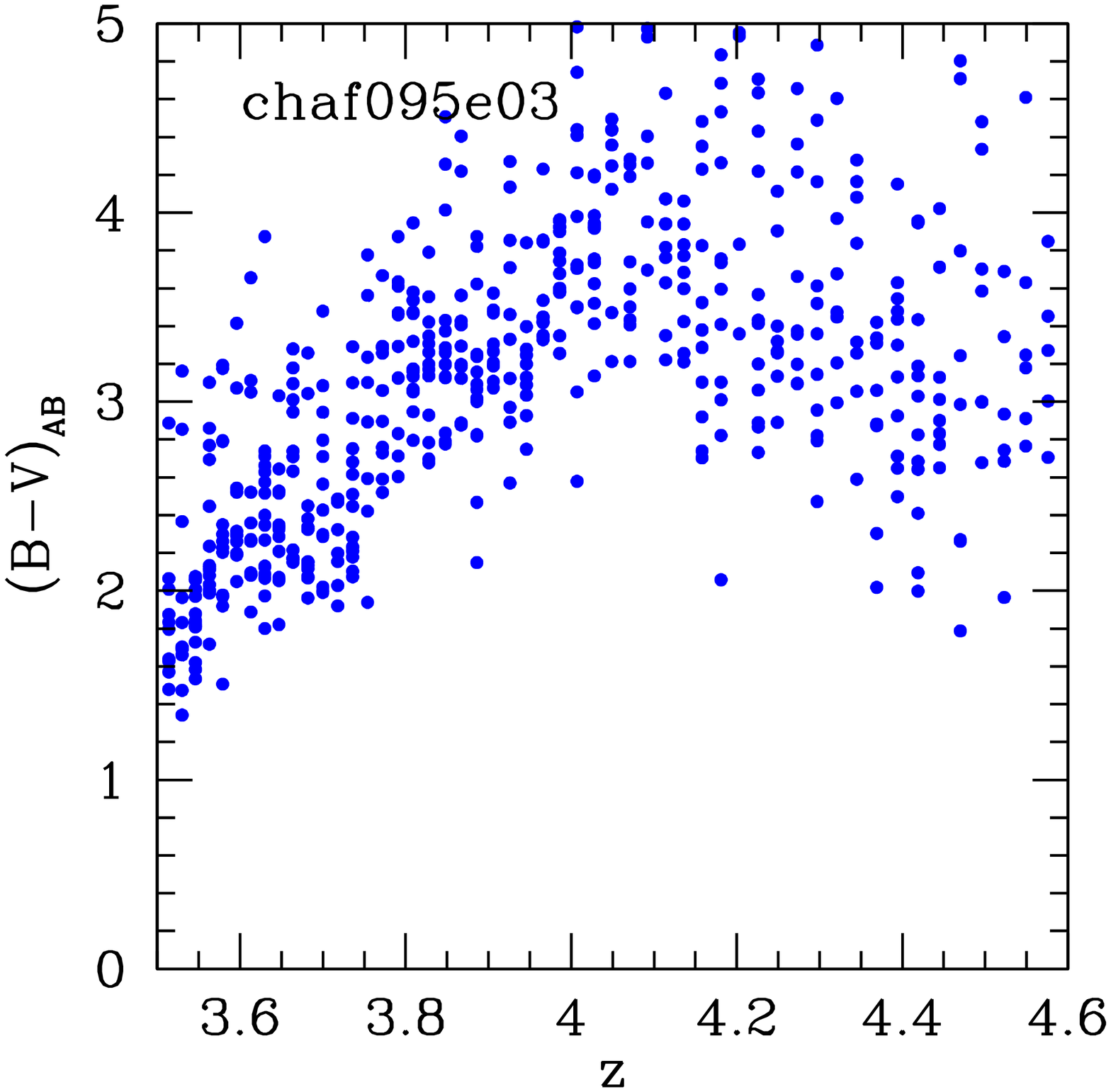} }
  \centerline{
    \includegraphics[width=6cm]{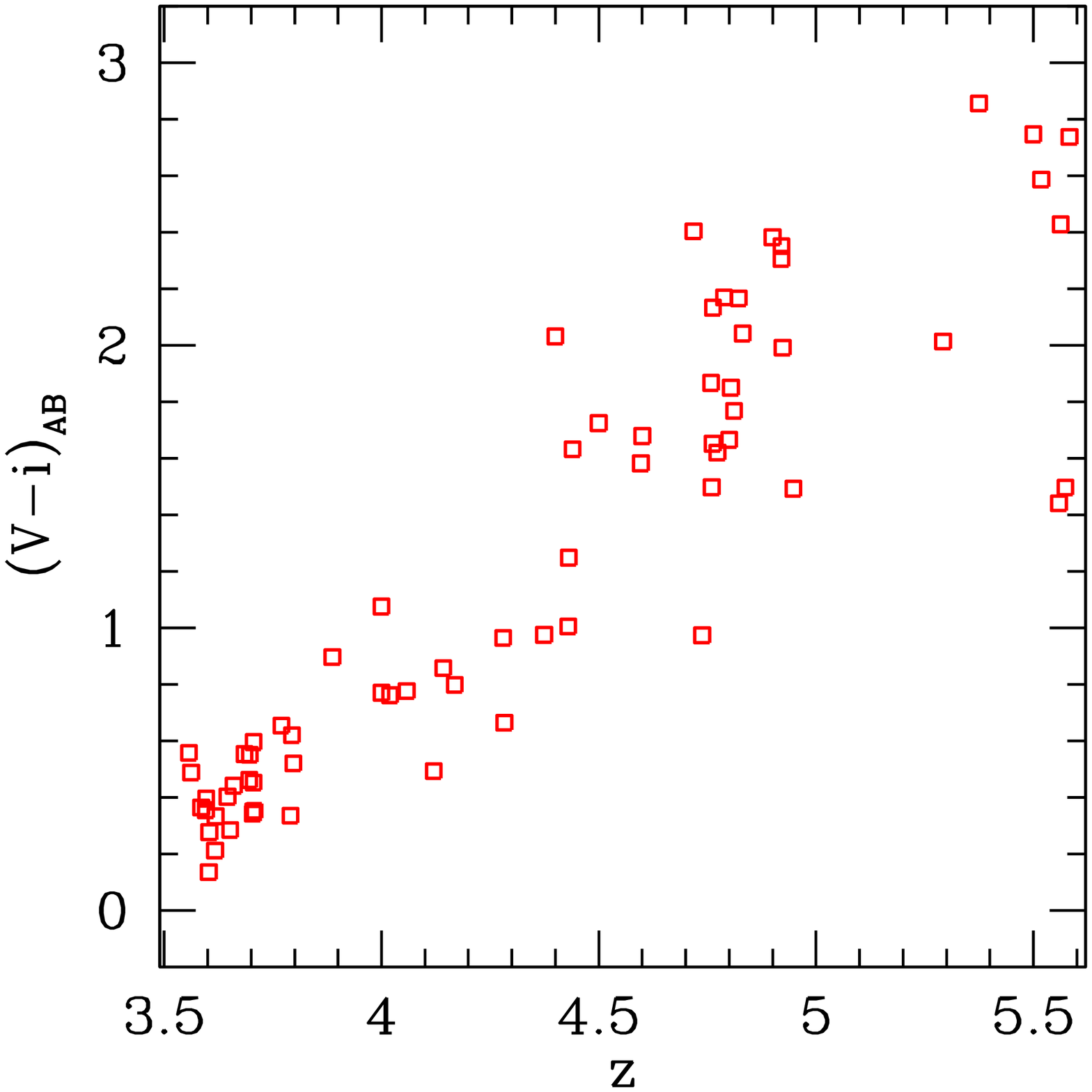} 
    \includegraphics[width=6cm]{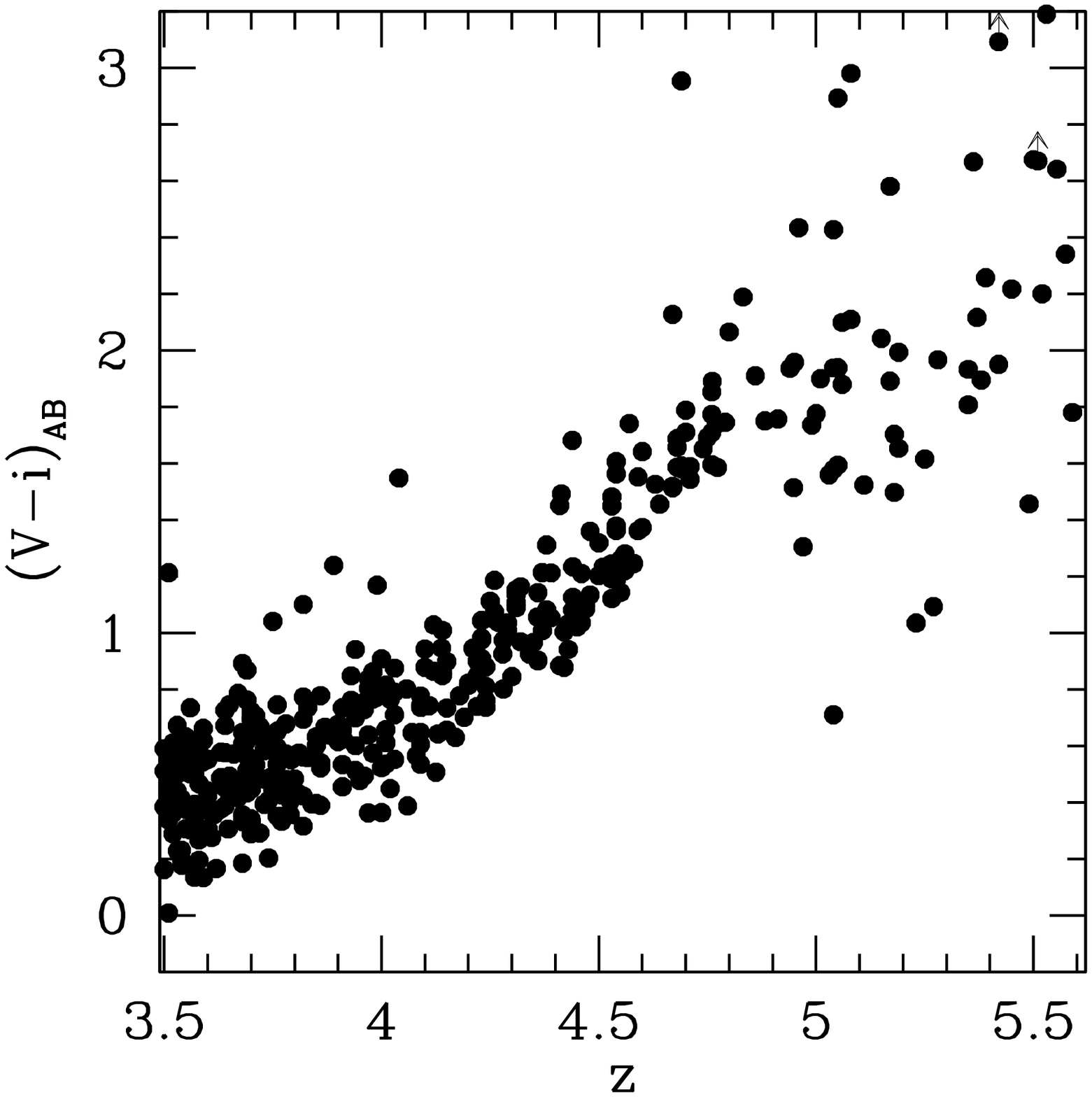} 
    \includegraphics[width=6cm]{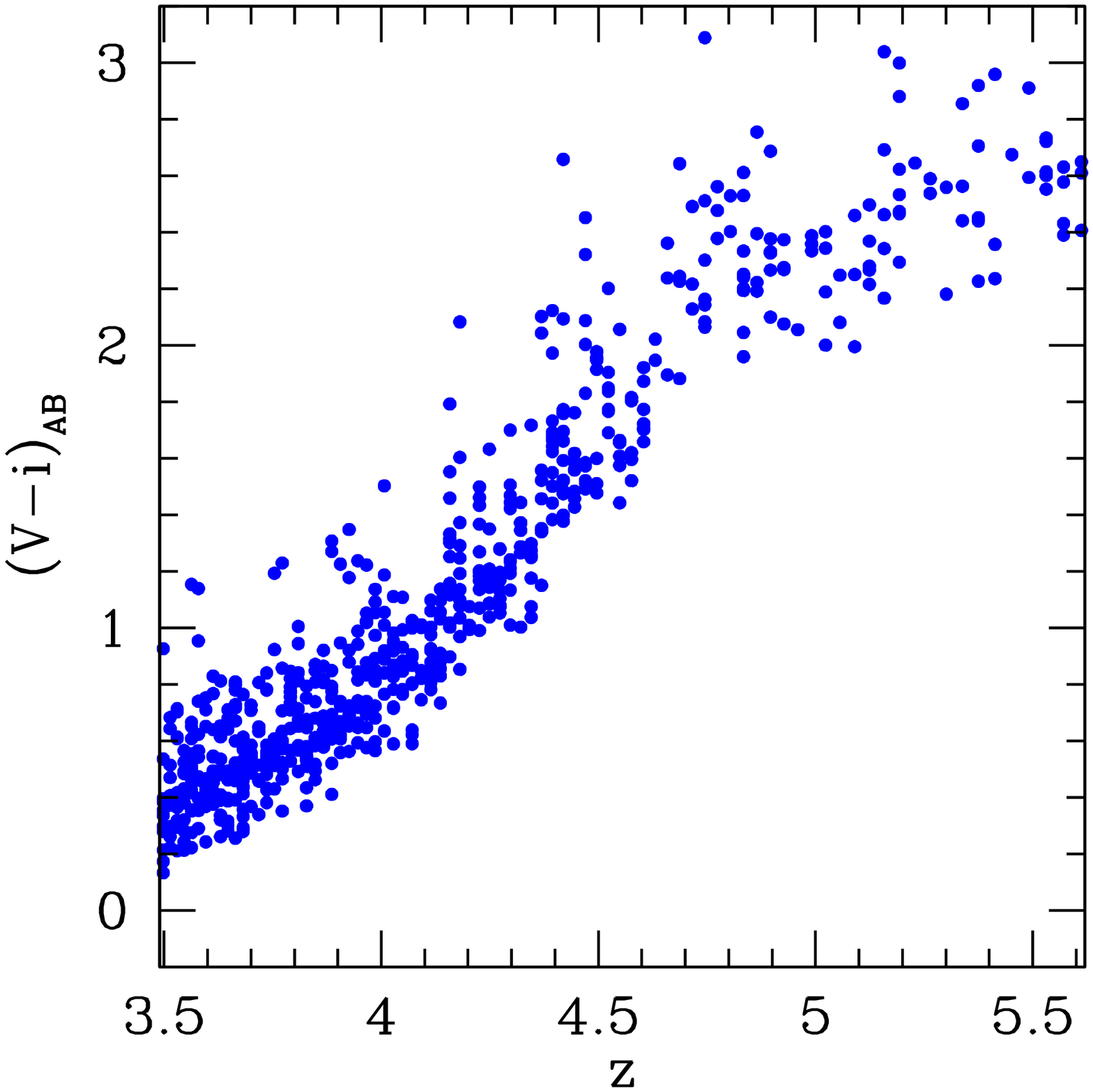} }
  \centerline{
    \includegraphics[width=6cm]{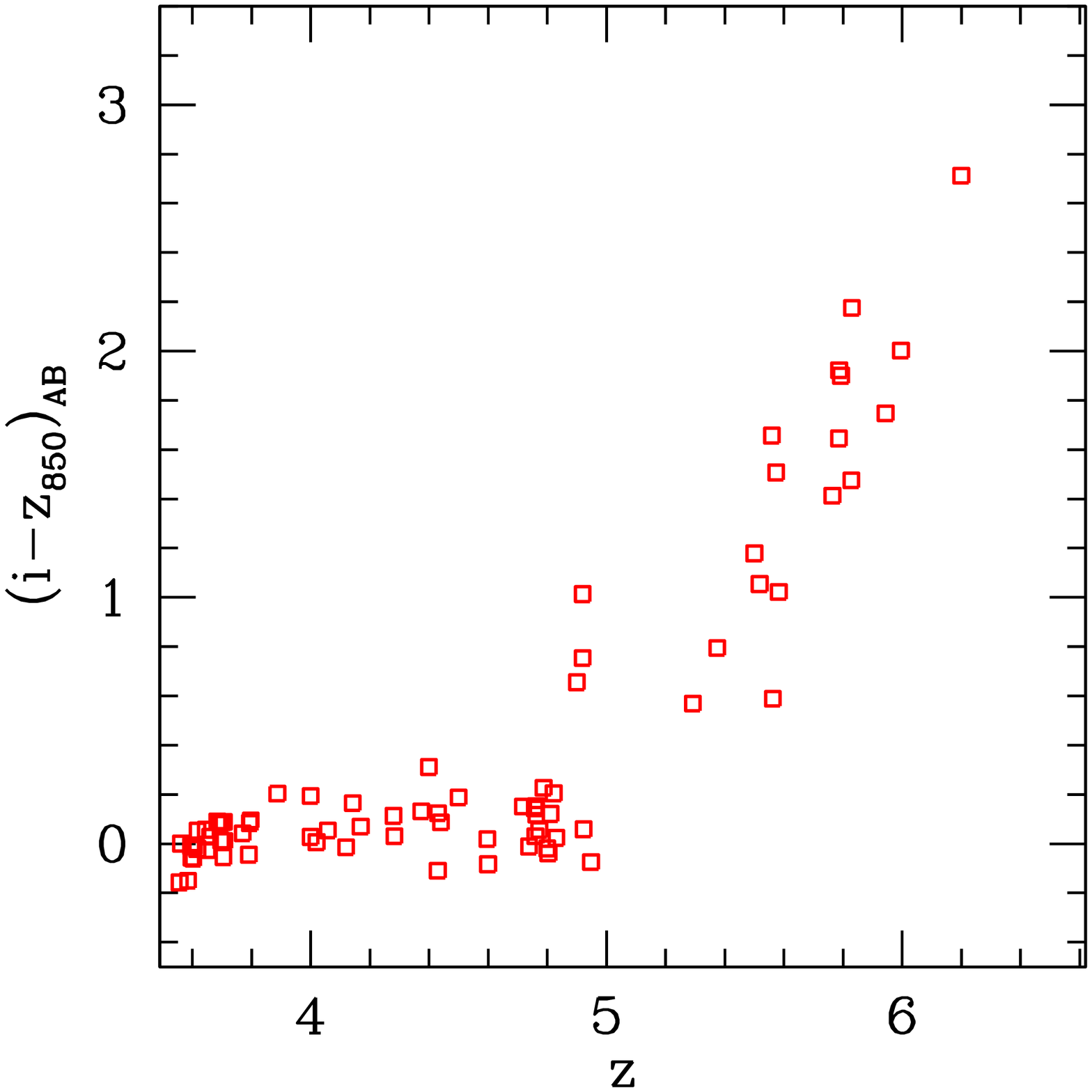} 
    \includegraphics[width=6cm]{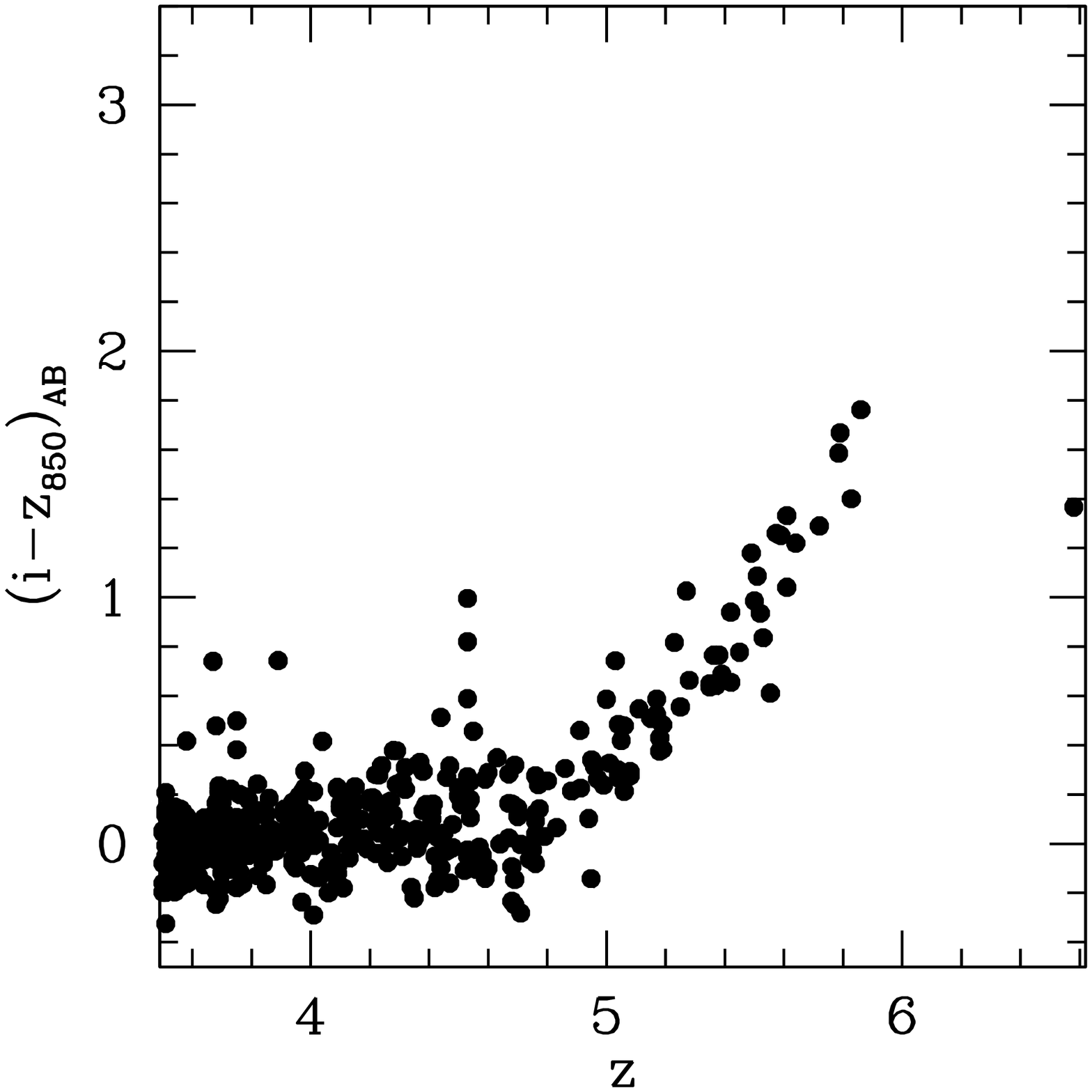} 
    \includegraphics[width=6cm]{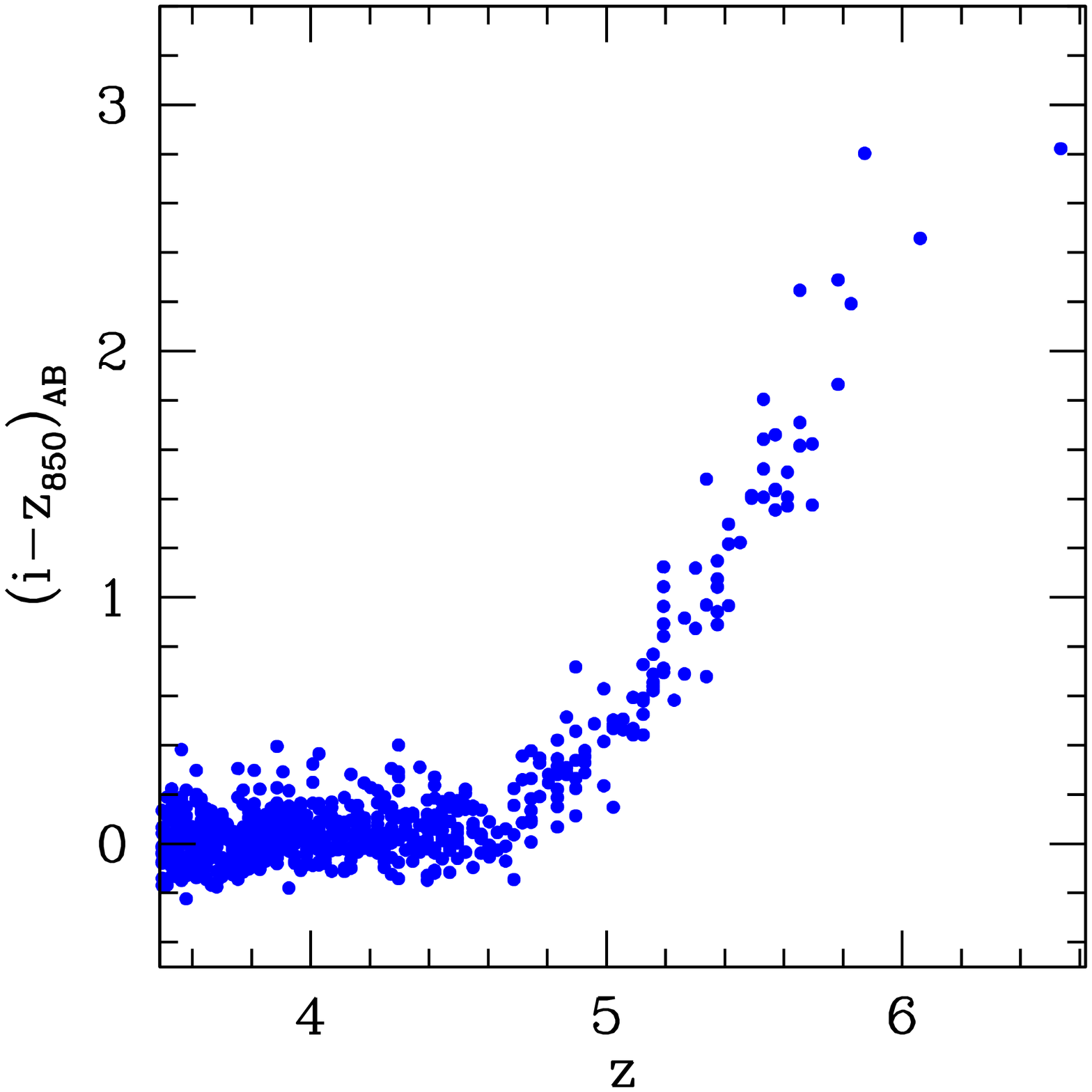} }
   
  \caption{$B$-$V$, $V$-$i$ and $i$-$z_{850}$ colors of galaxies with
    $z_{850}<26.2$ as a function of redshift.  Left panels: GOODS-S
    galaxies with spectroscopic redshift.  Mid panels: GOODS-MUSIC
    galaxies with photometric redshifts. Black arrows indicate that
    the magnitude estimation is based on an upper limit at
    1-$\sigma$. Right panels: model galaxies ({\tt std.f095.e03}).}
  \label{fig:colours}
\end{figure*}

The inclusion of the Lyman-$\alpha$ line does not change much the
predicted number counts of Lyman-break galaxies, with the exceptions
of the $i$-dropouts, which are especially sensitive to the line.
Figure~\ref{fig:lya} shows the effect of including the Lyman-$\alpha$
line, as explained in Section \ref{s:lya}, to the {\tt std.f095.e03}
best-fit model.  As expected, the effect of the Lyman-$\alpha$
emission line is to make the $i$-$z_{850}$ color bluer, thus shifting
the entrance in the selection criterion to slightly higher redshift
and decreasing the number counts.  This decrease is significant at the
bright end, which is then underestimated by the model (in one case it also
  removes the kink at bright luminosity, which is due to the presence
  of few, high-weight galaxies - see Section~\ref{s:deepfields}),
while the faint end is not strongly influenced.  This
conclusion does not change much with the choice of the $EW$.  The
inclusion of the Lyman-$\alpha$ emission line performed here is
clearly tentative, but this test shows that our results are robust
with respect to the inclusion of Lyman-$\alpha$ emission line under
reasonable assumptions.

We then concentrate on the {\tt std.f095.e03} model, to understand
to what extent model Lyman-break galaxies resemble the real ones.

Figure~\ref{fig:colours} shows the $B$-$V$, $V$-$i$ and $i$-$z_{850}$
colours of model and real galaxies as a function of redshift.
Observed galaxies are shown both from the GOODS-S catalogue, with
spectroscopic redshift, and for the GOODS-MUSIC catalogue, with
photometric redshift, which clearly has many more galaxies.  Only the
galaxies brighter than $z_{850}< 26.2$ are here considered, because of
the 90 per cent completeness limit of the GOODS-MUSIC catalogue.  As
already noticed in Fig. \ref{fig:colo}, model galaxies present colours
that closely resemble the observed ones, though they tend to be
slightly redder, especially in $B$-$V$ at z$\sim$4 (at least compared to
GOODS-MUSIC).  Analogous conclusions can be drawn by considering the
other best-fit model, {\tt std.f090.e01}.  This means that colours
cannot be used to constrain the parameters.  This conclusion is
strenghtened by noting that the strong reddening of all colors with
redshift is driven by IGM absorption.

\begin{figure*}
  \centerline{
    \includegraphics[width=6cm]{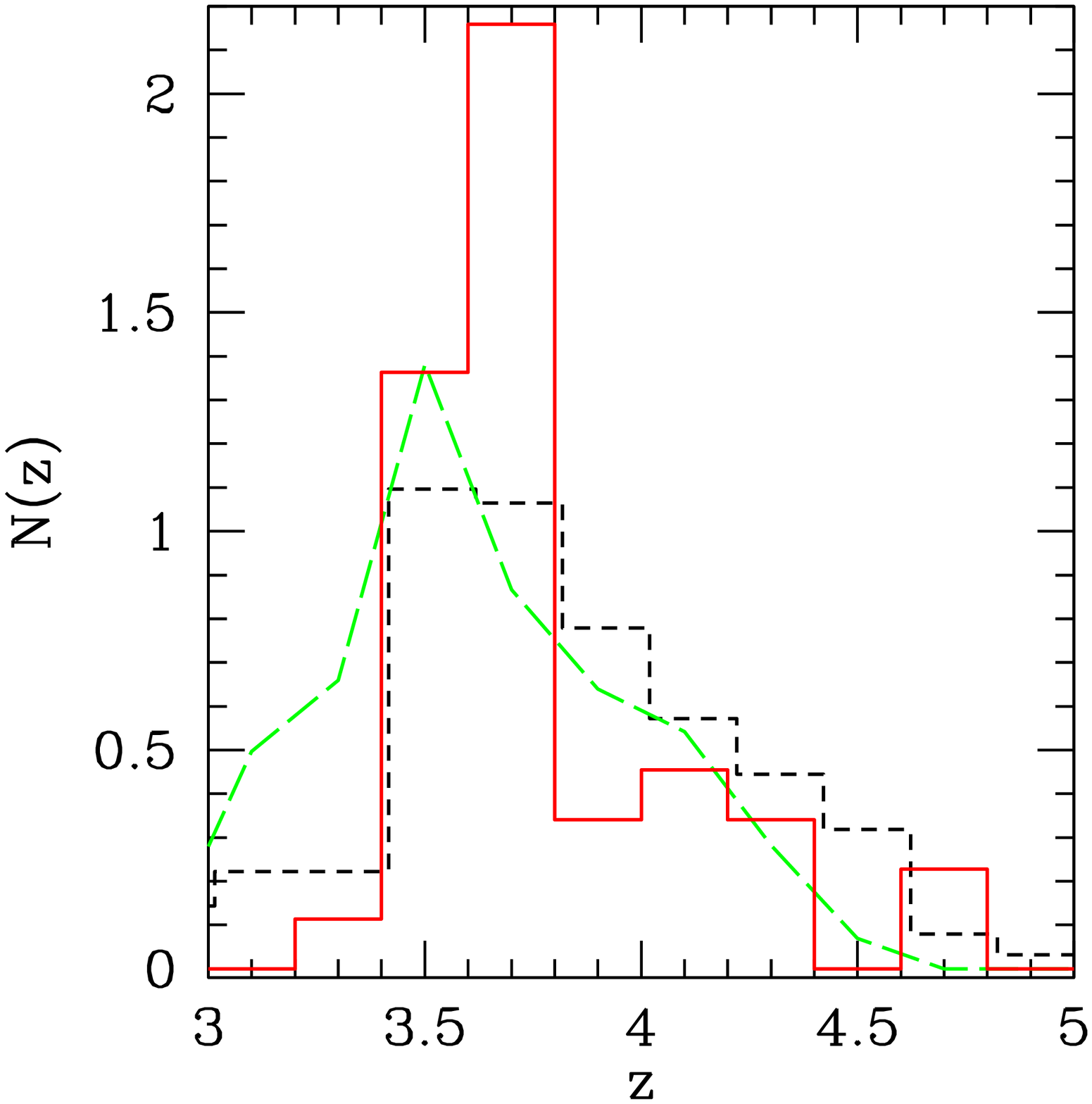} 
    \includegraphics[width=6cm]{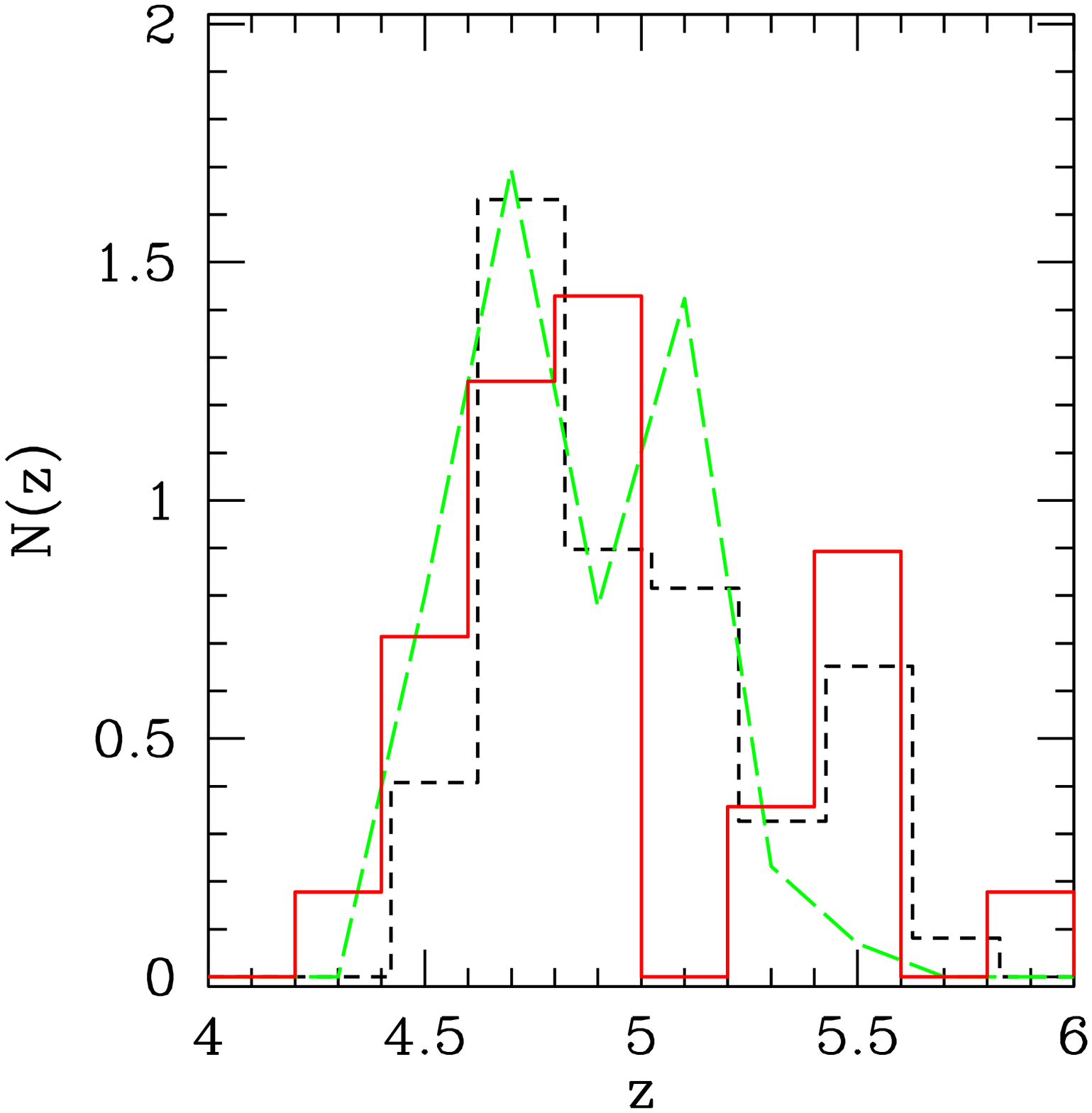} 
    \includegraphics[width=6cm]{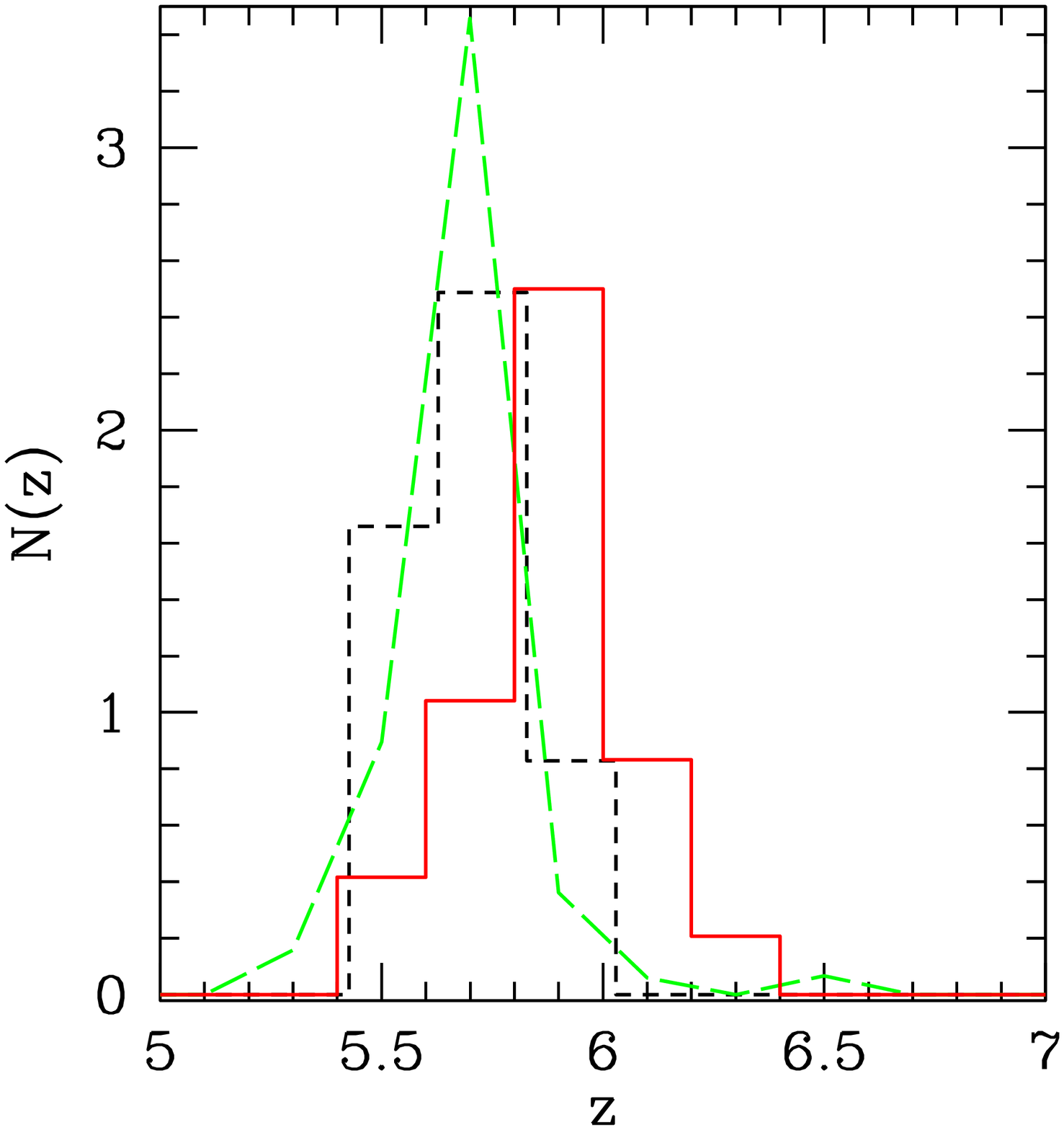} }
    
  \caption{Redshift distribution for $B$-, $V$- and $i$-dropouts in the
    GOODS-S (red solid lines) and GOODS-MUSIC (black dashed lines) data
    samples ($z_{850}<26.2$) and in the {\tt std.f095.e03} model
    (green long-dashed lines).}
  \label{fig:zdist}
\end{figure*}

Figure~\ref{fig:zdist} shows the redshift distributions of model
Lyman-break galaxies. We compare these predictions with the
spectroscopic redshift distribution of GOODS-S galaxies (red solid
lines) which constitute an incomplete sample \citep[see][for
details]{Vanzella08,Vanzella09} and with the photometric redshift
distributions of GOODS-MUSIC (black dashed lines) and consider only
galaxies brighter than $z_{850}=26.2$, the range where the GOODS-MUSIC
sample is complete. The agreement is good though not perfect; this can
be seen by comparing the average and standard deviation of the
predicted redshift distributions, ($3.71 \pm 0.35$; $4.88 \pm 0.27$;
$5.67 \pm 0.22$; for $B$-,$V$- and $i$-dropout respectively), with
those derived from \cite{Giavalisco04}, ($3.78 \pm 0.34$; $4.92 \pm
0.33$; $5.74 \pm 0.36$; for the three candidates respectively), the
GOODS-S spectroscopic sample ($3.78 \pm 0.37$; $4.98 \pm 0.39$; $5.92
\pm 0.20$) and GOODS-MUSIC photometric sample ($3.81 \pm 0.40$; $4.95
\pm 0.37$; $5.74 \pm 0.10$); however sample variance in the 150
arcmin$^2$ GOODS-S field is not negligible, so the level of agreement
is considered satisfactory, especially if we consider that a
confirmation of the validity of the redshift distribution is
represented by the consistency between number counts and luminosity functions.  Anyway,
since the uncertainty related to the redshift measurements increases at
higher redshifts where the detection is more difficult, this
comparison should be taken with grain of salt.  In particular, the
number of available photometric redshifts of $i$-dropout galaxies is
very low, because only few such galaxies are bright enough to be
detected in $J$ and $K$ bands.

\begin{figure*}
  \centerline{
    \includegraphics[width=6cm]{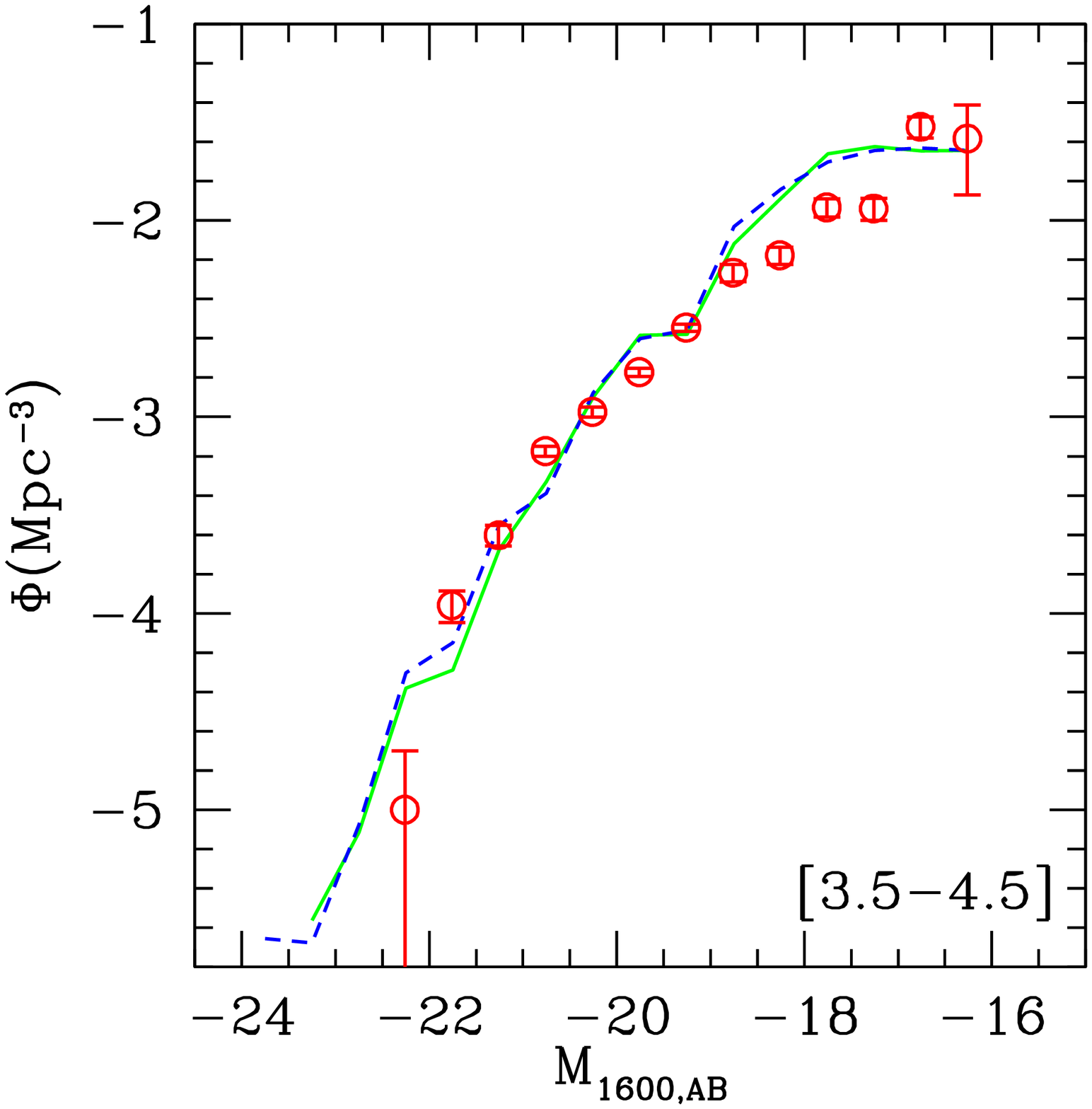} 
    \includegraphics[width=6cm]{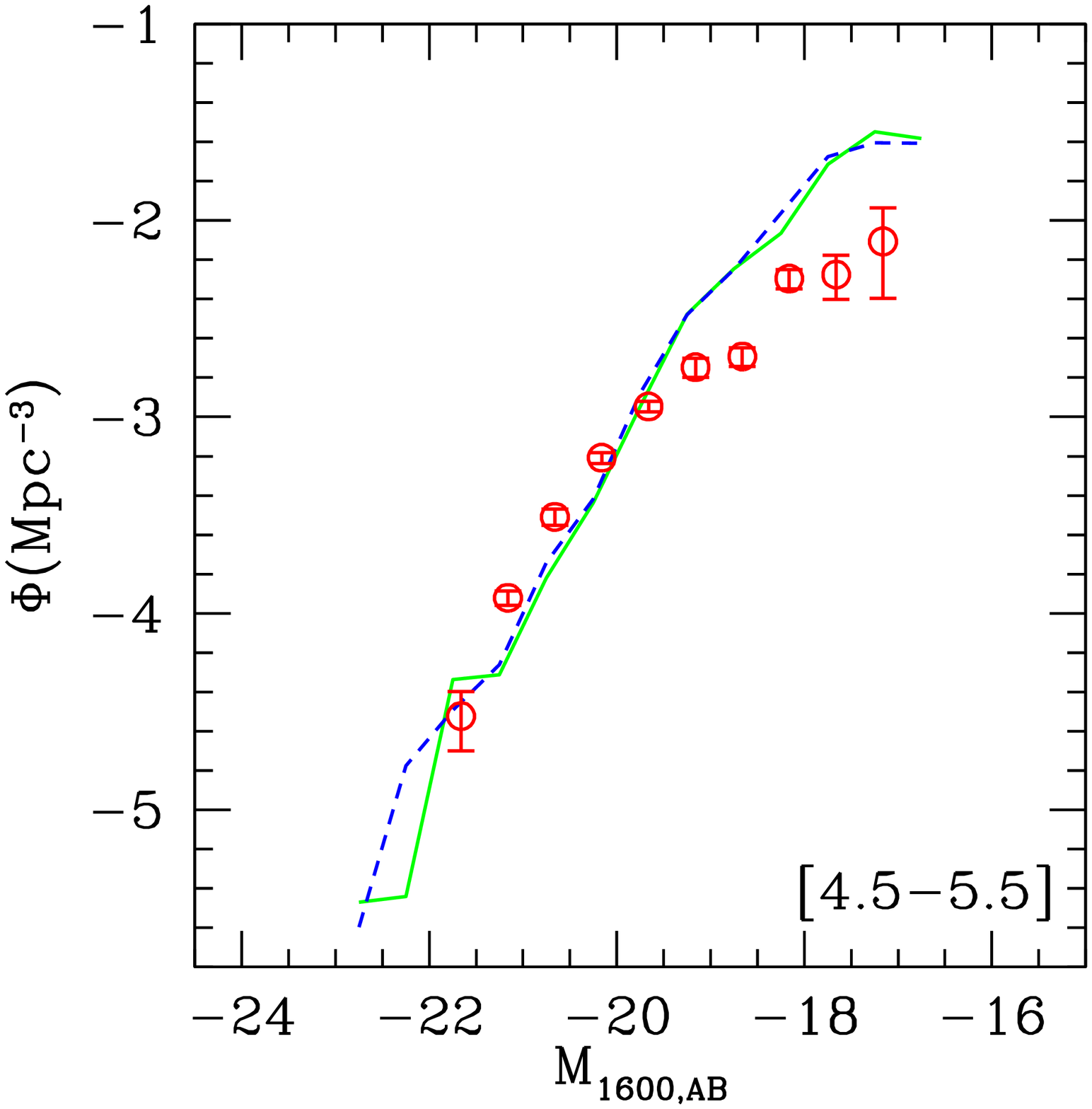} 
    \includegraphics[width=6cm]{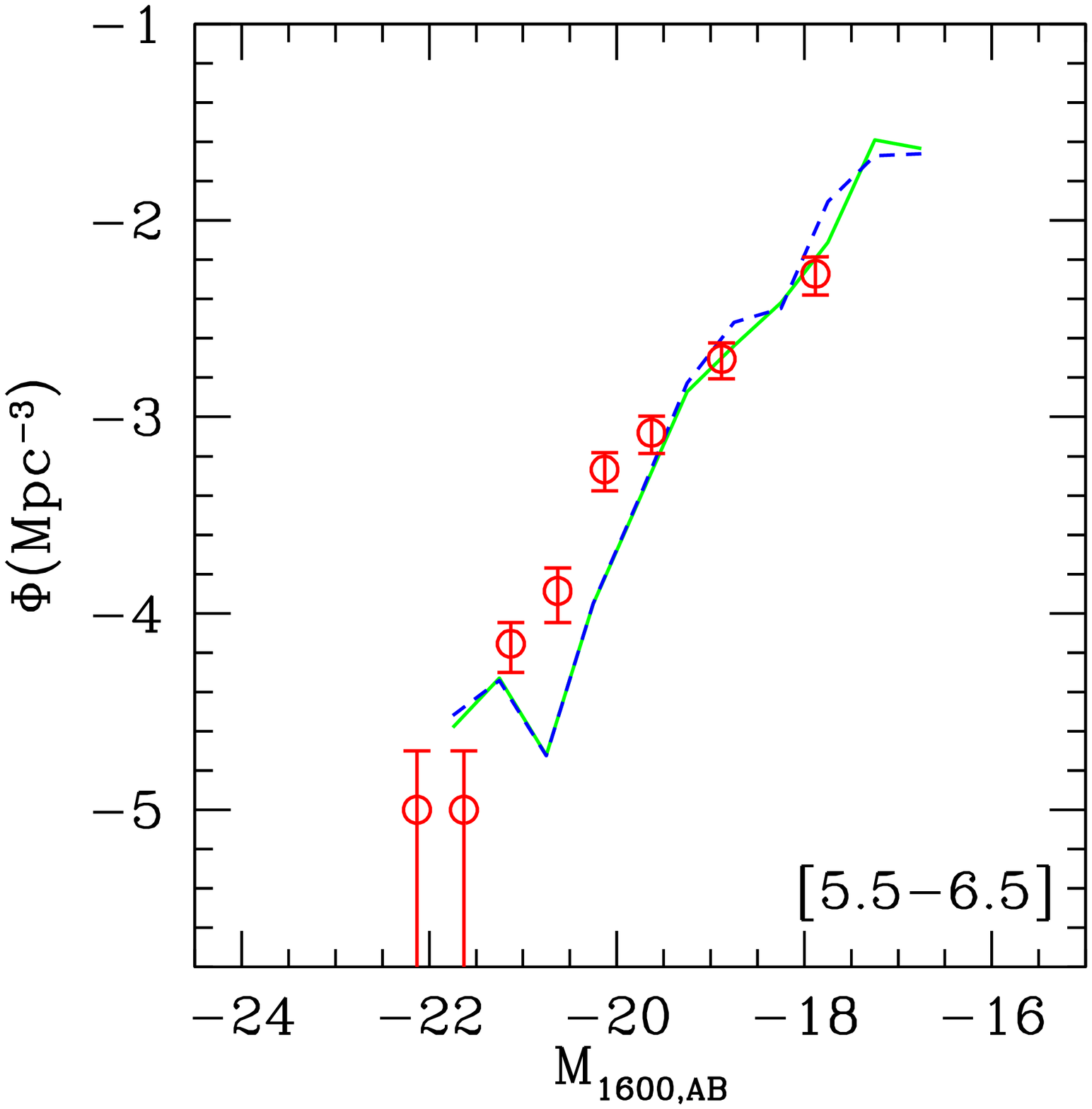} }

  \caption{Luminosity function of $B$-, $V$- and $i$-dropouts.  (Red)
    points are the estimate by Bouwens et al. (2007), (green) solid
    and (blue)
    dashed lines give predictions for the {\tt std.f095.e03} 
    and {\tt std.f090.e01} models.}
  \label{fig:lf}
\end{figure*}

\cite{Bouwens07} estimated the luminosity function of Lyman-break
galaxies by scaling all observations to the HUDF ($5$-$\sigma$) depth of 29 in the
$z_{850}$ band.  Assuming that the $B$-dropout technique selects all
star-forming (and not severely obscured) galaxies in the redshift
interval $3.6<z<4.5$, they obtained selection functions of $V$- and
$i$-dropouts by degrading images of low-$z$ $B$-dropout galaxies to
reproduce observational biases at higher redshift, then estimating how
many of them would be selected.  As shown in
figures~\ref{fig:colo} and~\ref{fig:selection}, most model galaxies in that redshift range
with high enough $z_{850}$ flux are $B$-dropouts, so it is fair to
compare Bouwens et al.'s estimate to the luminosity function of all
model galaxies in the redshift range $3.5<z<4.5$, $4.5<z<5.5$ and
$5.5<z<6.5$.  Figure~\ref{fig:lf} shows the luminosity function of the
two best-fit models, compared to the \cite{Bouwens07} determination,
adjusted to take into account the slightly different cosmology.  The
agreement is again good at high luminosities (with a possible underestimate
of the $i$-dropout bright tail), while a clear excess is
visible in the faint tail, especially in the $V$-dropouts.  We stress
that this result is different from the one of
figure~\ref{fig:best-fit} relative to number counts, because the two
comparisons follow two different approaches: in the first case the
model tries to reproduce exactly the observed quantity, in the second
case the observers try to reconstruct a quantity more directly
comparable with models, i.e. the luminosity function.  The good
agreement of data and model in terms of both number counts and
luminosity functions cofirms that model galaxies closely resemble the
observed population and their selection function.  Moreover, the model
follows very well the positive evolution (with time) of the knee of
the luminosity functions, and this allows us to support Bouwens et
al.' conclusion that we are indeed witnessing the hierarchical
build-up of the population of young galaxies.

Summing up, the {\sc morgana} model reproduces the main observational
properties (number counts, colours, selection functions, redshift
distributions and estimated luminosity functions) of bright
Lyman-break star-forming galaxies at $z>3.5$, (though a possible
underestimate of the number of bright $i$-dropouts may be present);
fainter galaxies, with $z_{850}\ga27$, have acceptable properties but
are over-produced.  However, dust attenuation heavily depends on
uncertain parameters of the ISM like $t_{\rm esc}$ and $f_{\rm mol}$,
so that multiple solutions are equally acceptable.

\section{A population of excess galaxies}
\label{s:inxs}

\begin{figure*}
  \centerline{
    \includegraphics[width=6cm]{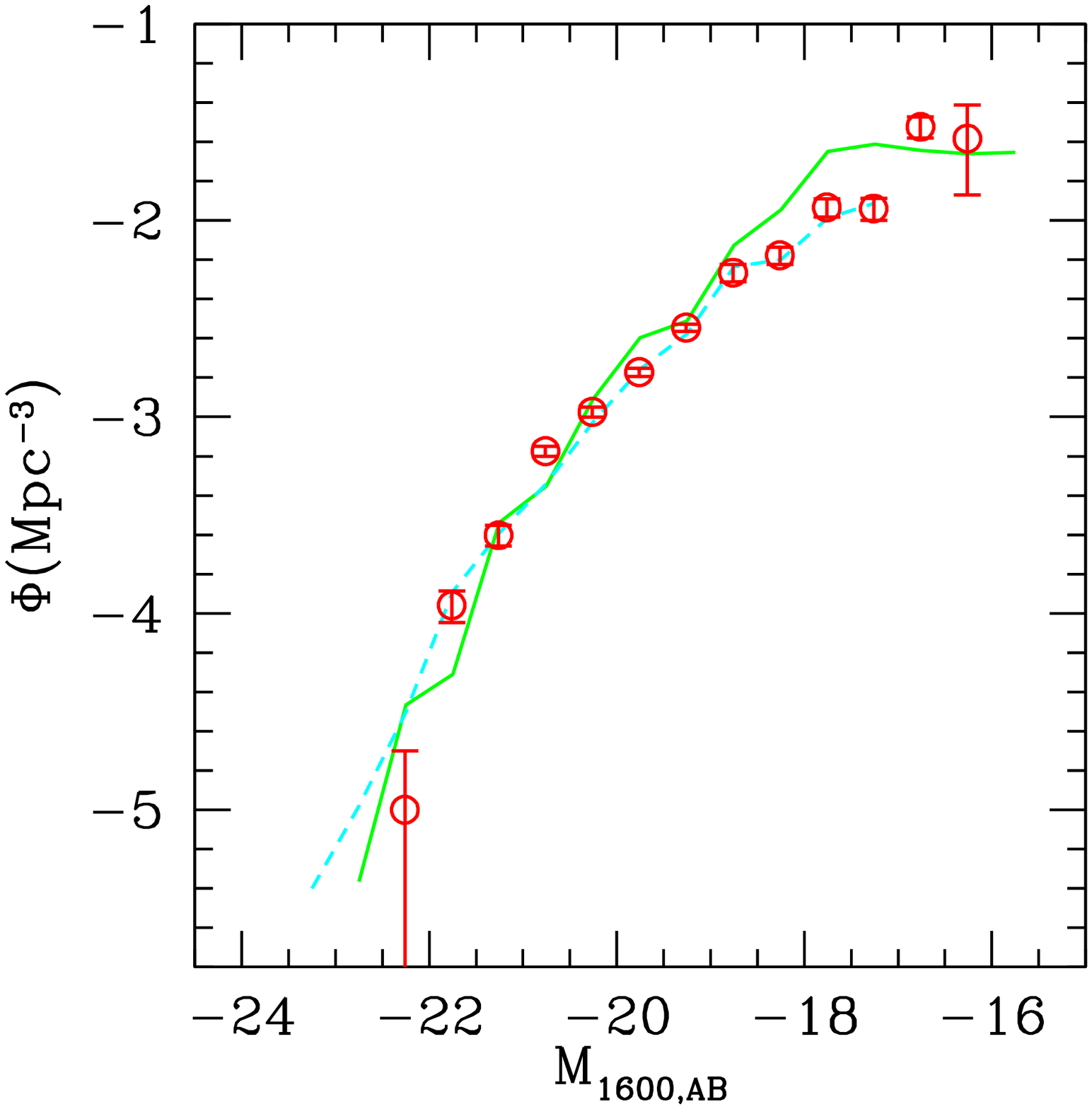} 
    \includegraphics[width=6cm]{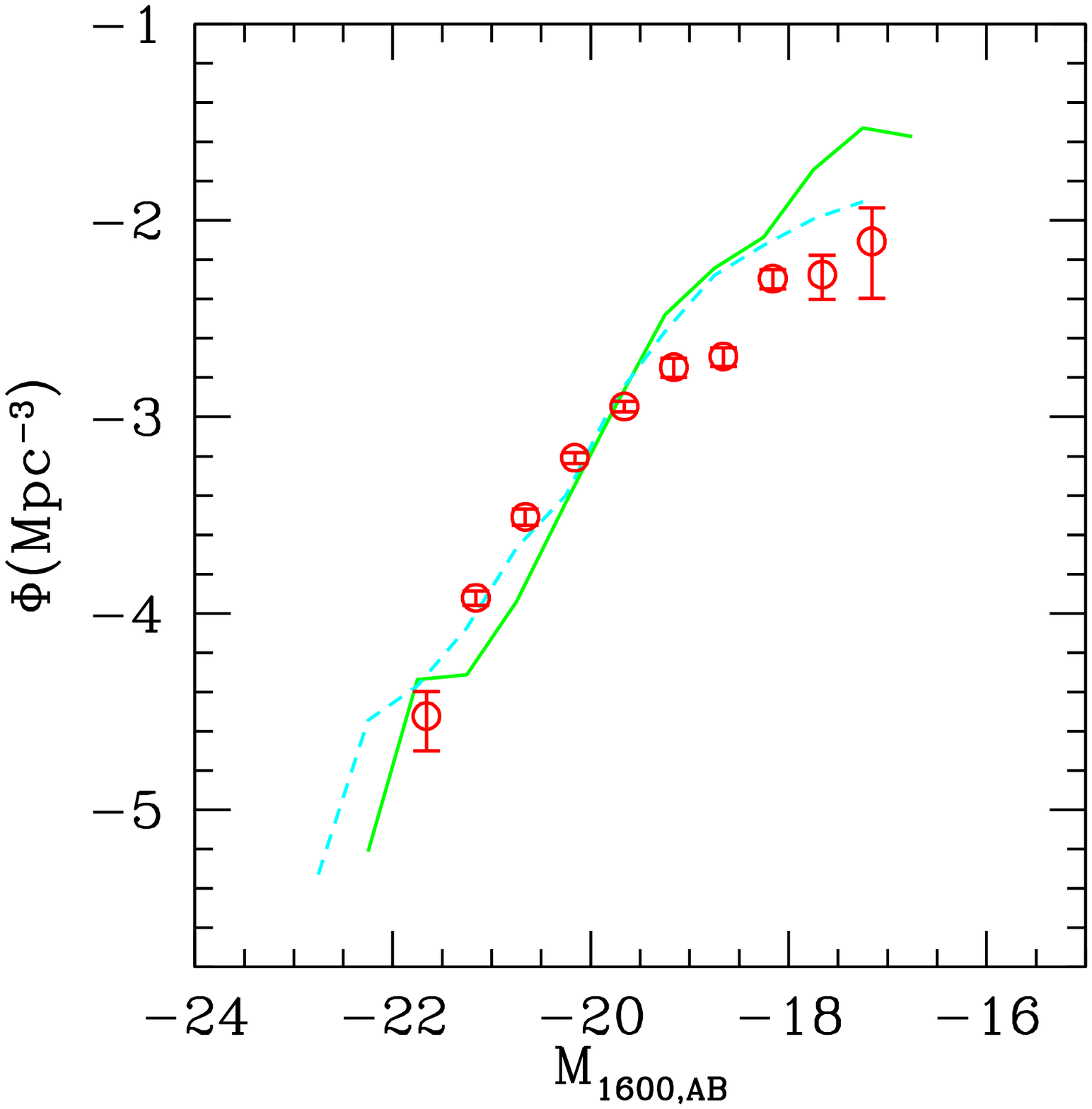} 
    \includegraphics[width=6cm]{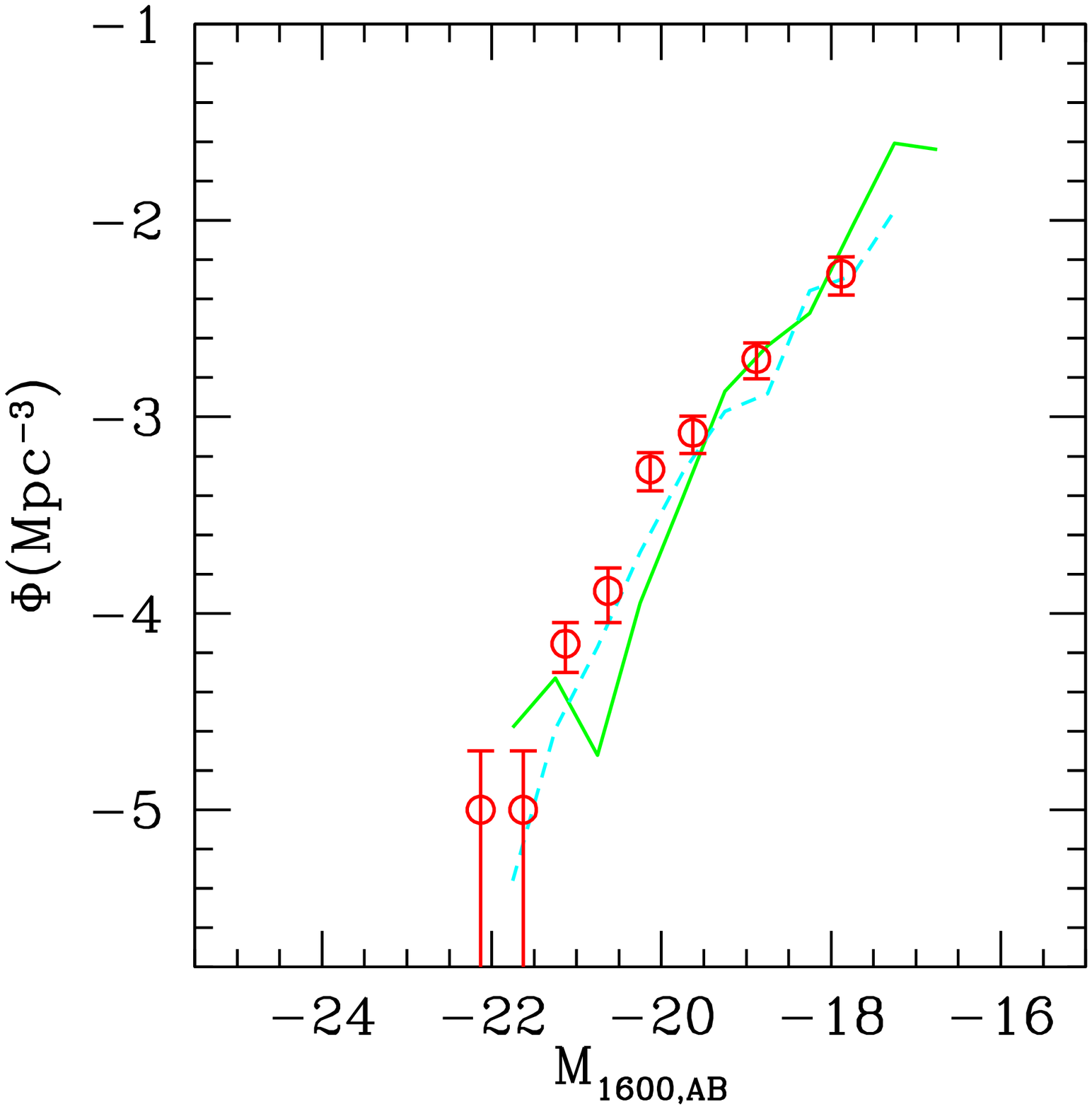} }

  \caption{Luminosity function of $B$-, $V$- and $i$-dropouts.  (Red)
    points are the estimate by Bouwens et al. (2007), (green) solid
    and (cyan) dashed lines give predictions for the {\tt
      nowind.f095.e03} and {\tt old.f050.e03} models.}
  \label{fig:oldlf}
\end{figure*}

Degeneracies in model predictions do not allow us to draw strong
conclusions from the good agreement shown above between model and real
Lyman-break galaxies at the bright end of the luminosity function.
Indeed, parameter degeneracy is not due only to {\sc grasil}; two
{\sc morgana} parameters, namely $n_{\rm dyn}$ and $n_{\rm quench}$ (see
section~\ref{s:improvements}) influence the star-formation rate
density at $z>3$ but give small difference at lower redshift, where
more constraints are available.
To strengthen this point, we show in Figure~\ref{fig:oldlf} the
prediction for the luminosity function of a model run without
quasar-triggered galaxy winds ({\tt nowind.f095.e03}) and the {\tt
  old.f090.e03} used in \cite{Fontanot07}, where a Salpeter IMF is
used in place of Chabrier, and a slightly different cosmology with
$\sigma_8=0.9$ is adopted.  The last model is run on a $512^3$ box
with slightly worse mass resolution.  The fit is very similar to our
best fits in all cases, and this confirms that bright Lyman-break
galaxies cannot give strong constraints to models as long as dust
attenuation is so uncertain.
The reason why quasar winds do not affect this observable is the following.
In our implemented scheme for quasar-triggered winds
  \citep{Fontanot06}  radiation from an accreting black hole can
  evaporate some cold gas in the star-forming bulge; if the putative
  evaporation rate is higher than the star formation rate, then all
  the cold mass in the host bulge is removed. Quasar winds are then
  able to quench star formation in bulges (mergers), and this could
  lead to a suppression of star formation.  However, due to the
  (modeled) delay between star formation and black-hole accretion,
  quasar winds take place when the star formation episode that has
  triggered accretion is almost over.  The independence of the
  Lyman-break luminosity function on winds is a further demonstration
  of this.

Previous papers \citep{Fontana06,Cirasuolo08,Marchesini08,Fontanot09b}
have shown that {\sc morgana} presents a deficit of massive galaxies
at $z>3$, when compared to the stellar mass function or to the K-band
luminosity function.  At the same time, {\sc morgana} is able to
reproduce the sub-mm counts of $z\sim2$ starburst galaxies
\citep{Fontanot07}, and even shows a modest excess of bright
$B$-dropouts (fig. \ref{fig:best-fit}).
  The two evidences of low stellar masses and a correctly high SFR are not contradictory: compared with similar
  models, {\sc morgana} produces stronger cooling flows and then
  stronger bursts of star formation that are readily quenched by
  stellar feedback, so that the high levels of star formation are
  maintained for short times and {\it do not} produce higher amounts
  of stellar mass.
  Moreover, the apparent underestimate of stellar masses by {\sc
    morgana} may be at least in part due to systematics in stellar
  mass estimates.  \cite{Tonini08} showed that inclusion of TP-AGBs
  in the Simple Stellar Population libraries give a significant
  downward revision of stellar masses at high redshifts or,
  equivalently, a boost to the $K$-band magnitude of model
  galaxies. According to \cite{Marchesini08}, our stellar mass
  function at $z\sim3$ remains low at the $\sim2\sigma$ level, but
  when we compare stellar masses of our model Lyman-break galaxies
  with estimates of \cite{Stark09}, which take into account TP-AGBs,
  we find a slight over-prediction, in contrast with the evidences
  mentioned above.  It is then difficult at this stage to assess to
  what extent a good match of the bright end of the luminosity
  function of Lyman-break galaxies at $z\sim4-5$ is in contrast with
  the apparent underestimate of the high end of the stellar mass
  function at $z\sim3$.

\begin{figure*}
  \centerline{
    \includegraphics[width=8cm]{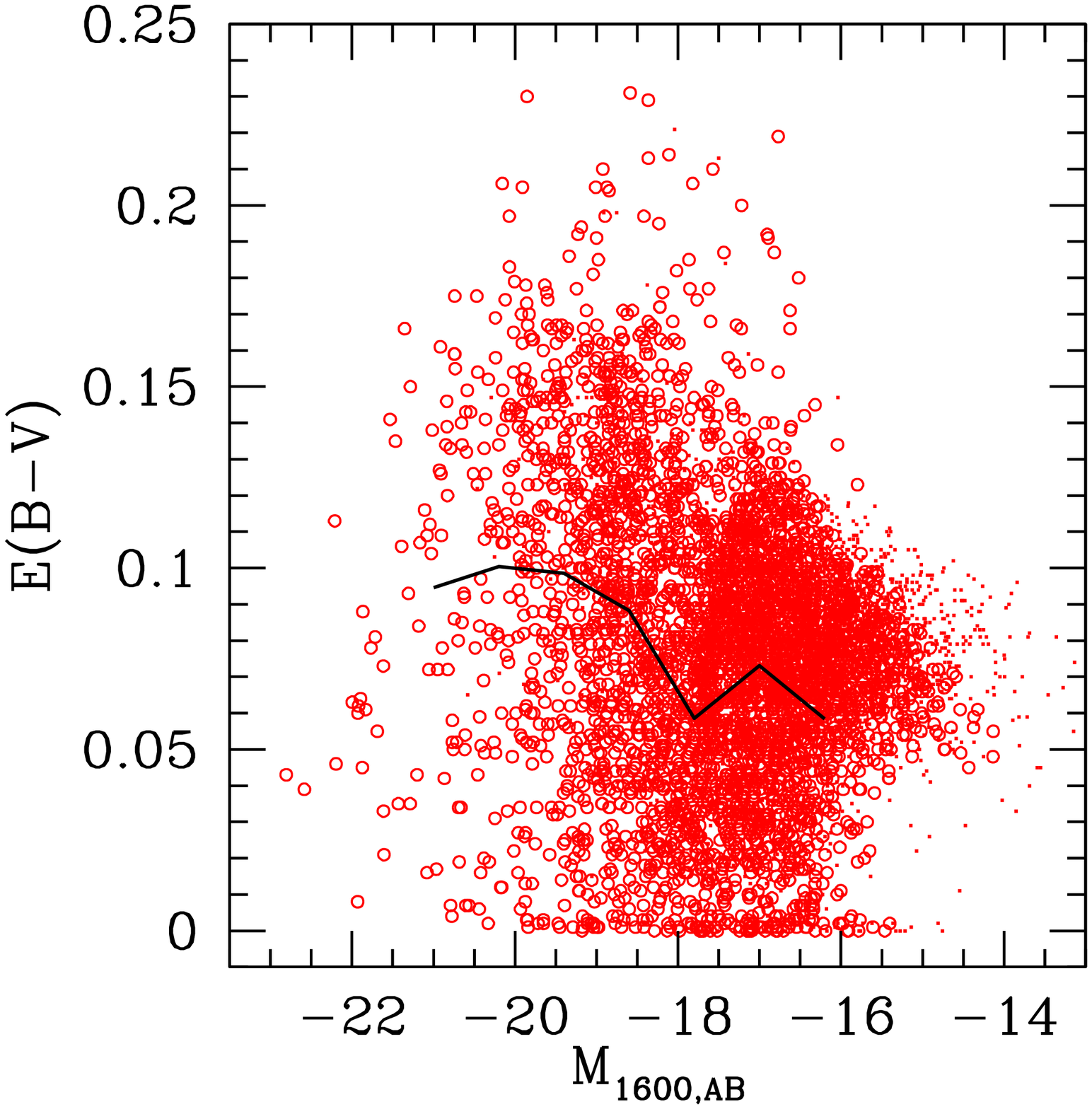} 
    \includegraphics[width=8cm]{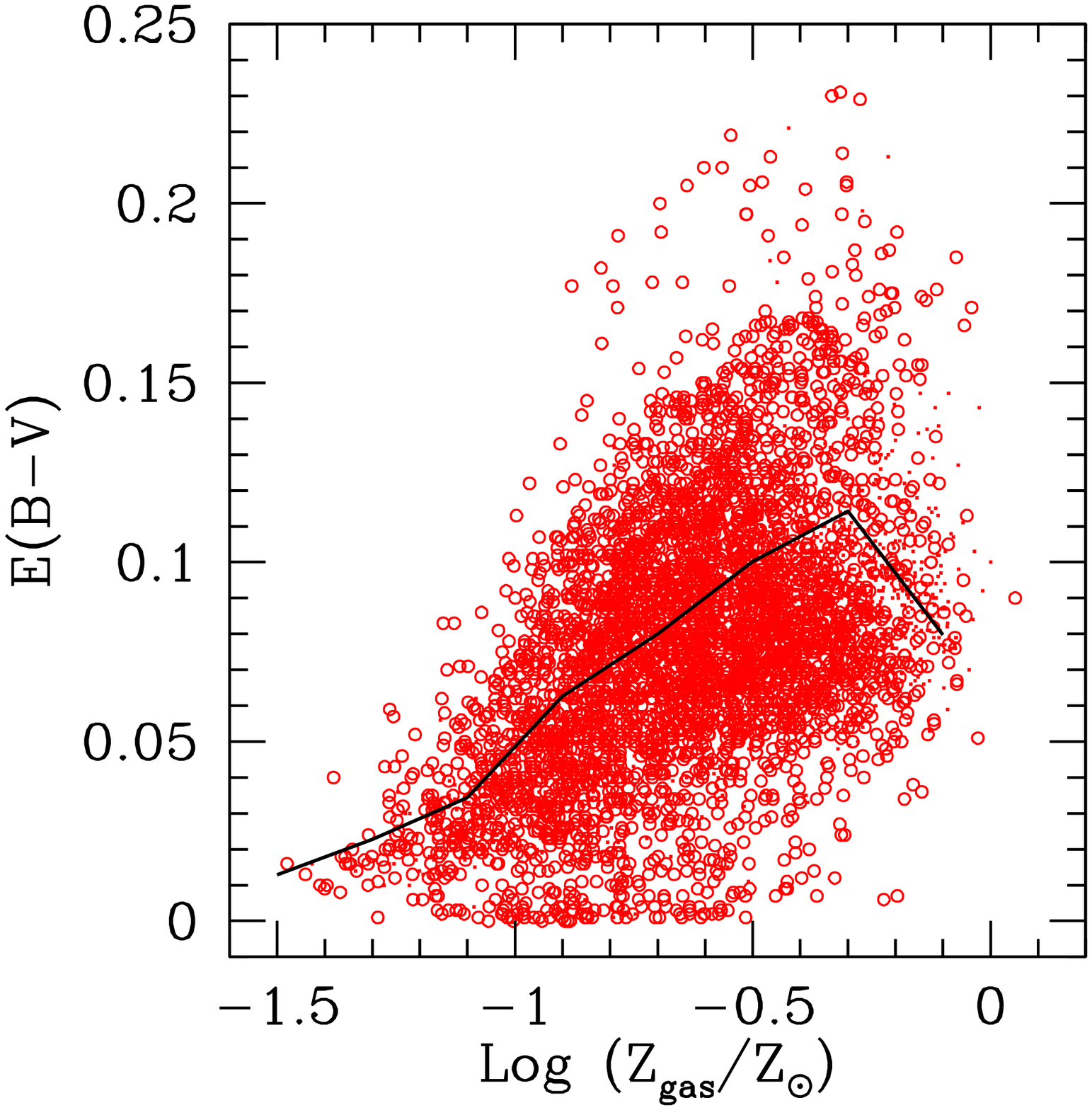}}
  
  \caption {$E(B-V)$ attenuation as a function of the UV rest-frame luminosity $M_{1600,AB}$ (left panel) and gas-metallicity (referred to solar metallicity)(right panel). All model $V$-dropout galaxies at redshift about $5$ and with $Log(M_{\star})>7.75$, are here considered. Open circles refer to central galaxies. The black solid lines represent the `weighted' mean values of $E(B-V)$ computed, respectively, for each bin of magnitude and gas-metallicity. $B$- and $i$-dropouts show a similar trend.}

  \label{fig:vdropebvandmetal}
\end{figure*}

\begin{figure*}
  \centerline{
    \includegraphics[width=6cm]{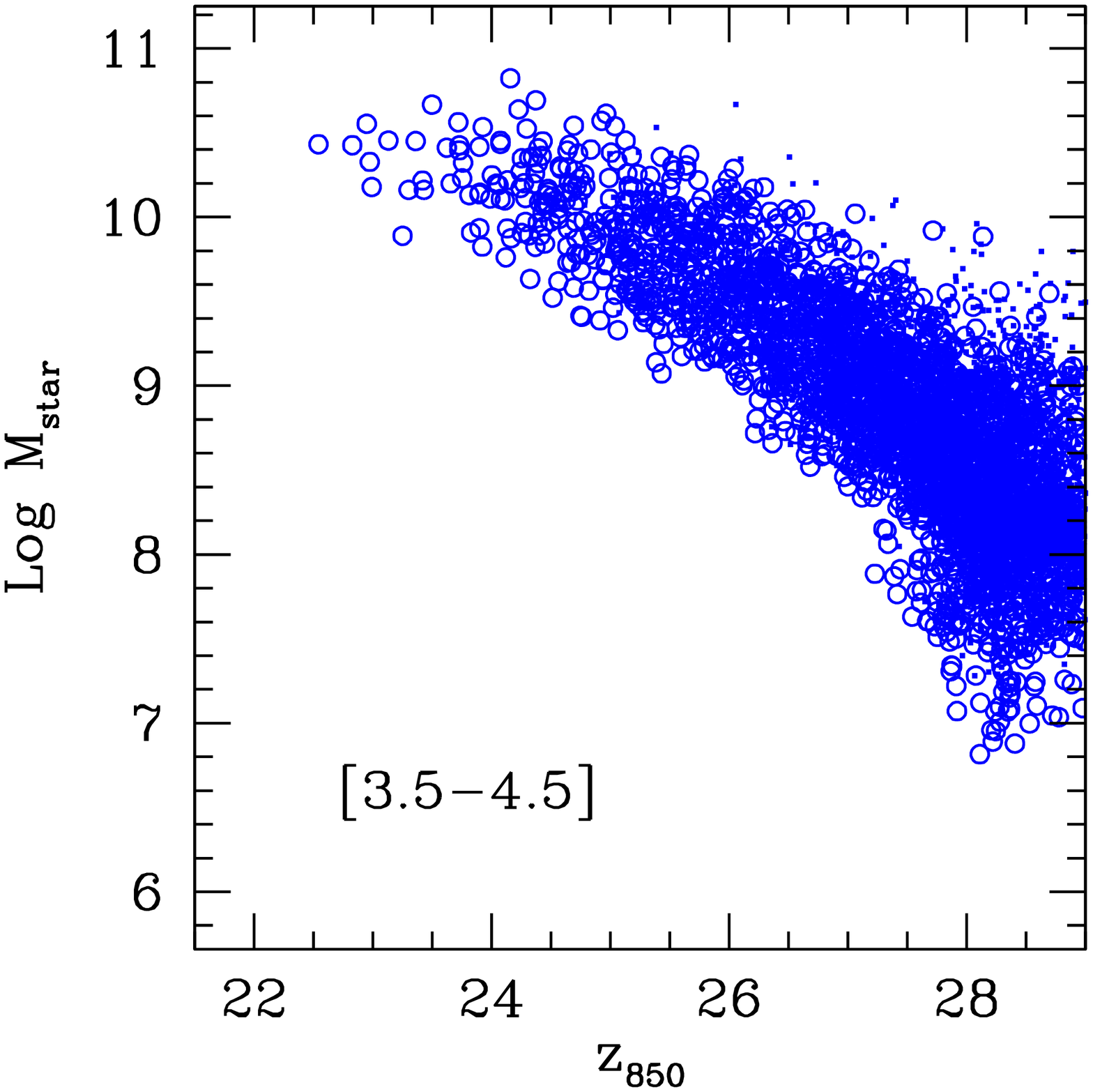} 
    \includegraphics[width=6cm]{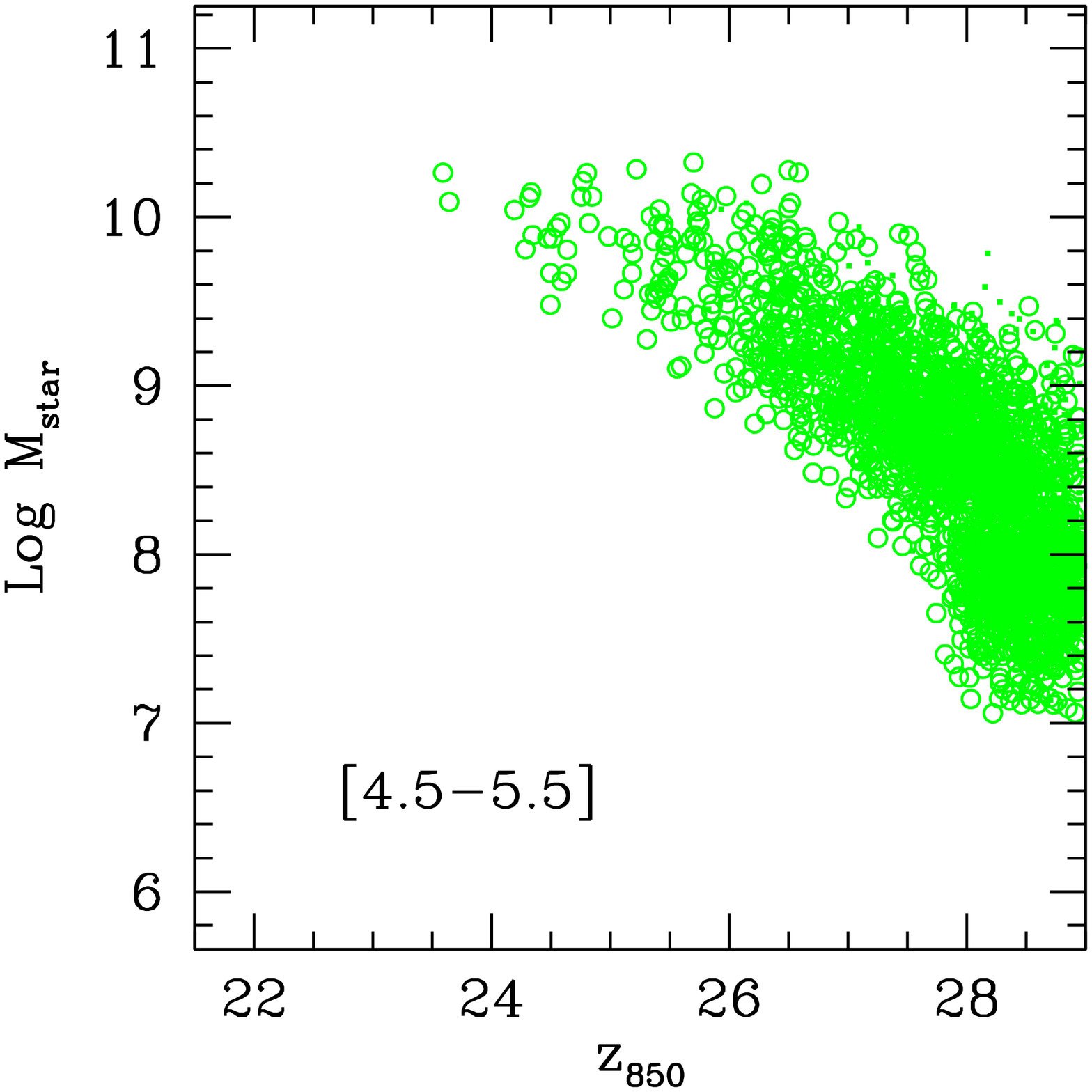} 
    \includegraphics[width=6cm]{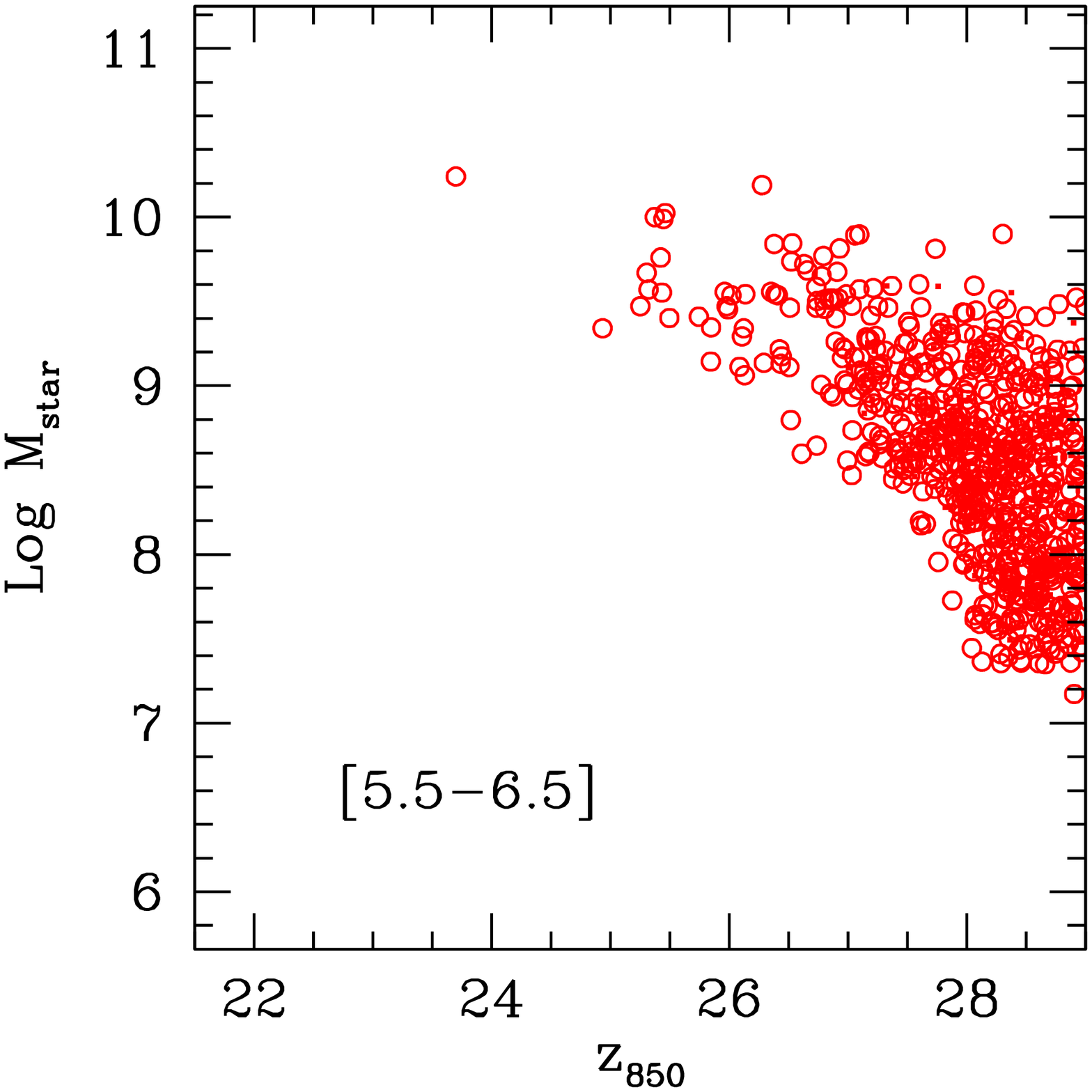} }
  \centerline{
    \includegraphics[width=6cm]{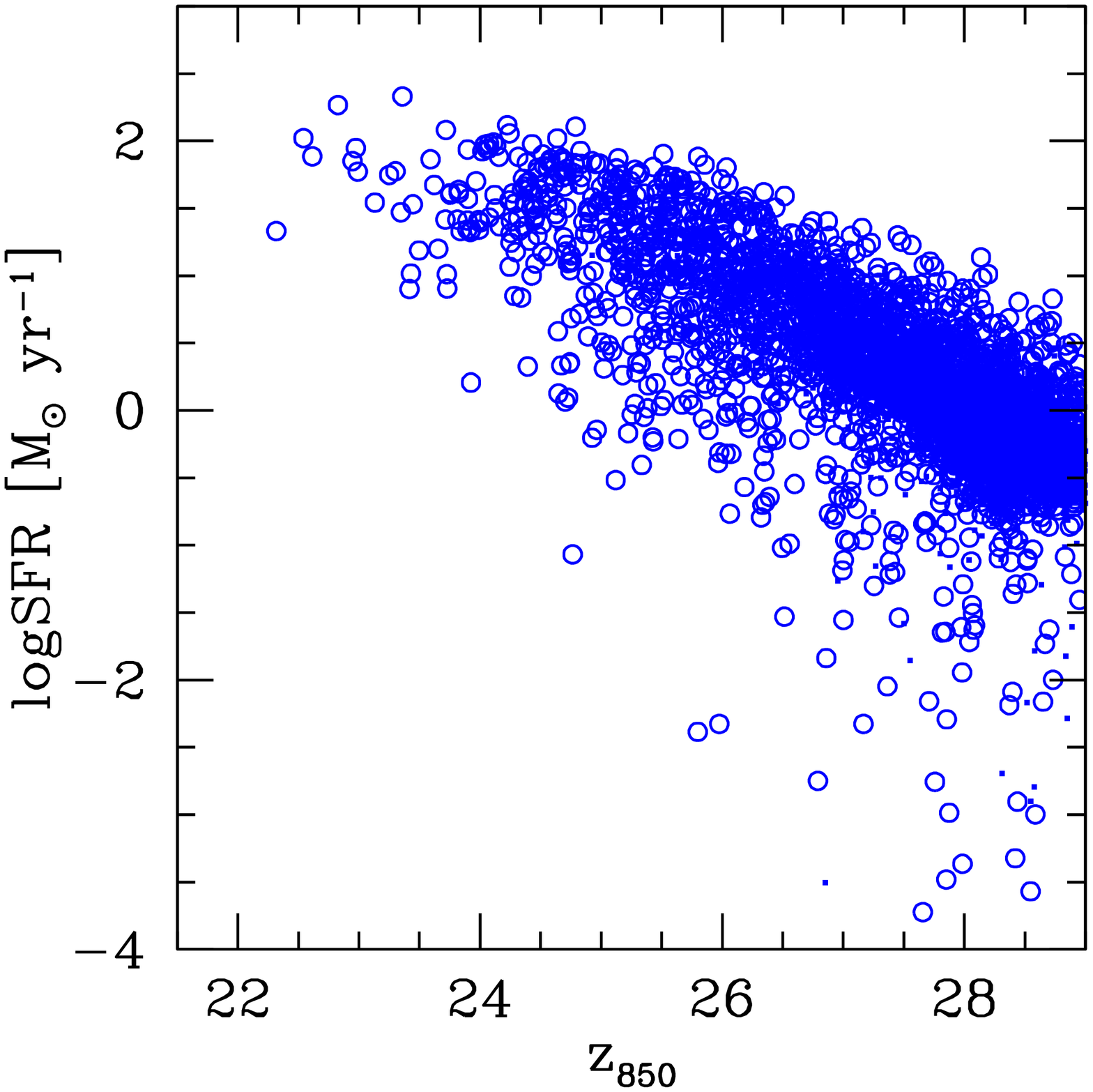} 
    \includegraphics[width=6cm]{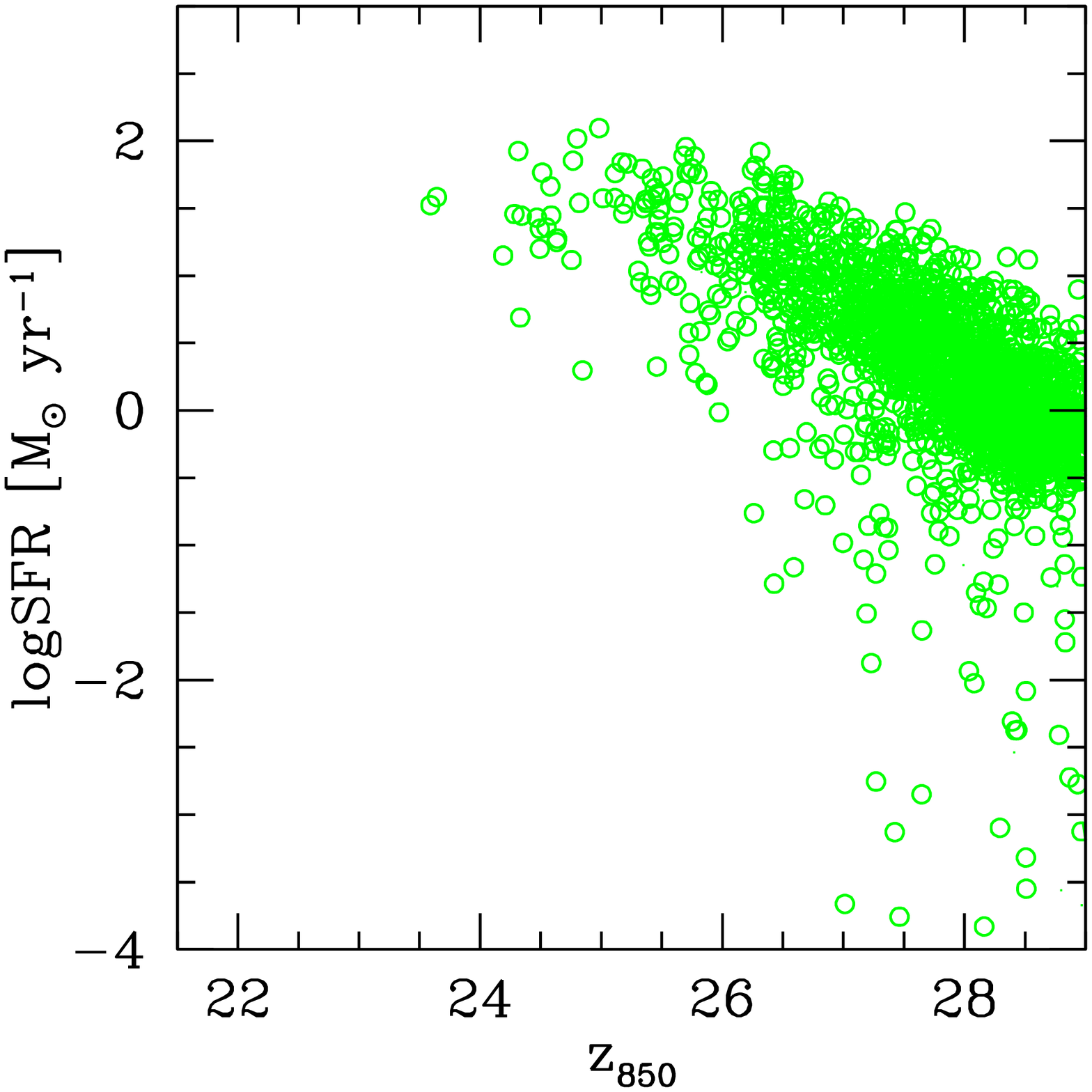} 
    \includegraphics[width=6cm]{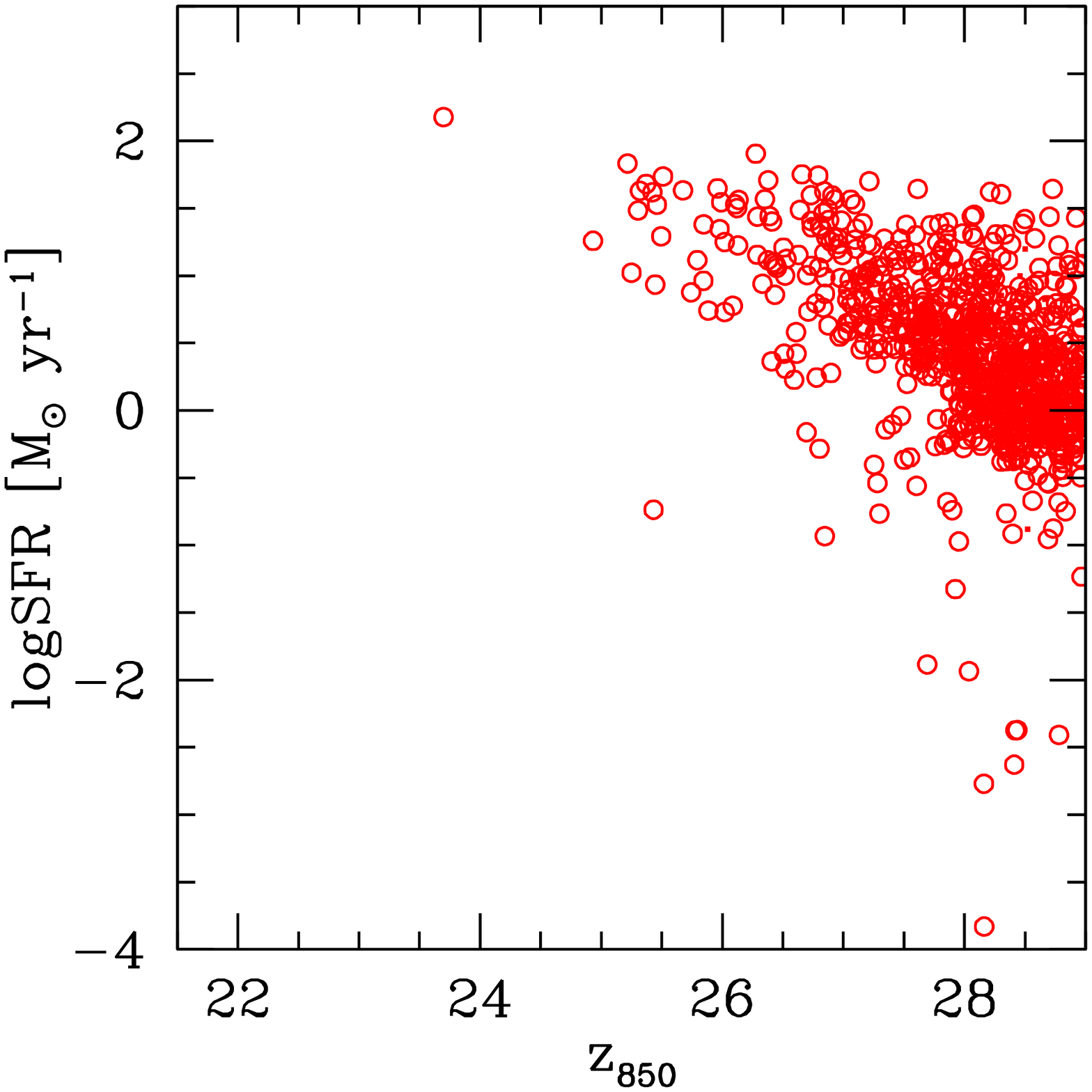} }
  \centerline{
    \includegraphics[width=6cm]{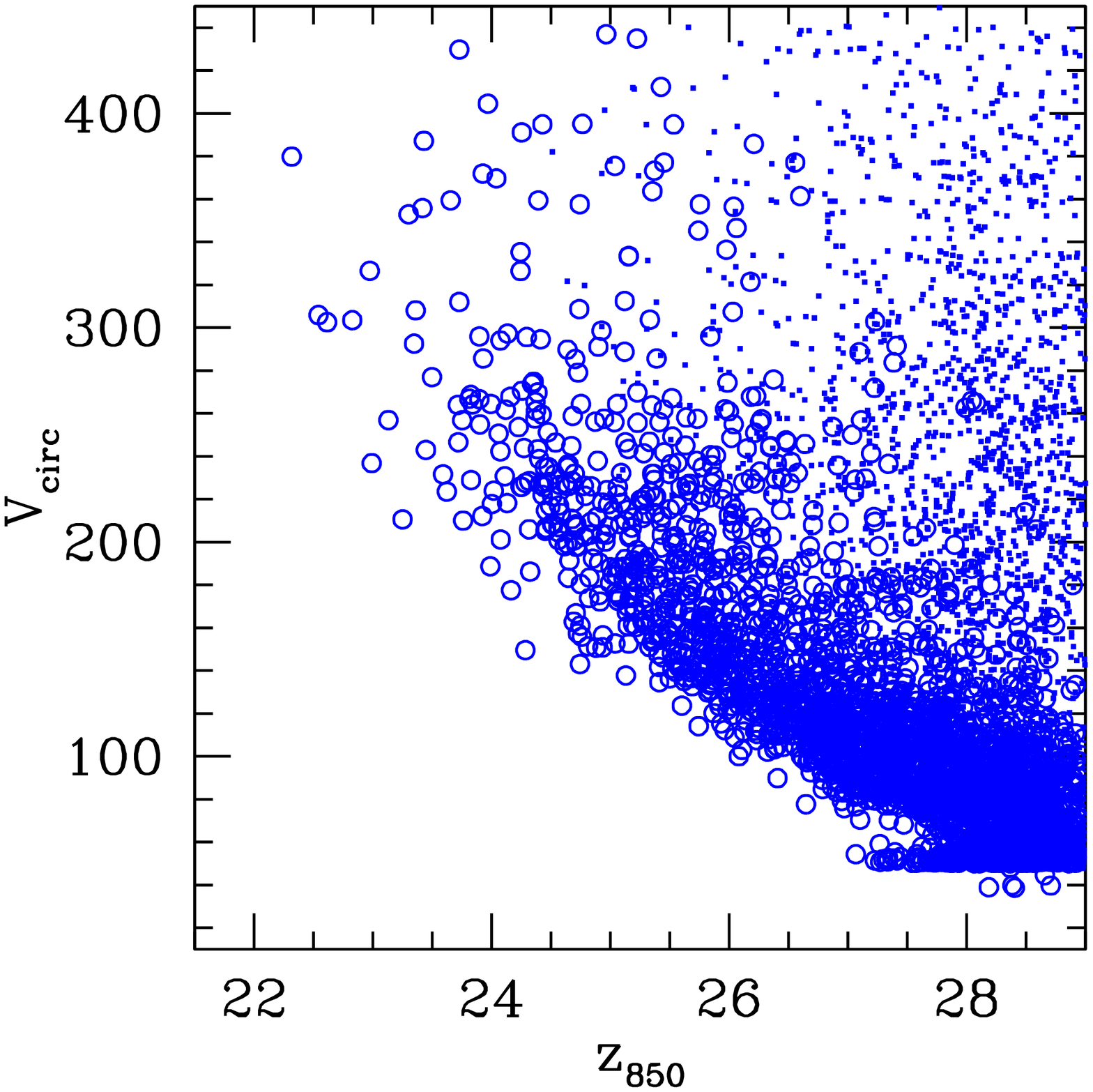} 
    \includegraphics[width=6cm]{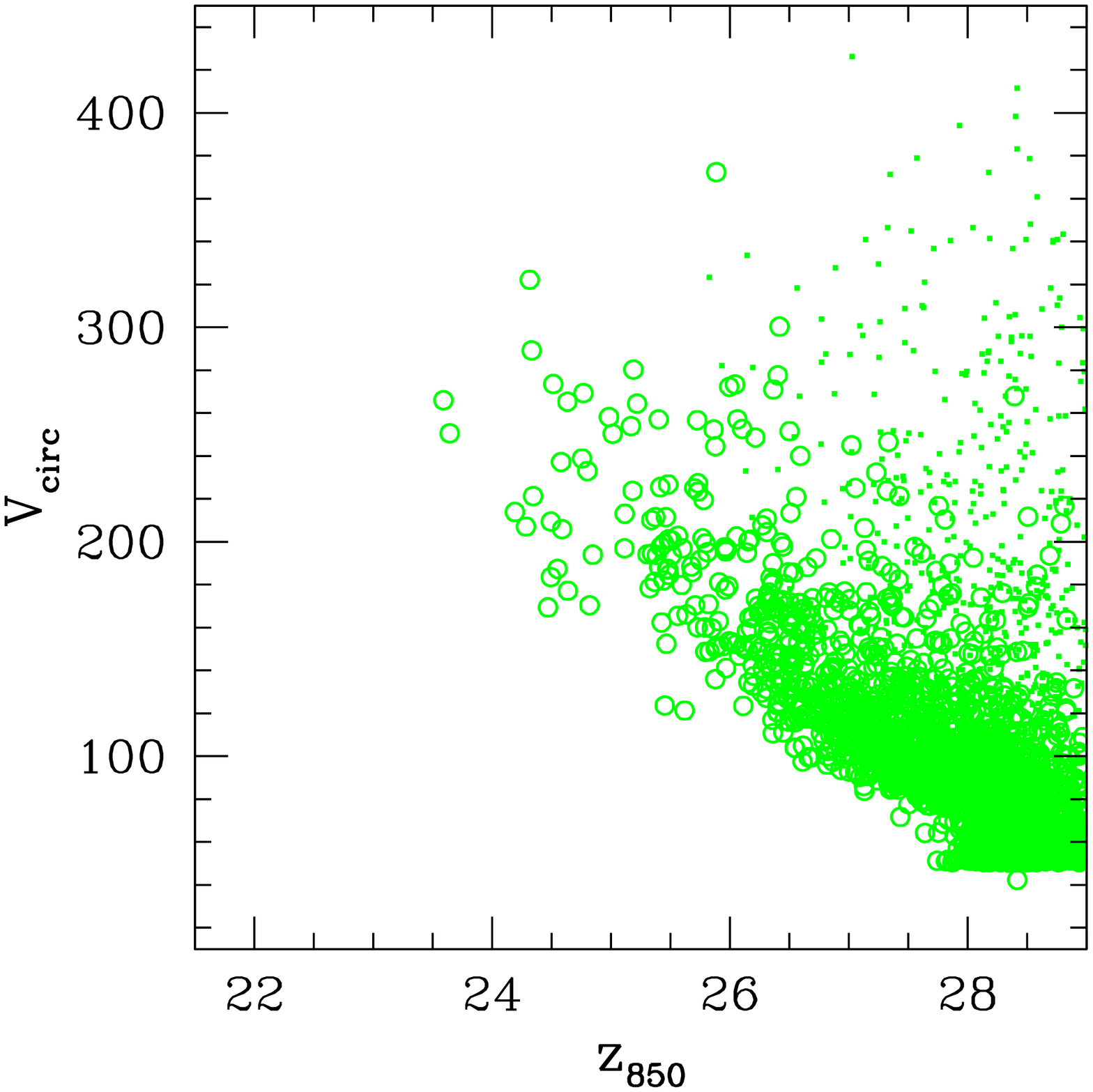} 
    \includegraphics[width=6cm]{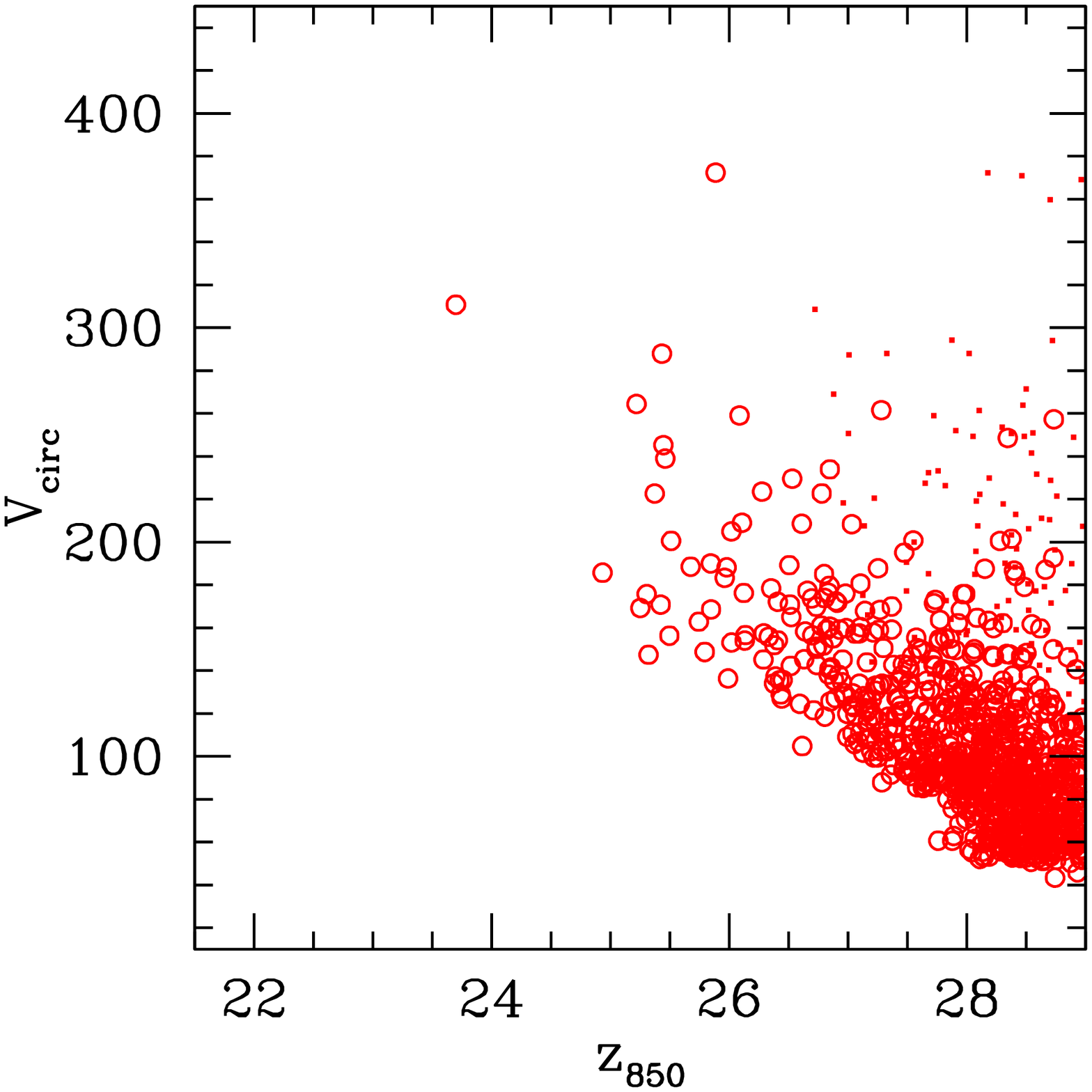} }

  \caption{Properties of model Lyman-break galaxies as a function of
    $z_{850}$ magnitude, for the model {\tt std.f095.e03}.  Upper
    panels give stellar masses, mid panel give star formation rates,
    lower panels give dark matter halo circular velocities.  Bigger
    open points refer to central galaxies.}
  \label{fig:relations}
\end{figure*}

Once the bright tail is matched, the model shows a clear excess of
faint dropouts.
Is this excess robust?  From the one hand, as explained in the
Introduction, the excess is expected: \cite{Fontanot09b} demonstrated
that galaxies with stellar masses (at $z=0$) in the $10^9-10^{10}$
{\msun} range are formed too early and are too passive since $z\sim3$,
and this is the main cause of the lack of (stellar mass and
archaeological) downsizing in models.  It follows that an excess of
small star-forming galaxies must be present at higher redshift, and
because the data go deeper than the completeness limit of the model,
this excess must be visible.  On the other hand, number counts at such
deep magnitudes may well be incomplete.  The use of the HUDF, and the
careful analysis of Bouwens et al. to determine the selection
functions, makes it unlikely that the model excess is purely due to
sample incompleteness.  Moreover, the reconstructed faint slope of the
luminosity function is $\sim-1.7$, in agreement with many other
determinations both at $z\ge3$ \citep{Yoshida06, Beckwith07, Oesch07}
and at $z\sim2$ \citep{Reddy08}, while the slope of model luminosity
functions is significantly steeper, in excess of $-2.0$.

Is this excess an artifact of dust extinction?  in other words, is
  it possible that {\sc morgana} is underestimating the star formation
  rate of bright galaxies (as well as stellar masses) but our {\it
    ad-hoc} tuning of dust parameters absorbs this discrepancy,
  creating an excess of faint galaxies? Figure~\ref{fig:vdropebvandmetal} shows the $E(B-V)$ attenuation as a function of UV rest-frame luminosity $M_{1600,AB}$ (left panel) and gas-metallicity (right panel) for model $V$-dropout galaxies at $z \sim 5$ and with $Log(M_{\star})>7.75$; central galaxies are highlighted as empty circles. The solid black lines represent the `weighted' mean values of $E(B-V)$ computed, respectively, for each bin of magnitude and gas-metallicity. For $B$- and $i$-dropouts we obtain a similar trend.
  The figure shows that the attenuation we obtain 
  is only slightly lower ($E(B-V)\sim0.1$ in place of 0.15) than what found by 
  \cite{Shapley01} at lower redshift, (but
  we are not using the same SED template of those authors because
  every galaxy in our model has its own SED), while, as mentioned above, the attenuation in the observed $i$ and $z_{850}$ bands is very similar to
  that used by \cite{Bouwens07}; the excess remains if we
  attenuate our galaxies as suggested in that paper.  Moreover, 
  the faint-end slope of model luminosity functions, in excess of
  $-2.0$, is inconsistent with the latest and very stable $-1.7$ value; with
  such a difference, an excess of faint galaxies will be obtained even
  if the bright tail is underestimated. This steep slope is in part
  due to to the fact that faint galaxies are less absorbed than bright
  ones, something that is well known in literature \citep[see,
    e.g.,][]{Sanders96}, and is consistent with the trend found by
  \cite{Shapley01} at $z\sim3$.  We conclude that it is unlikely that
  this excess is due to a peculiar behaviour of dust attenuation as
  predicted by {\sc grasil}.

\begin{figure}
  \centerline{
    \includegraphics[width=8cm]{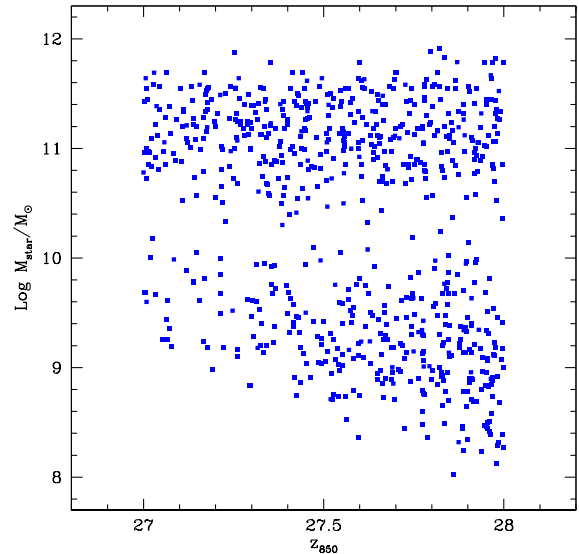} }
  \caption{Stellar masses of the $z=0$ descendant galaxies of
    $V$-dropouts with $27<z_{850}<28$.}
  \label{fig:desc}
\end{figure}

Other theoretical groups have compared their predictions of
Lyman-break galaxies to data (see the reference list in the
Introduction), but no group has clearly claimed to find a
similar excess, with the exception of \cite{Poli01,Poli03}, who
compared the predictions of a semi-analytical model \citep{Menci02}
with the reconstructed luminosity function of $I$ and $K$-selected
samples ({\em not} color selected). They reported an excess of faint
$I\sim27$ galaxies at lower redshift, $z\sim3$, which is even stronger
than the one we find here. 
 Also, \cite{Night06} and \cite{Finlator06}, extracting catalogues
  of Lyman-break galaxies from N-body simulations, find rather steep
  luminosity functions, with slopes $\sim -2$. \cite{Night06} discuss
  in some detail the implications of this steep slope, but claim broad
  consistency with available data; compared to the more recent
  determination of Bouwens et al., such steep slopes are nominally
  ruled out.
There are two reasons why many groups may
have not noticed such an excess: first, very deep fields like GOODS
and the HUDF are required to reliably sample the $z_{850}\sim27-28$
magnitude range, and these have been made available only recently.
Second, a simple (and over-simplistic) magnitude-independent
attenuation makes the excess less visible, so a more advanced tool as
{\sc grasil} is needed to notice it.

In the following of this section we try to assess whether the
excess can be fixed simply by a better fine-tuning of model
parameters, without spoiling the good match with local galaxies. To
answer this question, we focus on the properties of the excess
population.  Figure~\ref{fig:relations} shows stellar masses, star
formation rates and dark matter circular velocities of $B$-, $V$- and
$i$-dropout galaxies as a function of apparent $z_{850}$ magnitude,
for the best-fit model {\tt std.f095.e03}; central galaxies are
highlighted as empty circles.  Excess galaxies with $z_{850}\sim
27-28$ are characterized by star formation rates of order of $\la10$
{\msunyr}, stellar masses growing from $10^8-10^9$ {\msun} at $z\sim6$
to a factor of $\sim3$ higher at $z\sim4$, and the central ones are
hosted in dark matter halos with masses in the range from $10^{11}$ to $10^{12}$
{\msun}, whose circular velocities are in excess of 100 km
s$^{-1}$.  Figure~\ref{fig:desc} shows the stellar mass of the $z=0$
descendant galaxies of $V$-dropouts with $27<z_{85}<28$. Roughly half
of them do not grow much in mass since $z\sim5$, thus falling in the
stellar mass range of galaxies that are too old in models.  The other
half merge with larger galaxies to end up at the knee of the stellar
mass function.  In many papers, like e.g. \cite{Santini09}, the
observed specific star formation rates of massive galaxies is found to
be higher than model predictions by a factor at least of $\sim4$,
which is apparently at variance with the good match of the growth of
stellar mass in the same galaxies since $z\sim3$ \citep{Fontana06}.
The puzzle may be solved if the excess population of small galaxies
contributes to inflate the growth of massive galaxies by minor mergers, thus
compensating for the lower star formation rates.  We propose this as a
plausible explanation, although more stable estimates of star
formation rates in galaxies are necessary to assess this problem.

\begin{figure}
  \centerline{
    \includegraphics[width=8cm]{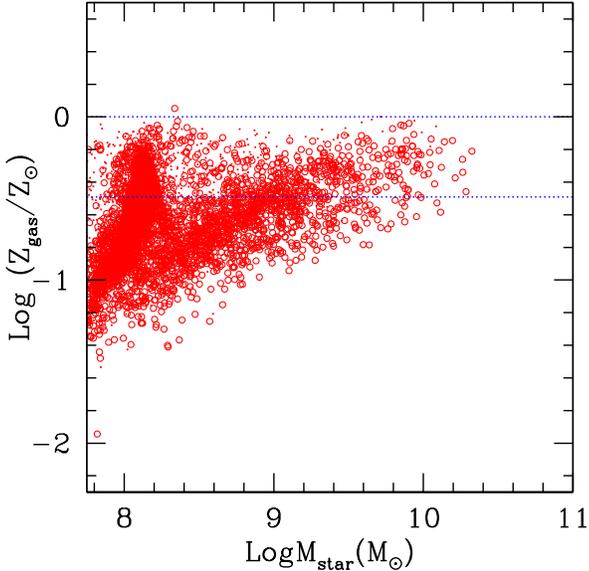} }
  \caption{Stellar mass -- gas metallicity relation for model galaxies
    at $z=5$.  In this plot all galaxies in the box are used. Bigger
    open points refer to central galaxies. The two horizontal lines
    mark solar and $1/3$ solar metallicities. We removed galaxies with
    less than 10 per cent gas fraction.}
  \label{fig:metals}
\end{figure}

An important hint on the mechanism responsible for the excess of faint
galaxies comes from their metallicities $Z$.  Observations show that
metallicities measured from emission lines decline with redshift, and
this decline is stronger for smaller galaxies.  This has been measured
at $z\sim2$ by \cite{Savaglio05} and by \cite{Erb06}, \cite{Erb07},
and at $z\sim3$ by \cite{Maiolino08}, but some measures exist at
higher redshift \citep{Ando07} that seem to confirm the trend up to
$z\sim6$.  Besides, \cite{Pettini01} report a value of $1/3$ solar as
the typical metallicity of $z\sim3$ Lyman-break galaxies.  We show in
figure~\ref{fig:metals} the predicted metallicities of the gas
component of galaxies at $z=5$ as a function of their galaxy stellar
mass.  We show gas metallicities because these are measured from
emission lines.  A mass-metal relation is visible, but with a very
high normalization, for which $10^{10}$ M$_\odot$ galaxies are already
enriched above $1/3$ solar, the typical gas metallicity of $z\sim 3$
Lyman-break galaxies.  Moreover, a blob of small, highly enriched
galaxies is visible, that is probably absent in the real world.  These
high metallicities justify the slightly redder colours of model
galaxies and the shift toward lower redshift of the selection function,
and may be at the origin of the underestimate of the number of bright
$i$-dropouts.
High metallicities at high redshift are a common problem of
  galaxy formation models, as pointed out in \cite{Maiolino08};
  relatively high metallicities at $z\sim2-3$ are reported by
  \cite{Guo08}, and studying model predictions for the Milky Way
  \cite{DeLucia08} noticed a dearth of low metallicity halo stars, due
  to excessive enrichment of small galaxies (before they become Milky
  Way satellite) at high redshift.
  Can this mismatch in gas metallicity create the excess of faint
  galaxies? Observations \citep{Maiolino08} suggest a downsizing trend for
  metal enrichment, with the stellar mass-gas metallicity relation
  steepening at high redshift. Our relation is rather flat, so that we
  are probably overestimating gas metallicity more in fainter objects.
  If more metals give higher attenuation, this mismatch would lead to
  an artificial {\it flattening} of the LF, thus decreasing the excess
  of faint  Lyman-break galaxies rather than creating it.

There is a broad consensus among modelers on the fact that AGN feedback is most
effective in massive galaxies at low redshift, so the excess galaxies
should be suppressed by stellar feedback. This should work in the
direction of creating strong gas and metal outflows, so as to keep low
both stellar mass and metallicity.  Coming back to the question on
whether this excess can be suppressed by fine-tuning model parameters,
it is easy to do it by increasing the efficiency of stellar feedback,
but at the cost of destroying the good fit of the stellar mass
function at low redshift.  Indeed, as clear in
Figure~\ref{fig:relations}, such galaxies, though small, are hosted in
relatively large dark matter halos with circular velocities well in excess of 100
km s$^{-1}$.  Because the efficiency of gas removal mainly scales with
halo circular velocity \citep[as found since the seminal work
  of][]{Dekel86}, this would be done at the cost of suppressing also
star formation in those $10^{12}$ {\msun} halos at $z=0$ in which star
formation efficiency is highest, thus destroying the good agreement
with the local stellar mass function (and luminosity functions as
well).

Stellar feedback is manily driven by SNe, and each SN produces roughly
$10^{51}$ erg of energy, which translates into $10^{49}$ erg per {\msun}
of stars formed if roughly 1 SN explodes each 100 {\msun} of solar
masses. If a fraction $\varepsilon$ of SN energy is injected in
  the ISM instead of being quickly radiated away ($\varepsilon\sim0.3$
  in our model),
the velocity $v_{\rm eject}$ at which a mass $M_{\rm eject}$
can be accelerated by a mass $M_\star$ of stars formed is  $v_{\rm
  eject}\sim 500 \sqrt{(\varepsilon/0.3) M_\star/M_{\rm eject}}$ km
s$^{-1}$.  As a result, one
{\msun} of stars can not eject much more than one {\msun} of gas out of a
halo with circular velocity of $\sim200$ km $s^{-1}$ (and escape
velocity higher by the canonical factor of $\sqrt{2}$).  
In order to limit the number and metal content of faint 
Lyman-break galaxies hosted  in 100 km/s halos at
  $z\sim5$, 
$v_{\rm eject}$ must be higher than it is at $z\sim0$, 
and this can be
obtained as follows: (i) increasing energy feedback efficiency
$\varepsilon$ (if not the SN energy), (ii) increasing the number of
SNe per solar mass of stars formed, (iii) decreasing halo circular
velocities (or their number at fixed $V_c$).  The first solution would
imply that feedback works differently in such starburst, with the
paradoxical conclusion that feedback can be most efficient in the
densest environments.  Anyway, the gain factor is unlikely to be very
high, because if 32 per cent of SN energy (in discs) is available for
feedback, as assumed in this paper, the gain cannot be higher than a
factor of three in energy (or $v_{\rm eject}^2$).  Things would change
if such SNe were more energetic than their low-redshift and
high-metallicity counterparts. Solution (ii) would be obtained by
assuming a different, more top-heavy IMF in these galaxies, while
solution (iii) would require that dark matter is not completely cold
and collisionless as assumed.

In all cases, a solution of this discrepancy, and a proper
reproduction of stellar mass and archaeological downsizing, requires
some deep change in the models, be it a more sophisticated feedback
scheme, more energetic SNe at high redshift, a varying IMF or a
different dark matter theory.  Moreover, this question is of great
interest because, due to their steep luminosity function, such
galaxies would contain most produced metals, so they are very
important contributors to metal pollution of the IGM.

While theoretical progress can be achieved by exploring the ideas
given above, observational breakthroughs will have to wait the next
generation of telescopes.  Indeed, while the photometric selection of
Lyman-break galaxies is feasible with the current facilities up to
magnitude $z_{850}\sim$28, even though on a relatively small field like GOODS,
the spectroscopic observations of relevant features like the break of
the continuum blueward the $Ly\alpha$ line and the detection of the
ultraviolet absorption lines is still challenging for sources fainter
than $z_{850}\sim25.5$.  Beyond this limit the main spectral feature is
the Lyman $\alpha$ emission line (if present), often observed without
continuum; only composite spectra allow to increase the signal-to-noise.
Future observations in the next decade with ground-based 30-40m class
telescopes (e.g. TMT, E-ELT), with ALMA and from the space with JWST
will be crucial to make an observational breakthrough;
one of the main goals of these
facilities is to perform spectroscopic surveys of faint sources,
reaching easily $z_{850}\sim27$. Not only the precise estimation of the
redshift will be possible, but the validation of the dropout technique
at that level and information on the total mass (from the stellar
velocity dispersion), mass distribution and rotation curve,
metallicity, dust content, galactic wind energetics and stellar
population distribution will be straightforward to derive from 30-40m
spectra of the galaxies already discovered in the 8-10m surveys. The
rich phenomenology revealed by these spectra will give us new insight
on the formation of galaxies, that will complement the theoretical
efforts made in the next years.

\section{Summary}
\label{s:conclusions}

We have compared the predictions of {\sc morgana}$+${\sc grasil}
\citep{Monaco07,Silva98,Fontanot07} with data of Lyman-break galaxies
at $z\sim4-6$.  Model predictions have been produced by simulating
deep ACS-HST fields, taking into account photometric errors 
as in the GOODS-S survey \citep{Giavalisco04},  and the effect of Lyman-alpha
emission line.  They have been compared to number counts of $B$- $V$-
and $i$-dropouts, scaled to the HUDF sensitivity by \cite{Bouwens07},
colour evolution, redshift distributions and selection function of
GOODS-S \citep{Vanzella09} and GOODS-MUSIC \citep{Grazian06} galaxies,
and luminosity functions estimated by \cite{Bouwens07}.

Model galaxies closely resemble the observed ones in terms of sizes,
stellar masses, colors, dark matter halo properties and spatial
distribution but have significantly higher metallicities. However
for such galaxies, observed at rest-frame UV wavelengths, dust
attenuation is as important as uncertain.  We find that reasonable
tuning of {\sc grasil} parameters, namely escape time $t_{\rm esc}$ of
stars from the parent molecular cloud and fraction $f_{\rm mol}$ of
gas in the molecular phase, can result in more than a magnitude
difference.  The two parameters are also degenerate, i.e. decreasing
the first has a similar effect as increasing the second and {\it vice
  versa}.  These parameters are also degenerate with model parameters
that regulate the amount of star formation in high-redshift halos.

Fixing the parameters it is possible to reproduce the bright end of
the number counts and luminosity functions, with 
a modest overestimate of the number of bright $B$-dropouts 
and a possible
underestimate of bright $i$-dropouts.  We then notice an
excess of faint ($z_{850}\ga 27$) star-forming galaxies, especially
when considering $V$-dropouts at $4.5<z<5.5$.  This excess is unlikely
to be due to sample incompleteness or to some peculiar behaviour of
{\sc grasil} dust attenuation. \cite{Fontanot09b}, found a serious
discrepancy between models of galaxy formation (including {\sc
  morgana}) and data for galaxies in the mass range $10^9-10^{10.5}$
{\msun} at $z=0$.  These form too early in the models, are too passive
at $z<3$ and host too old stellar populations at low redshift.  This
clearly implies an excess of small star-forming galaxies at high
redshift that forces modelers to suppress star formation at later
time, so as to recover the correct number of small galaxies at $z=0$.

We define a class of galaxies which are over-produced by the model.
These faint Lyman-break galaxies, forming stars at $\la 10$ {\msunyr},
have masses that grow in stellar mass from $z\sim6$, where their mass
is $\sim3\times10^8-3\times10^9$ {\msun}, to $z\sim4$, where their
mass has grown to $\sim10^9-10^{10}$ {\msun}.  They are hosted in
relatively massive dark matter halos, with masses in the range $10^{11}-10^{12}$
{\msun} and circular velocities in excess of 100 km s$^{-1}$.
These galaxies show a stellar mass -- gas metallicity relation which
is too high compared to the (sparse) available evidence. At $z=0$
roughly half of them are found in small galaxies, the other half
having evolved into more massive objects.

We conclude that some feedback mechanism should be at play, causing
star formation in these galaxies to be suppressed by strong gas and
metal outflows.  These metals would then be important polluters of the
IGM, and because this pollution takes place before the peak of cosmic
star formation, constraining this process is of great importance also
for IGM studies.  As originally shown by \cite{Dekel86}, the amount of
ejected mass mainly scales with the halo circular velocities, and this
is true for most semi-analytic galaxy formation models.  In this way,
under very simple assumptions, a suppression of these galaxies implies
a suppression of star formation in those $10^{12}$ {\msun} halos for
which the peak of efficiency of star formation is expected at $z=0$.
Solving this discrepancy requires either some form of feedback that,
surprisingly, is most effective at the highest gas densities, or some
more exotic solution, like higher SN energy in dense environments with
low metallicity, a top-heavy IMF or a not perfectly cold and
collisionless dark matter which suppresses such compact halos.
Observational breakthroughs able to clarify the phisical processes
responsible for the suppression of these ``excess'' galaxies may be
obtained with the future 30-40m optical telescopes, ALMA and JWST; a
strong theoretical and numerical effort on this topic should be
carried out in the next future to produce predictions that can be
tested by future observations.

\section*{Acknowledgments}

We thank Mario Nonino, Gabriella De Lucia, Adriano Fontana and Nicola
Menci for many discussions, Andrea Grazian for his help with the
GOODS-MUSIC sample. We acknowledge financial contribution from contract
ASI/COFIN I/016/07/0 and PRIN INAF 2007 ``A Deep VLT and LBT view of the Early
Universe''. Computations were carried out at the ``Calcolo
Intensivo'' unit of the University of Trieste, and partly at the
``Centro Interuniversitario del Nord-Est per il Calcolo Elettronico''
(CINECA, Bologna), with CPU time assigned under University of
Trieste/CINECA grants.

\bibliographystyle{mn2e}
\bibliography{morgana_lyb}

\bsp

\label{lastpage}
\end{document}